\documentclass{emulateapj}

\newcommand{\kms}{\hbox{km s$^{-1}$}}

\newcommand{\mdotyr}{\hbox{$M_\odot$ yr$^{-1}$}}

\shorttitle{Accretion onto very low-mass stars and brown dwarfs}
\shortauthors{Herczeg \& Hillenbrand}

\begin{document}

\title{UV excess measures of accretion onto young very low-mass stars and brown dwarfs}

\author{Gregory J. Herczeg\altaffilmark{1} \& Lynne A. Hillenbrand\altaffilmark{1}}

\altaffiltext{1}{Caltech, MC105-24, 1200 E. California Blvd., Pasadena, CA 91125}

\begin{abstract}
Low-resolution spectra from 3000--9000 \AA\ of young low-mass stars and
brown dwarfs were obtained with LRIS on {\it Keck I}.   The
excess UV and optical emission arising in the Balmer and Paschen continua  yields mass accretion rates ranging from $2\times10^{-12}$ to $10^{-8}$ \mdotyr.  These results are compared with {\it HST}/STIS spectra of roughly solar-mass accretors with accretion rates that range from $2\times10^{-10}$ to $5\times 10^{-8}$ $M_\odot$ yr$^{-1}$.  The weak photospheric
emission from M-dwarfs at $<4000$ \AA\ leads to a higher
contrast between the accretion and photospheric emission relative to higher-mass counterparts. 
The mass accretion rates measured here are systematically $\sim 4-7$ times larger than those from H$\alpha$ emission line profiles, with a difference that is consistent with but unlikely to be explained by the uncertainty in both methods.
 The accretion luminosity correlates well with many line luminosities, including high Balmer and many \ion{He}{1} lines.  Correlations of the accretion rate with H$\alpha$ 10\% width and line fluxes show a large amount of scatter.  Our results and previous accretion rate measurements suggest that $\dot{M}\propto M^{1.87\pm0.26}$ for accretors in the Taurus Molecular Cloud.
\end{abstract}
 

%
%


\section{INTRODUCTION}
In the magnetospheric accretion model for accretion onto young stars, the stellar
magnetosphere truncates the disk at a few stellar radii.  Gas
from the disk accretes onto the star along magnetic field lines and
shocks at the stellar surface.  The $\sim 10^4$ K optically-thick post-shock and optically-thin pre-shock gas produce emission in the Balmer and Paschen continuum and in many lines, including the Balmer and Paschen
series and the \ion{Ca}{2} IR triplet \citep{Har94,Cal98}.  The
accretion luminosity may be estimated from any of these accretion
tracers and subsequently converted into a mass accretion rate.

Historically, mass accretion rates onto young stars have been measured
directly using optical
veiling from the Paschen continuum and stronger Balmer continuum emission shortward of the Balmer jump.  Modelling of the Paschen and Balmer continuum emission with a generic isothermal, plane-parallel slab yields $\dot{M}\sim 10^{-7}-10^{-10}$ $M_\odot$ yr$^{-1}$ for accretors with masses of 0.4-1.0 $M_\odot$  \citep{Val93,Gul98}.
\citet{Cal98} obtained similar accretion rates by 
applying more realistic shock geometries with a range of temperatures to explain the observed
accretion continuum.  The accretion luminosity
can be measured instead at longer wavelengths with optical veiling measurements from high-dispersion spectra
\citep[e.g.][]{Bas90,Har91,Har95,Har03,Whi04}, albeit with a larger uncertainty in
bolometric correction.

Several deep surveys have since identified large populations of young
very low-mass stars and brown dwarfs with masses $<0.2$ $M_\odot$ 
\citep[e.g.][]{Bri98,Luh04,Sle06,Gui07}.  Excess IR emission indicates the presence of disks around some of these very low-mass objects$^2$ \citep[e.g.][]{Muz03,Jay03,Nat04}.
The excess Balmer continuum emission produced by accretion onto very low-mass stars and brown dwarfs has not
been observed previously. 
Instead, accretion has been been measured based on magnetospheric models of the
H$\alpha$ line profile (Muzerolle et al. 2000, 2003, 2005), by extending
relationships between $\dot{M}$ and line luminosities established at higher masses to brown
dwarf masses (Natta et al. 2004, 2006; Mohanty et al. 2005), or in a few cases from measurements of the Paschen continuum \citep{Whi03}.
\footnotetext[2]{The objects studied in this paper span the mass range from 0.024--0.2 $M_\odot$; hereafter all are referred to as ``stars'' for ease of terminology.  We refer to $\sim 0.3-1$ $M_\odot$ stars as ``higher-mass stars.''}

In this paper, we present the first observations of the Balmer
jump, and hence direct measurements of the accretion luminosity, for young very low-mass stars and brown dwarfs using data obtained with LRIS on {\it Keck I}.  We compare the Balmer and Paschen continuum emission from very low-mass stars to that from higher-mass accretors that were observed with {\it HST}.
The weak U-band photospheric emission of M-dwarfs leads to a high
contrast between accretion and photospheric emission at $\lambda<3700$
\AA.  The properties of the Balmer   
continuum and jump for very low-mass stars and brown dwarfs are mostly similar to those of the higher-mass accretors, although the sensitivity to low $\dot{M}$ improves to smaller stellar mass.  The Balmer jump tends to be larger for smaller mass accretion rates, indicating lower optical depth for the emitting gas.
The luminosity in many lines, including the Balmer series and \ion{He}{1} lines, correlate well with the accretion luminosity.
  Our $\dot{M}$ estimates are systematically $\sim 4-7$ times higher than those estimated from modelling the H$\alpha$ line profile.  We find that $\dot{M}\propto M^{1.9}$ for accretors in Taurus and discuss biases in the selection of those accretors that may affect this calculated relationship.

\section{OBSERVATIONS}

\subsection{{\it Keck I}/LRIS}

We used LRIS \citep{Oke95,McC98} on Keck I to obtain low-resolution optical spectra covering 3000--9000
\AA\ of 17 young, low-mass
Taurus sources and 2M1207-3932 on 23
November 2006.  We observed 2M0436+2351 and obtained a deeper observation of 2M1207-3932 on 7
February 2007. 
Our observation log (Table 1) lists the exposure time for each target and abbreviations for each target used throughout this paper.  

Most targets were observed with a 175$^{\prime\prime}\times1\farcs0$
slit.  
The light was split into separate red and blue beams with the D560
dichroic.  The red light illuminates a $2048\times2048$ CCD with $0\farcs21$
pixels and the blue light illuminates a
$2048\times4096$ CCD with $0\farcs135$ pixels.  We used the
400/8500 grism and the GG495 filter in the red channel and the 400/3400 grism in the blue channel.  Each red pixel covers $\sim1.85$ \AA, yielding a resolution
$R\sim1400$ over the 5400--9000 \AA\ wavelength range.  Each blue pixel covers  $\sim1.1$ \AA,
yielding a resolution $R\sim700$ over the 3000--5700 \AA\ wavelength range.  For each target we typically obtained  
multiple red spectra during a 1--2 long blue exposure.

We bias-subtracted the 23 November blue exposures and estimated a bias
subtraction for the 7 February 2007 blue exposure from images with no blue
emission.  An overscan subtraction was applied to all red images.  
The red images were flat-fielded with dome flats, the blue images from
November 2006 were flat-fielded with sky flats, and the blue images from February 2007 were flat-fielded
with halogen lamp flats.  We located each spectrum
on the detector as a function of position in the dispersion direction.  The counts in the red
and blue spectra were extracted with 11 and 13-pixel widths, respectively.  The sky background was measured from nearby windows and subtracted from the spectrum.  The spectral regions within 25 \AA\ of 4590 and 5270 \AA\ include a transient ghost and are ignored.

The relative wavelength solution was calculated from a HgNeArCdZn lamp
spectrum.  The absolute wavelength solution for each observation was
calculated by measuring the positions of sky emission lines.

All stars were observed at the parallactic angle.  On 23 November the seeing improved
from $0\farcs9$ to $0\farcs7$ from the beginning to the middle of the
night and remained stable at $0\farcs7$ for the second half of the night.
On 7 February the seeing was $0\farcs7$ during our observation of 2M0436+2351 and $1\farcs2$ during our observation of 2M1207-3932.

We obtained observations of the spectrophotometric standards  \citep{Mas88,Oke90}
PG0205+134 and PG0939+262 at the
beginning and end of the November 2006 run and G191B2B and PG1545+035 on February 7 2007 with the $1\farcs0$ 
aperture.    The November 23 standards
have published fluxes from 3100--8000 \AA.  We approximate the flux of PG0939+262 from 8000--9000
\AA\ by assuming that the white dwarf spectrum is a blackbody, and confirm this calibration by comparing the spectra
of 2M1207-3932 obtained on both nights.
The atmospheric extinction correction was obtained
from multiple observations of G191B2B on 7 February 2007.

Slit losses were $\sim70$\% at the beginning
of the night and $\sim22$\% at the end of the night in Nov. 2006.  
 In adjacent red spectra of the same target, the extinction-corrected count rate increased early in the night and remained stable during the second half of the night, mirroring similar changes in the seeing.  This increase in count rate can account entirely for the factor of $\sim 3$ in slit losses.  The count rates in the first half of the night are therefore multiplied by a factor between $1-3$, based on when the observation occurred.
  The small wavelength dependence in this factor is found by interpolating from the
observations of the photometric standards.

Table 1 lists synthetic UBVR magnitudes for each source and
our estimated uncertainty in absolute flux calibration.
The estimated photometry is within 0.1--0.5 mag of most existing
literature measurements, with the exception of several
strong accretors.   The initial flux calibration of 2M1207-3932 on the two nights differed by 0.27 mag.
 We estimate that the
absolute fluxes are calibrated to at worst a factor of 2 early in the
night and a factor of 1.3 in the second half of the night for the Nov. 2006 observations.

\subsection{Archival HST Spectra}
We supplement our new {\it Keck} data with existing blue {\it HST}/STIS spectra of late-K or early-M accretors and two young non-accretors taken from the MAST archive (see Table 2).  The spectra were obtained with the G430L grating and span 2900-5700 \AA\ with a spectral resolution of $\sim600$. Several of these observations were discussed in Herczeg et al.~(2004, 2005, 2006), including calculations of the mass accretion rate.  We re-analyze these data here to provide a small set of higher-mass accretors with spectra analyzed in a manner consistent with our {\it Keck I}/LRIS spectra to compare with results from our sample of low-mass accretors.  

\subsection{Description of our sample}
 Several targets (GM Tau, CIDA 1,
 2M0414+2811) in our {\it Keck I}/LRIS program were selected based on expectation of a high mass accretion rate inferred from strong excess continuum emission and optical line emission.  Other
 targets were selected based on a combination of spectral type, low extinction, existing
 $\dot{M}$ measurements from H$\alpha$, and visual magnitude.  MHO 5 and
 KPNO 11 were selected also for a full-width of the H$\alpha$ line at
 10\% the peak flux (hereafter H$\alpha$ 10\% width) of 150--200
 \kms\ \citep{Muz03,Moh05}, which is intermediate between definite accretors and definite non-accretors for their respective spectral types.
Three relatively-bright young stars in Taurus were chosen as photospheric templates based on spectral type and lack of strong accretion, as identified by  by weak H$\alpha$ emission and absence of IR excess emission.  Since the template stars are young, their surface gravity will be similar to the rest of the sample and provide a better match to the photospheric spectrum than could be obtained from older field dwarfs.   

To obtain a larger range in $\dot{M}$ for analyzing the properties of the accretion continuum and relationships between line and continuum emission (see \S5 and 6), we supplemented the {\it Keck I}/LRIS spectra with {\it HST}/STIS observations of stars with higher central masses.  This sample consists of most available {\it HST} observations of the 3000--5000 \AA\ region for accreting young stars with spectral types late K or early M.

Table 3 describes the properties of these targets (see also \S 9).  Multiple
 LRIS observations of 2M1207-3932 and {\it HST} observations of DF Tau and TW Hya are treated independently throughout this paper.

\section{Observations of Accretion Diagnostics}
Figure 1 compares three M5-M5.5 stars in our {\it Keck}/LRIS spectra from 3200--5600 \AA.  Strong TiO and CaH molecular bands \citep{Kirk93} dominate the photospheric emission of mid-M dwarfs.  At $>3700$ \AA, such features are strong from V927 Tau and S0439+2336 but shallow from S0518+2327.  This veiling of the photospheric emission from S0518+2327 is attributed to excess Paschen continuum emission.
S0439+2336 and S0518+2327 both exhibit strong excess emission shortward of jump at 3700 \AA, which is attributed to excess Balmer continuum emission.  Line emission, including in the Balmer series, is much stronger from S0439+2336 and S0518+2327 than from V927 Tau.  The excess Balmer and Paschen continuum and emission lines detected from young stars are commonly attributed to accretion.  In the following subsections we describe and measure these accretion diagnostics for each source.

\subsection{Measuring the Paschen Continuum}
Veiling$^3$ is measured by comparing the depth of absorption
features, which are filled in by emission from the accretion continuum, to photospheric templates.  These features include the TiO and CaH absorption bands and also the strong \ion{Ca}{1} $\lambda4226$ absorption line.  The \ion{Ca}{1} line depth may be related to chromospheric structure and may be affected by stellar rotation \citep{Mau96,Sho97}.
\footnotetext[3]{The veiling at wavelength $\lambda$, $r_\lambda$, is defined by the accretion continuum flux ($F_{observed}-F_{photosphere}$) divided by the photospheric flux.}

The young  M-dwarfs V927 Tau (M5), MHO 7 (M5.5), CFHT 7 (M6.5) and 2M0439+2544 (M7.25) serve as photospheric templates (see \S 9.1).   These stars provide a better match to the surface gravity and chromospheric emission of young accretors than would older field M-dwarfs.  2M0439+2544 is not an ideal template because of ongoing accretion, and V927 Tau may also be accreting (see \S3.4 and \S 9.4.2).  Use of 2M0439+2544 and V927 Tau as templates may cause us to underestimate the veiling by 0.05 and 0.02, respectively, at 5000 \AA\ and 0.01 each at 7000 \AA.  At $<4500$ \AA\ we use CFHT 7 instead of 2M0439+2544 as a photospheric template for late M-stars. 

Large veilings ($\gtrsim2$) are difficult to measure because the high ratio
of accretion-to-photospheric flux masks any photospheric
absorption features.  However, a large error for a high value of veiling leads to only a small error in the measurement of the underlying continuum flux.  Our lower detection limit of veiling measurements is $\sim 0.05-0.1$, depending on how well the photospheric template matches the spectrum.
The veiling is calculated from
regions that do not include obvious strong emission lines.  For weaker accretors, any such lines may fill in some of the absorption and thereby artificially increase our veiling estimate for some stars, particularly at 4227 \AA\ (see Fig. 2).  

Figures 2--4 show three spectral regions where veiling is measured.
Table 4 lists  veiling measurements and the corresponding flux in the
accretion continuum at distinct wavelengths between 4000-9000 \AA.  
Excess continuum emission is detected longward of 4000 \AA\ for many stars in our sample and is identified as the Paschen continuum.  
The accretion continuum is relatively flat at $<6000$ \AA\ for most targets, and increases at $\lambda>6000$ \AA\ for most sources where it is detected.   The continuum flux from 2M0414+2811 at 8600 \AA\ is substantially lower than that at 7800 \AA, which may be consistent with weaker Brackett continuum emission longward of the Paschen jump.  CIDA 1 shows no such flux reduction, which may suggest an alternate veiling source.  

The early-M and late-K stars from the STIS data do not have the strong TiO and CaH absorption bands for veiling measurements.  In several locations, the photospheric emission drops by a factor of 2--2.5 within $\sim 200$ \AA.  Table 5 list veilings at 4600 and 5400 \AA\ for this sample, measured by comparing the depth of these features with photospheric templates, and at 4000 \AA, measured from the slope of the photospheric emission.  We detect Paschen continuum emission from DF Tau, TW Hya, V836 Tau, DG Tau, and RU Lup.

\subsection{Measuring the Balmer Continuum}
When seen from young stars, excess Balmer continuum emission 
shortward of $\sim 3700$ \AA\ is typically
attributed to accretion (Figure 5).  The Balmer limit occurs at
 $3646$ \AA, but line blending in the Balmer series shifts the
apparent jump to 3700 \AA.  The observed Balmer jump, 
defined here as the ratio of flux at
3600 \AA\ to the flux at 4000 \AA, ranges from $\sim 0.35$ for the non-accretors to
5.47 for the accretor 2M1207-3932 (Table 6).

Figure 6 compares spectra of stars with the weakest Balmer jumps in our sample to that of photospheric templates from Table 4.
Excess Balmer continuum 
emission is clearly detected from 2M0455+3028, S0439+2336, and V927 Tau relative to
photospheric templates.
Some weak, noisy Balmer continuum emission is also detected
from MHO 5 and KPNO 11.  MHO 7 and CFHT 7 may each
show very mild increases in emission shortward of the Balmer jump, 
possibly from chromospheric emission (see \S 3.4).  

As defined here, the Balmer jump is the ratio of Balmer plus Paschen continuum flux at 3600 \AA\ divided by the Paschen continuum flux at 4000 \AA.  The intrinsic Balmer Jump ($J_{mod}$ in Table 6) is measured from the accretion continuum after dereddening the spectra (see \S 9.2) and subtracting a scaled photospheric spectrum.    Table 6 also lists the measured slope of the continuum emission from 3200--3650 \AA\ ($S_B$), scaled to the observed flux at 3600 \AA.

\subsection{Emission Line Measurements}
Tables 7--10 list equivalent widths and fluxes for H Balmer
lines, He lines, other strong lines associated with accretion, and lines usually associated with outflows.$^4$  These tables are not complete linelists.
We measure most emission line fluxes and equivalent widths by fitting
a Gaussian profile to the observed line.  The focus was poor at $\lambda>8000$ \AA\ and produced asymmetric emission lines.  Line strengths in that spectral region are measured by summing the flux across a continuum-subtracted window that includes the entire line. 
 The Balmer and \ion{Ca}{2} H \& K lines are bright from every source and are measured directly
from the spectra.  For all other lines we subtract a scaled 
photospheric template to increase our
sensitivity to small fluxes.  These line equivalent widths and corresponding fluxes are relative to the photospheric template and may be offset by a small amount in equivalent width if line emission is present but undetected in the photospheric template.  This method may also yield false weak detections if the depth of \ion{Na}{1} or \ion{Ca}{2} photospheric absorption features differ substantially from star-to-star.
 Weak \ion{Ca}{2} IR triplet emission
from several targets is only treated as real if all three of the lines are detected, unless the non-detection has a large upper limit on equivalent width. 
Our detection limits typically range from
$\sim 0.1-1.5$ \AA, depending on the match with a
photospheric template and the signal-to-noise.  The equivalent widths in low-resolution spectra are affected by absorption and may differ from equivalent widths measured in high-resolution spectra because of different spectral regions used to calculate the continuum emission.

The sources with strongest veiling also show the strongest line emission.
For example, \ion{Ca}{2} IR triplet emission is detected from only CIDA 1,
2M0414+2811, GM Tau, 2M0441+2534, and S0518+2327.    Similarly, the Paschen series is detected
from only CIDA 1, 2M0414+2811, and GM Tau.  Several Paschen lines blend with and contribute $\leq 10\%$ of the flux in the \ion{Ca}{2} IR triplet lines.
Other lines, including the
\ion{Na}{1} D doublet, \ion{He}{2} $\lambda4686$, and \ion{He}{1} lines are
detected from most stars with excess Balmer emission. 

Optical forbidden lines, including the [\ion{O}{1}]
$\lambda6300,6363$, [\ion{S}{2}] $\lambda6716,6731$, and [\ion{S}{1}] $\lambda4069,4076$ doublets, are detected
from many sources.  [\ion{N}{2}] $\lambda6584$ is detected from 2M0444+2612 but is difficult to resolve from the bright H$\alpha$ emission for many other stars in the sample.  The [\ion{O}{1}] lines are unresolved in the cross-dispersion direction, which constrains the emission to have a FWHM of $<0.2-0.3$.

\subsection{Designation of Accretors from the Balmer continuum}
Many emission line diagnostics have been used to identify accretion onto young very low-mass stars and brown dwarfs.  An H$\alpha$ equivalent width of $>10$ \AA\ is the traditional criterion above which a young star is classified as an accretor.  Against the weaker photospheric continua of redder M-dwarfs, chromospheric emission can produced H$\alpha$ emission with equivalent widths larger than 10 \AA.  In lieu of this diagnostic, the full-width at 10\% the peak emission in the H$\alpha$ line has been used to diagnose accretion, with values greater than 180--200 \kms\ indicating accretion for very low-mass stars \citep[e.g.][]{Whi03,Muz03,Nat04,Moh05}.  Accretion onto low-mass stars has also been identified by emission in permitted optical lines.   Optical forbidden lines are produced by outflows and can therefore be used as indirect accretion diagnostics because outflows require ongoing or recent accretion \citep[e.g.][]{Ham94,Har95}.  The presence of excess Balmer and Paschen continuum emission is also associated with accretion \citep[e.g.][]{Val93,Gul98}.  In this subsection we describe the Balmer jump and excess Balmer continuum emission as an accretion diagnostic for the individual stars in our sample.  In general, the classification of stars as accretors or non-accretors based on the presence or absence of Balmer continuum emission is consistent with previous classifications using alternate accretion diagnostics.

Two mid-M dwarfs in our sample, CFHT 7 and MHO 7, were previously classified as non-accretors.  For CFHT 7, \citet{Gui07} measured only weak H$\alpha$ emission and found an absence of excess IR emission shortward of and including the {\it Spitzer} 24$\mu$m MIPS bandpass.  
For MHO 7, \citet{Muz03} measured weak H$\alpha$ emission with a 10\% width of
115 \kms\ and no excess K-band emission.  These two stars have Balmer jumps of 0.36 and 0.34, respectively.  Both stars are chromospherically active, with equivalent widths in \ion{Ca}{2} H \& K and \ion{He}{1} lines that are larger than is typically measured from chromospheres of older, magnetically-active M-dwarfs \citep[e.g.][]{Gia82,Giz02a,Rau06,All06}.  The earlier-type non-accretors LkCa 7 and V819 Tau are bluer and have Balmer jumps of 0.44 and 0.46, respectively.

Many stars in our LRIS sample have much larger observed Balmer jumps than those non-accretors. The stars GM Tau, CIDA 1, 2M0414+2811, 2M0441+2534, and S0518+2327 have Balmer jumps ranging from 2.0--3.4 and excess Paschen continuum emission detected at $>6000$ \AA.  These stars are also associated with strong emission in the Balmer series, \ion{He}{1} lines, the \ion{Ca}{2} IR triplet, \ion{O}{1} $\lambda7773,8446$, and many forbidden lines

The stars MHO 6, 2M0439+2544, 2M0444+2512, S0439+2336, 2M0436+2351, and 2M1207-3932 have excess Paschen continuum emission detected at 4227 \AA\ but not at $>5500$ \AA, and five have Balmer jumps between 0.7--1.4.  The two observations of 2M1207-3932 yield outlier Balmer jumps of 3.9 and 5.7.  These six targets are associated with some strong permitted line emission, including bright Balmer continuum emission and \ion{He}{1} line equivalent widths of $>1$ \AA, and some forbidden line emission.  Any \ion{Ca}{2} IR triplet emission is faint.

Each of these stars can be identified as accretors based on their high Balmer jump.  Similarly, the higher-mass stars TW Hya, DF Tau, RU Lupi, and DG Tau have Balmer jumps that are well above that of non-accretors and can be classified as accretors based on either the Balmer jump or the presence of bright emission lines.  As with other diagnostics, small Balmer jumps with values near those of non-accreting templates may be inconclusive for classifying stars as accretors.
V836 Tau has an observed Balmer jump of 0.60, which is higher than non-accretors with similar spectral type.  Many young late K and early M stars that are not identified with accretion show no signs of excess Balmer continuum emission.  Thus, this excess emission from V836 Tau is identified as a clear indication of accretion.

MHO 5, 2M0455+3028, CIDA 14, KPNO 11, and V927 Tau all have observed Balmer jumps consistent with or slightly above the non-accretors.  Figure 6 shows that some excess Balmer emission relative to photospheric templates is present from each of these stars, with veilings at 3600 \AA\ that range from 0.14-0.54 (see Table 11).  Emission line diagnostics suggest that MHO 5, CIDA 14, and V927 Tau are actively accreting (see \S9.4), while the status of 2M0455+3028 and KPNO 11 is uncertain from other diagnostics.  

The weak excess Balmer continuum emission from these five stars could in principle be produced by the stellar chromosphere.    Younger M-dwarfs are bluer and more chromospherically active than older M-dwarfs \citep{Card07}, which is consistent with enhanced Balmer and Paschen continuum emission.   However, the Balmer jump and elevated Balmer continuum emission is not detected during stellar flares \citep{Haw91,Eas92}, which may suggest that stellar chromospheres do not produce any Balmer continuum emission.  In this case, any weak excess emission could be attributed to accretion.  A larger sample of U-band spectra from non-accreting young M-dwarfs is needed to definitively rule out a chromospheric origin. 
For the remainder of the paper we treat the excess emission from these five stars as upper limits.  We also suggest the criterion that any observed Balmer jump of $>0.5$ for a mid-M dwarf should be considered an accretor.

\section{Mass Accretion Rates}

\subsection{Description of Accretion Models}
The accretion emission consists of a strong Balmer continuum, a weaker
Paschen continuum, and many lines.  The flux in the accretion
continuum is measured for the entire 3000--9000 \AA\ wavelength range from only the strongest three accretors (see Table 4).  For the weakest accretors the veiling is  measurable only shortward of the Balmer jump.  Calculating the
total luminosity of the accretion shock requires modelling the
observed accretion SED to obtain a bolometric correction that accounts
for unseen emission.

The H Balmer and Paschen continua and series can
can be approximated by an isothermal,
plane-parallel, 
pure hydrogen slab model following \citet{Val93} and \citet{Gul98}.
These simplistic models were developed based on the outdated
boundary layer accretion paradigm but are robust to different geometries for
the emitting gas.
  We use these isothermal plane-parallel model from \citet{Val93}
because it does not rely on a specific morphology of the
accretion flow and it provides measurements of the accretion luminosity 
calculated with a methodology consistent with
\citet{Val93} and \citet{Gul98}.

The accretion slab models of \citet{Val93} approximate the broadband hydrogen accretion
continuum with the following free parameters: temperature $T$, density $n_e$, path length $L$ of the emitting gas (an
opacity parameter), turbulent velocity, and
filling factor $\delta$ of the accretion slab.  At low density ($n_e<10^{13}$
cm$^{-3}$) the  Balmer jump is large and determined by the temperature
of the emitting gas.  At higher density stronger H$^{-}$ continuum emission  reduces the size of the Balmer jump and makes the continuum shape at $<3700$ \AA\ bluer.  The 
Balmer and Paschen continua become redder and the size of the Balmer jump increases with decreasing temperature, while both continua become bluer and the size of the Balmer jump decreases with increasing optical depth ($n_e$ and $L$). 
 In our fits we vary the temperature, density, and path length while setting the turbulent velocity to $\sim 120$ \kms, which 
is smaller than the 150 \kms\ used by
\citet{Val93}.  Larger turbulent velocities yield Balmer line profiles that are
broader than is observed.  Substantially 
smaller turbulent velocities are unable to reproduce
the shoulder of the Balmer jump, located between the real Balmer jump at 3646 \AA\ and the observed Balmer jump at $\sim 3700$ \AA, that is attributed to blending of high
Balmer lines.
The synthetic spectrum is then scaled to the measured flux accretion continuum.  The filling factor of the emission on the stellar surface is
 calculated from the accretion luminosity and temperature.

\citet{Cal98} developed more advanced
models of the pre- and post-shock gas at the footpoint
of the magnetospheric accretion column on the star.  In their models the Balmer continuum is produced in the optically-thin pre-shock gas while the Paschen continuum is produced in the optically-thick post-shock gas.
The optically-thin pre-shock gas has an intrinsically large Balmer jump.  The optically-thick post-shock region has a high density and small Balmer jump.  The observed Balmer jump is therefore not density-sensitive but instead depends on the relative amount and temperature of the pre- and post-shock gas.
Our primary goal for modelling the measured accretion continuum is estimating
bolometric corrections to convert the continuum flux to an accretion
luminosity.  \citet{Cal98} find that the two model variants lead to 2--5\% differences in the slope of the Balmer continuum and 5--10\% differences in the slope of the Paschen continua, so that the continuum emission from shock models is redder.  However, our models and the models of \citet{Cal98} both successfully explain continuum emission at $<5600$ \AA\ and both underestimate continuum emission at $>6000$ \AA\ (see \S 4.2).  We therefore suggest that any differences between the two models in the bolometric correction are modest.

\subsection{Application of Models to the Observed Spectra}
We fit the synthetic accretion spectrum plus a photospheric template to
the unreddened blue spectra between 
3200--5600 \AA.  The photospheric template is scaled to match our
estimate for photospheric emission between 3200--5600 \AA, which is
informed by veiling estimates (see Tables 4--5).

The stars with measurable accretion continuum emission longward of the
Balmer jump allow us to constrain some parameters of the plane
parallel slab.
Figure 7 shows that the slope of the Balmer continuum constrains $T$ while the Balmer jump constrains $n_e$.  Values of $T$ and $n_e$ for each star are adopted from the center of the acceptable contours, with $L$ is subsequently tuned to match the strength of the high Balmer lines.  Since $L$ and $n_e$ are not independent, a lower $L$ may be consistent with a larger $n_e$.  In several cases $L$ could not be tuned to yield the observed flux in the high Balmer lines.
In the context of the \citet{Cal98} shock models, the larger Balmer jump may be caused instead by fainter Paschen continuum emission from the optically-thick post-shock gas relative to the Balmer continuum from the optically-thin pre-shock gas.

The lower Balmer lines are easily affected
by emission or absorption in the wind and are consistently 
underestimated by our models.  In magnetospheric accretion models of Muzerolle et al.~(2001) and Kurosawa et al.~(2006), emission in the Balmer lines is dominated by the funnel flow.  If the high Balmer lines are also produed in the accretion flow, then the slab optical depths in our models should be larger.
The slab optical depth may be smaller if instead the wind is optically thick and absorbs a large percentage of 
emission in all of the Balmer series lines.

The stars S0439+2336, 2M0455+3028, MHO 5, CIDA 14, KPNO 11, and V836 Tau have weak Balmer continuum emission and
undetected Paschen continuum emission.  Since the size of
the intrinsic Balmer jump for these targets is uncertain, we assume values for $n_e$, $T$, and $l$ based on
our other data.  We adjust the parameters to match the synthetic
Balmer series and scale the emission so that the spectrum of a
photospheric template plus accretion matches the total emission spectrum.

Table 11 lists approximate accretion slab parameters for each star, the
estimated Balmer jump of the accretion continuum, the slope of the Balmer continuum, the veiling at 3600 and 4000 \AA\ measured from these fits, 
and the total 
accretion slab luminosity.   
The continuum emission from each target is well
characterized by temperatures of $\sim 7000-9500$
K, densities of $(0.3-2)\times10^{14}$ cm$^{-3}$ and lengths of $\sim (0.5-50)\times
10^{7}$ cm.  In Figure 8 we show our fits to each spectrum.  The synthetic hydrogen continuum is consistent with veiling measurements  at $<5600$ \AA\ but 
underestimates the veiling at $>6000$ \AA\ from 2M0414+2811, 2M0441+2534, CIDA 1, GM Tau, and S0518+2327 by a factor or 1.2--2.  The intrinsic Balmer jump of 2M1207-3932 ($15\pm4$ and $7\pm2$ for our two observations) is particularly large, which indicates a low H$^-$ opacity.  A few outliers in the \citet{Val93} and \citet{Gul98}
data also have large Balmer jumps.

\subsection{Calculating Mass Accretion Rates and Errors}
The accretion luminosity, $L_{acc}$, can be converted to the mass
accretion rate, $\dot{M}$,
by assuming that the accretion energy is reprocessed entirely into the
accretion continuum. 
The mass accretion rate is then
\begin{equation}
\dot{M}=\left(1-\frac{R_*}{R_{in}}\right)^{-1}\frac{L_{acc}R_*}{G M_*}\sim 1.25 \frac{L_{acc}R_*}{G M_*}
\end{equation}
%
where the factor $(1-\frac{R_*}{R_{in}})^{-1}\sim 1.25$ is estimated by assuming the
accreting gas falls onto the star from the truncation radius of the
disk, $R_{in}\sim 5$ $R_*$ \citep{Gul98}.  The stellar mass and radius are listed in Table 3 (see also \S 9.3).  We
calculate mass accretion rates between $2\times 10^{-12}$ to $10^{-8}$ $M_\odot$ yr$^{-1}$ for the very low-mass stars and brown dwarfs in our {\it Keck}/LRIS sample and $2\times 10^{-10}-5\times 10^{-8}$  $M_\odot$ yr$^{-1}$ in the {\it HST}/STIS sample of higher-mass stars (Table 11).

The parameters in Equation 1 suffer from systematic and random uncertainties related to the data itself, interpretation of the data, and geometrical assumptions.  In the following subsections and in Table 12, we describe for purposes of explicit clarity these sources of uncertainty and how they affect our measurements of $\dot{M}$.

\subsubsection{Errors in Assumed Stellar Properties}
Distance enters into calculations of $\dot{M}$ (Eqn. 1) because $L_{acc}\propto d^2$ and $R\propto d$, yielding $\dot{M}\propto d^3$.  
Based on the kinematic distances calculated by \citet{Ber06}, the
distance to any individual Taurus member can differ by $\sim
15$ pc from the standard 140 pc distance \citep{Ken94}$^5$.
The $\sim 11$\% distance uncertainty in Taurus
leads to a 0.13 dex in $\dot{M}$.
The actual deviation in distance may
not be random for our Taurus sample because the selection of targets here and 
in previous work is influenced by
the optical brightness.  Targets located in front of the cloud will be brighter 
because of a smaller distance modulus and lower extinctions.

Extinction also affects measures of $L_{acc}$ and $R$.  The adopted 0.5 mag 
uncertainty in $A_V$ (see \S 9.2) leads to a 0.32 dex uncertainty in $L_{acc}$ and a 0.08 dex 
uncertainty in $R$ for a total of 0.4 dex uncertainty in $\dot{M}$.  If $R_V>3.1$, 
then the extinction curve would be flatter and our $A_V$ would be underestimated.

The $\dot{M}$ measurements also depend on the ratio $\frac{R}{M}$,
which is calculated from the effective temperature and theoretical 
evolutionary tracks (\S 9.3).  
The temperature is uncertain by $\sim 75$ K, including both the
uncertainty of roughly 0.5 spectral type subclasses and the conversion of spectral
type to temperature (\S 9.1).  In the Baraffe et al. (1998)
pre-main sequence evolutionary tracks at $2$ Myr and with
temperatures from \citet{Luh03}, a difference of 0.5 spectral classes
corresponds to $25\%$ larger mass and a $10\%$ smaller radius.  
Alternate pre-main sequence evolutionary tracks from
\citet{Dan98} would increase our mass estimates 1.1--1.7 times larger than the \citet{Bar98} tracks.  \citet{Whi04} suggest that evolutionary tracks underestimate stellar masses by 30--50\%.

All sources except V927 Tau and DF Tau are assumed to be single.  If any sources are 
binaries, the lower luminosity per star would imply that we have overestimated $R$.  
The shallower potential well would result in $\dot{M}$ being 1.4 times lower than is 
calculated here.   In cases where both sources are accretors, the accretion 
luminosity and mass accretion rate is the total rate onto both stars.

\subsubsection{Uncertainty in Bolometric Corrections}
Converting the measured accretion flux into an accretion luminosity requires a bolometric correction for unseen emission (\S 4.2).  However, the shape of the broadband accretion continuum is not well understood.  
Our single-temperature slab models underestimate the veiling at $6000-9000$ \AA\ by factors of 1.2--2, as which also occurs for the multi-temperature shock models \citep{Cal98}.
The veiling at even longer wavelengths \citep[e.g.,][]{Whi04,Edw06} also seems larger than can be explained by the isothermal slab that fits well the Balmer continuum and short-wavelength Paschen continuum.

If we hypothetically double the accretion continuum flux at $\lambda>6000$ \AA, the total accretion luminosity increases by $0.15$ dex.  If instead
the flux per \AA\ is constant between 6000--20000 \AA\ (despite the
presumed presence of the Paschen jump), then the accretion luminosity increases
by $0.3$ dex.  A smaller uncertainty of $\sim 0.1$ dex in $L_{acc}$ is introduced by the uncertainty in temperature ($\sim 1000-2000$ K) of the plane-parallel slab.

\subsubsection{Exclusion of line emission}
Our $\dot{M}$ estimates rely on accurate calculations of the total accretion energy that escapes from the shock.
Such emission is seen in 
both continuum and line emission, but we include only the total
continuum emission from the accretion slab in estimating $L_{acc}$.  The observed Balmer line fluxes (Table 7) total 0.2 times $L_{acc}$ for most of the accretors but are roughly equivalent to $L_{acc}$ for the M7-M8 accretors (see Tables 7 and 11).
The formation of and flux in lower Balmer
lines is complicated by emission and absorption in the accretion
funnel flow and in outflows.  
In principle other lines (e.g., \ion{Ca}{2} IR triplet, Lyman
series, FUV and X-ray lines) may also need to be included.  For TW Hya, X-ray and UV lines account for $\sim 30-45$\% of the total accretion flux, most of which is in \ion{H}{1} Ly$\alpha$, but most have
fluxes that when summed are likely $<20\%$ of the accretion flux \citep{Kas02,Her04}.   Our exclusion of line emission from the accretion luminosity is consistent with previous $L_{acc}$ estimates and is essential for simplifying these calculations and  comparisons with existing $\dot{M}$ estimates.

\subsubsection{Geometric assumptions}
In the magnetospheric accretion paradigm, gas accretes from the inner disk truncation radius, $R_{in}$, onto the star along magnetic field lines, where $R_{in}$ is determined by where the magnetic field intercepts the disk.
\citet{Gul98} noted that infall energy depends on $R_{in}$ and therefore included the factor of $(1-\frac{R_*}{R_{in}})$ in Equation 1.  The $R_{in}$ ranges from 3--10 $R_*$ based on magnetic field strengths of higher-mass CTTSs \citep{Joh07}.  We assume $R_{in}=5 R_*$ for all targets in this sample for consistency with the same assumption made by \citet{Gul98}.  A factor of $2$ error in $\frac{R_*}{R_{in}}$
leads to a $\sim 0.05-0.12$ dex error in the mass accretion rate.

As the gas accretes along the magnetic field lines and shocks at or near the stellar photosphere, some of the emission from the hot spot will be directed at the star.  We follow \citet{Val93} and \citet{Gul98} in assuming that all of the accretion energy escapes from the star as hydrogen continuum emission.  By this assumption, any accretion emission directed at the star is either coherently re-radiated by the star or provides some heating to the nearby gas, which then cools by hydrogen continuum emission.
\citet{Har95} and \citet{Har03} instead assumed that half of the Balmer and Paschen continuum emission is absorbed by the star, which increases the calculated $\dot{M}$ by a factor of two.  The accretion luminosity may also not be isotropic.

The accretion hot spot is assumed to be the source for all Balmer and Paschen continuum emission from our sample.  No excess emission is seen in a sample of 
non-accreting K-dwarfs that serve as templates for the higher-mass accretors 
discussed here.  Thus, the weak excess Balmer emission from V836 Tau that is attributable to accretion can be accurately measured.  Whether the faint excess Balmer emission from 
2M0455+3028, CIDA 14, MHO 5, KPNO 11, and V927 Tau is produced by accretion or may be chromospheric 
is uncertain.  Using V927 Tau as a template may lead us to underestimate the upper limit of $\dot{M}$ 
onto 2M0455+3028 by 0.2 dex.  Weak Balmer continuum emission in the template spectrum has a negligible effect for targets with larger $L_{acc}/L_{bol}$.

\subsubsection{Cumulative Effect of Uncertainties on Estimates of Mass Accretion Rates}
The uncertainty in bolometric correction, exclusion of emission lines, and the re-radiation of H Paschen and Balmer continuum emission by the star can only increase the estimated accretion rate, while the uncertainty in binarity and pre-main sequence tracks can only decrease the estimated accretion rate.  
When combined, these errors suggest that our accretion rates have a relative uncertainty of $\sim 0.6$ dex and underestimate the accretion rate by a factor of 1.2 (or 2.4, if one assumes no re-radiation of the accretion luminosity by the central star).

  This error analysis is not a rigorous assessment of the uncertainty in $\dot{M}$ at a specific confidence interval.  The uncertainty differs for various targets depending on the accuracy of stellar parameters, and for many targets the error is dominated by uncertainty in $A_V$.  For 2M1207-3932, which has an accurate $A_V=0$, the $\dot{M}$ is likely underestimated by a factor of 2 (or 4, if the accretion emission is not re-radiated), primarily because we exclude line emission in calculating $\dot{M}$.  The standard deviation of $\sim0.3$ dex about this factor of 2 is dominated by systematic uncertainties in the bolometric correction.  Systematic uncertainites of this magnitude are also present in all previous UV-excess or optical veiling estimates of $\dot{M}$.  
Several uncertainties, including in the extinction measurements, could be reduced with the addition of near-UV spectra, better photospheric templates across the entire wavelength region, a better understanding of veiling at long wavelengths.

Intrinsic variability in accretion rate can also affect non-simultaneous obserations.  \citet{Val03} describe variability of near-UV emission in repeated {\it IUE} spectra of young stars.  The average near-UV variability is 29\% for the 15 K1-M2 accretors that were observed at least 5 times with {\it IUE}.  The near-UV emission from these stars is dominated by and should correlate directly with the accretion rate \citep{Gul00}.  Variability of 0.1--0.2 dex is also commonly seen in repeated veiling measurements of DF Tau, TW Hya, and AA Tau \citep{Joh97,Bat02,Bou07}.  \citet{Scho06} confirm that variability in accretion rate is also seen for brown dwarfs.  We infer that the accretion rate is variable by $\sim 0.2$ dex.

\section{Relationships between line and continuum accretion diagnostics}
The Balmer continuum can be difficult to detect because of large extinctions or instrument limitations.  Other accretion diagnostics are therefore usually more accessible and can be used to infer accretion rates.  In the following subsections we derive and evaluate empirical relationships between emission lines and accretion luminosities and rates.  In \S 5.1 we calculate accretion rates from published data to incorporate additional objects in our analysis in \S 5.2--5.4.  We then compare the accretion luminosity and resulting accretion rate with H$\alpha$ line profile measures of accretion (\S 5.2), line luminosities (\S 5.3), and line surface fluxes (\S 5.4).

\subsection{Calculating accretion luminosity from optical veiling measurements}

Although ours are the first Balmer jump measurements of young brown dwarfs, detections of excess Paschen continuum emission have been published.  Calculating $\dot{M}$ from these objects increase the overlap of $\dot{M}$ measured using excess continuum emission and H$\alpha$ line profile modelling (\S 5.2).
\citet{Kra06} use broadband {\it HST} photometry to measure excess V-band emission from several young brown dwarfs.  The V-band excess is significant  when $r_V>1.3$, which restricts usage to high mass accretion rates relative to the given stellar mass.   We use $\log L_{acc}=-29.59+\log F_{5450}$ (calculated from Tables 4 and 11) to measure $\dot{M}$ of $10^{-9}-10^{-11}$ \mdotyr\ for GM Tau, KPNO-Tau 4, KPNO-Tau 6, and KPNO-Tau 12 (Table 13).  The $\dot{M}$ from GM Tau based on V-band excess is nearly identical to that measured more directly from our low-resolution optical spectra.  However, the true $\dot{M}$ was likely 20\% larger during the ${\it HST}$ observation because the V-band excess was $\sim 0.2$ mag brighter than was estimated here.

\citet{Har03} analyzed {\it HST}/STIS spectra of binaries with early M spectral types and masses mostly within the $\sim 0.25-0.55$ $M_\odot$ range, including veiling measurements at 6100 \AA\ and luminosities in [\ion{O}{1}] $\lambda6300$, H$\alpha$, and \ion{Ca}{2} $\lambda8542$ lines.  
   Using $\log L_{acc}=-29.67+\log F_{6190}$ calculated empirically from our {\it Keck I}/LRIS spectra (see Tables 4 and 11), we find  $L_{acc}$ that are $0.28-0.75$ times those calculated by \citet{Har03}.  We adopt these lower $L_{acc}$ for further analysis.

\subsection{Comparing UV-excess Accretion Rates to H$\alpha$ Line Profile Models}

Strong H$\alpha$ emission is usually attributed to accretion.  At high-resolution the line profile often includes P-Cygni, inverse P-Cygni, or self-reversed absorption that corrupts equivalent width measurements \citep[e.g.][]{Rei93,Muz03}.  Muzerolle et al.~(1998, 2001) applied magnetospheric accretion models to calculate the formation and radiative transfer of H$\alpha$ lines.  These models treat the line profile, scaled by stellar parameters, as a proxy for $\dot{M}$.  Muzerolle et al.~(2003, 2005) applied these models to very low-mass stars and brown dwarfs to estimate accretion rates.  Figure 9a and Table 14 compare mass accretion rates for
stars with $\dot{M}$ measured from both H$\alpha$ line profile models and UV-excess measurements.  

The UV excess accretion rate measurements are $\sim 0.84$ dex larger than the H$\alpha$ accretion rate measurements with a standard deviation of 0.3 dex, if MHO 5 and CIDA 14 are ignored.  If both the H$\alpha$ and weak excess Balmer emission from those two stars are attributable to accretion, then the UV excess measures are 0.6 dex larger than the H$\alpha$ accretion measures but with a larger (0.5 dex) source-to-source variation.  

This difference is consistent with the systematic methodological differences between the two methods.  However, that the systematic errors are in opposite directions for the two methods is unlikely.   For the UV excess method several uncertainties (see \S 4.3) would serve to increase our $\dot{M}$ estimates and therefore the discrepancy between the two methods.  
The uncertainty in the H$\alpha$ $\dot{M}$ calculations of $\sim 3-5$ are dominated by uncertainty in the disk truncation radius of the inner disk, assumed to be 2.2--3 $R_*$ \citep{Muz03}.  The model H$\alpha$ line width is proportional to $M_*^{0.5}$, so overestimating $M_*$ will lead to underestimating $\dot{M}$ for both methods.  Intrinsic variability in $\dot{M}$ can explain much of the scatter once the systematic variation between these two methods is corrected.

In the Muzerolle et al. H$\alpha$ line profile models, the accretion rate is estimated from the line profile along with the stellar mass and disk inclination.  Although $\dot{M}$ is correlated with the H$\alpha$ line width in the models, different $\dot{M}$ may vary by more than an order of magnitude for equivalent 10\% widths \citep[see also][]{Kur06}.  \citet{Nat04} found an empirical relationship between $\dot{M}$ and H$\alpha$ 10\% width, which is used in the absence of H$\alpha$ line profile modelling.   
We find a 0.86 dex scatter between the $\dot{M}$ measured here from UV excess and $\dot{M}$ from the \citet{Nat04} H$\alpha$ relationship (Fig. 9b) and a similar amount of scatter for a similar best-fit line$^4$ from Table 15. 
2M1207-3932 is a particular outlier on this plot, with an $\dot{M}$ estimated from the H$\alpha$ 10\% width that is two orders of magnitude larger than that measured from the UV excess.  
\footnotetext[4]{We find that $\dot{M}=-14.0+0.014$(H$\alpha$ 10\% width) from the data in Table 14, excluding upper limits, and $\dot{M}=-12.6+0.010$(H$\alpha$ 10\% width) if we combine this data with additional $\dot{M}$ calculated from the \ion{Ca}{2} $\lambda8662$ line fluxes from Mohanty et al.~(2005, see also \S 5.4).}

\subsection{Empirical relationships between line and continuum luminosities}

In \S 3.4  we qualitatively describe that excess continuum and line
 emission are produced by accretion processes.  Lines from accretors exhibit a wide range of profiles in high-resolution spectra, with equivalent widths that are correlated with veiling \citep[e.g.][]{Ham92,Bat96,Muz98,Ber01,Moh05}. 
Figure 10 shows that
 the luminosity in many lines ($L_{line}$) is tightly correlated with
 accretion continuum luminosity, although this relationship does not require
  that the emission lines and continuum are
 produced by the same gas.  We improve the fits for the H$\alpha$, [\ion{O}{1}] $\lambda6300$, and \ion{Ca}{2} $\lambda8542$ lines by supplementing our data with results from \citet{Har03}.  All line and accretion luminosities used in this subsection were measured from simultaneous data.

Linear fits of the relationship between $\log L_{acc}$ and $\log$ $L_{line}$ are calculated using the statistical 
package {\it asurv} \citep{Fei86}, with upper limits ignored (see Table 15 and Figure 10).
 Most upper limits are not significant and their inclusion does not substantially change the results.
 2M0455+3028, KPNO 11, CIDA 14, MHO 5, or V927 Tau (squares in Figure 10) are not included in these fits because
 the source of the excess Balmer emission is uncertain.  
For several lines at $>5600$ \AA, the high $\dot{M}$ range for these fits is dominated by CIDA 1 and GM Tau, both of which have large uncertainty in $A_V$ and therefore $\dot{M}$.  The fits based on few data points should be used with caution.

The H$\alpha$-$L_{acc}$ correlation has more scatter than the correlation between higher Balmer lines and $L_{acc}$ because the larger oscillator strength of H$\alpha$ can lead to stronger wind absorption.  The Balmer and \ion{Ca}{2} H \& K line luminosities from KPNO 11, 2M0455+3028, MHO 5, and CIDA 14 lie above the linear fit to accretion luminosity.  Since these four stars are tentative weak accretors, such line emission is likely produced by the chromosphere.  
The relationship between the Balmer and \ion{Ca}{2} K lines and the
accretion luminosity derived from our data is consistent with the results
from \citet{Gul98}.  The luminosities in these lines are systematically $\sim$0.5 dex
lower in the \citet{Val93} sample than in the \citet{Gul98} sample.$^5$  The nature of this discrepancy is not known.
\footnotetext[5]{In \citet{Val93} and here, the Balmer and \ion{Ca}{2} H \& K lines are measured directly from the spectrum while \citet{Gul98} measures line fluxes after the subtraction of a photospheric template.  These two different methods would slightly exacerbate the discrepancy between the line luminosities.}

 A large amount of scatter in the relationship between [\ion{O}{1}] line luminosity and accretion luminosity arises because the [\ion{O}{1}] emission line forms in an accretion-powered outflow \citep{Ham94,Har95}.  The comparison of the \citet{Har03} sample and our sample may be corrupted because [\ion{O}{1}] emission is typically spatially-extended \citep[e.g.][]{Har04}.  Any extended [\ion{O}{1}] emission could lie outside the narrow $0\farcs2$ slit width used by \citep{Har95}, but our $1\farcs0$ slit width includes any forbidden line emission related to the star.
The outflow line luminosities are correlated with accretion rate, though strong accretors sometimes drive weak outflows while some weak accretors may drive powerful outflows.
For example, the outflow
lines from GM Tau and 2M0414+2811 are substantially stronger than those from CIDA 1,
despite the higher mass accretion rate onto CIDA 1.  Similarly, many bright
outflow lines are detected from 2M0444+2512 despite a modest mass accretion
rate.  
The \ion{He}{1} and \ion{Na}{1} D lines are usually associated with both infall and outflow \citep{Har95,Ber01,Edw06} but here are more tightly correlated with the accretion luminosity than [\ion{O}{1}].

\subsection{Comparison with $\dot{M}$ measurements from emission line fluxes}

In \S 5.3 we derive empirical relationships between line and accretion continuum luminosity.  Lacking flux-calibrated spectra, \citet{Muz98} and \citet{Moh05} converted  the \ion{Ca}{2} equivalent width into line flux per stellar surface area and then correlated the line flux directly with $\dot{M}$.
Such a correlation naturally corrects for radius
uncertainties introduced by errors in extinction or distance.  Compared with equation 1, correlating $\frac{L_{line}}{R_*^2}$ with $\dot{M}$ will be in error in $\dot{M}$ by a factor of $\frac{M}{R^3}$.

Figure 11 compares the \ion{Ca}{2} $\lambda8542$, \ion{Ca}{2} $\lambda8662$, and \ion{He}{1} $\lambda5876$ line fluxes with accretion rate for the low-mass accretors presented here, high-mass accretors from Muzerolle et al. (1998 for the $\lambda8542$), and low- and high-mass accretors from Mohanty et al. (2005 for the $\lambda8662$ line)$^6$.  Those data have $\dot{M}$ and \ion{Ca}{2} line fluxes measured from non-contemporaneous observations and will suffer from variability. 
\footnotetext[6]{We calculate $\log \dot{M}=1.28 \log F($\ion{Ca}{2} $\lambda8542) -16.6$, $\log \dot{M}=1.03 \log F($\ion{Ca}{2} $\lambda8662) -15.2$, and $\log \dot{M}=1.63 \log F($\ion{He}{1} $\lambda5876) -17.9$
 from line fluxes and UV excess measures of accretion.  We also note that the relative fluxes in the three \ion{Ca}{2} IR triplet lines are usually consistent with the optically-thick limit of 1:1:1 \citep{Her80,Ham90}, whether the emission is in a broad or narrow line \citep{Bat96}.  Even at high-resolution all three \ion{Ca}{2} IR triplet lines blend with Paschen lines.}
 The $F$(\ion{Ca}{2} $\lambda8542)-\dot{M}$ and $F$(\ion{He}{1} $\lambda5876)-\dot{M}$ relationships are shallower than was previously measured from only the high-mass sample, while the $F($\ion{Ca}{2} $\lambda8662)-\dot{M}$ relationship is similar to that derived by Mohanty et al.(2005).   The standard deviations between $\dot{M}$ predicted from these line correlations and the $\dot{M}$ measured from UV continuum excess are 1.0, 0.71, and 1.1 dex for \ion{Ca}{2} $\lambda8542$, \ion{Ca}{2} $\lambda8662$, and \ion{He}{1} $\lambda5876$ line fluxes, respectively.

\section{DISCUSSION}
\subsection{Selecting methods to measure $\dot{M}$}
In the magnetospheric accretion paradigm, the Paschen continuum emission is produced in optically-thick post-shock gas, the Balmer continuum emission is produced in optically-thin pre-shock gas, and the Balmer line emission is mostly produced primarily in the magnetospheric accretion column \citep{Har94,Cal98,Muz01,Kur06}.  In the simplistic plane-parallel models all three accretion diagnostics are produced by the same gas.
In either case, the accretion energy is reprocessed into Balmer and Paschen continuum emission, which can be measured and subsequently converted into accretion rate.
Previously,
the accretion continuum measurements were applied only to accretors with $\sim 0.3-2$ $M_\odot$.  \citet{Muz03} developed models to measure $\dot{M}$ from H$\alpha$ line profiles for only the very low-mass stars and
brown dwarfs.  At higher mass accretion rates the H$\alpha$ emission line profile is difficult to use for estimating $\dot{M}$ because emission and absorption in strong
stellar outflows corrupt the line profile  \citep{Muz01,Kur06}.  
We provide the first measurements of the Balmer continuum emission from very low-mass stars and brown dwarfs and describe this emission as qualitatively similar to that from higher-mass accretors.  The UV excess measures of $\dot{M}$ are $\sim 4-7$ times larger than those from H$\alpha$ line profile modelling with some source-to-source differences that may be
methodological or related to intrinsic source variability. 

 In the context of magnetospheric accretion models, that the excess accretion continuum and the H$\alpha$ emission yield different $\dot{M}$ may not be surprising because they form in different locations.
The primary advantage of the H$\alpha$ models is that the line profile is
independent of the intrinsic line luminosity and only moderately
dependent on the stellar mass.  According to \citet{Muz03}, the uncertainty in $\dot{M}$ from H$\alpha$ modelling is a factor of 3--5 and is dominated by uncertainty in the size of the magnetosphere.  The factor of $\sim 4$ random error in UV excess measures of accretion is dominated by errors in extinction and distance.    The UV excess method relies predominantly on directly observable parameters, with modelling required only for calculating a moderate bolometric correction to derive $L_{acc}$.  The accretion energy is converted to $\dot{M}$ by equation 1.  In contrast, accretion
rates from the H$\alpha$ line profile rely on a correct interpretation and modelling of the
line profile in the magnetospheric funnel flow, and may be complicated by outflows, emission from hotspots, complex magnetic fields, or unexpected geometries.

Figures 12-13 compares the spectral properties of the excess hydrogen continuum emission versus stellar mass and mass accretion rate.  The size of the Balmer jump tends to increase with decreasing $M$ (and $\dot{M}$), consistent with lower $n_e$ ($\sim 3\times10^{13}$ cm$^{-3}$ for 2M1207-3932 compared with $\sim 2\times10^{14}$ cm$^{-3}$ for higher mass accretors) in single-temperature slab models or with a larger contribution of emission from the optically-thin pre-shock gas relative to optically-thick post-shock gas in the magnetospheric accretion models.   The slope of the Balmer continuum, which may be affected by our atmospheric correction, tends to increase with decreasing $M$ (and $\dot{M}$).  This is consistent with a lower temperature ($\sim 7500$ K for 2M1207-3932, compared with $\sim 8000-12000$ K for higher-mass accretors) in the plane-parallel slab models.
 Muzerolle et al.~(2003) describes that the temperature in the magnetospheric accretion flow must be at least $10^4$ K to have sufficient opacity in H$\alpha$ to produce the broad line profiles.
The percent of the stellar surface subtended by the accretion hot spot ($\delta$, see Table 11)$^7$ is typically 0.01-1\% for low-mass stars and brown dwarfs, relative to 0.1-10\% for higher-mass stars \citep{Val93}.  These results should be fully accounted for as tests of the magnetospheric accretion models.
\footnotetext[7]{The estimates of $\delta$ are inversely correlated with $L$, the length of the accretion slab, and $T$, the temperature of the accretion slab.  The calculated $L$ may be non-physical, and in this work are larger here than the $\sim 10^7$ typical for \citet{Val93}.  As a result, the difference in $\delta$ between the two works may be underestimated.}

For regions with low extinction the UV excess method is the most accurate measure of accretion luminosity.  The accretion luminosity can also be measured from the Paschen continuum at 3700--8000 \AA, albeit with larger uncertainties and reduced sensitivity to small accretion rates.  UV excess measures of accretion can be applied to even lower mass objects than studied here, limited only by telescopic sensitivity.  H$\alpha$ line modelling may be a sufficient substitute for low-mass stars and brown dwarfs when UV excess measures are not possible, either due to large or uncertain extinction or distance, or instrumentation limitations.   The H$\alpha$ models at present cannot be applied to large $\dot{M}$ due to complex line profiles.  They also cannot be applied to $\dot{M}<10^{-12}$ \mdotyr, below which the opacity in the H$\alpha$ line wings may be insufficient to measure accretion \citep{Muz03}.    However, H$\alpha$ models may be better suited than UV excess measurements for small $\dot{M}$ onto higher mass/luminosity accretors, where the contrast between the Balmer continuum and photospheric emission is small.  For example, the excess Balmer emission is difficult to detect onto the objects CIDA-14, V927 Tau, and 2M0455+3028, but the 10\% width of the H$\alpha$ line is large enough to suggest accretion. Additional comparisons of $\dot{M}$ between these two methods are required to better characterize the differences.  

Secondary diagnostics, such as the \ion{Ca}{2} IR triplet,
Pa$\beta$, or Br$\gamma$ lines, are related to $L_{acc}$ or, less accurately, $\dot{M}$ by
empirical relationships and can provide useful $\dot{M}$ estimates \citep[see also][]{Muz98,Nat04,Moh05}.  
The many random uncertainties inherent in both the UV excess and
H$\alpha$ methods are reduced if the sample size is large enough.
However, systematic uncertainties in the both accretion measures are propogated into these correlations.  Some uncertainty is also introduced into these relationships because the line emission may be produced in different gas than the accretion continuum and because wind absorption can reduce the line flux.
Of the line luminosity-accretion luminosity relationships analyzed in Table 15, the accretion luminosities estimated from the high Balmer lines and the \ion{He}{1} $\lambda5016$ line have the least amount scatter.  The line luminosity-accretion luminosity relationships are preferable to line flux-$\dot{M}$ relationships (Fig. 11), which have scatters of $\sim 0.75-1.0$ dex.  That this amount of scatter is similar to the scatter in the $M-\dot{M}$ relationship (see \S 6.2) suggests that the line flux-$\dot{M}$ relationships are not accurate.  

Many of the commonly used lines (\ion{Ca}{2} IR triplet, Pa$\beta$, Br$\gamma$) are less sensitive to $\dot{M}/M$ than either the H$\alpha$ line profile or excess UV emission.  They therefore provide a less complete picture of the full range of accretion.  The completeness depends on mass if detection limit of each diagnostic varies with spectral type.  These diagnostics are often more accessible than modelling H$\alpha$ profiles or measuring the UV excess and are particularly useful for variability studies of a individual objects \citep[e.g.][]{Ale05,Aze06} or in large samples \citep{Nat06}.


\subsection{Accretion rate versus mass}
Figure 14 shows $\dot{M}$ versus $M$ for Taurus accretors
with $A_V<2.5$ from \citet{Gul98}, \citet{Har03}, \citet{Cal04}, \citet{Muz05}, 
and from the low-resolution {\it Keck I} and {\it HST} spectra analyzed in this work$^8$.  We use {\it asurv} \citep{Fei86} to calculate that 
\begin{equation}
\log \dot{M}=(1.87\pm0.26) \log M - (7.73\pm0.17),
\end{equation}
which is similar to previous estimates of 
$\dot{M}\propto M^{1.8-2.3}$ \citep{Muz05,Moh05,Nat06}.  The 
standard deviation of the fit to $\dot{M}$ is 0.8 dex.  This standard deviation is too large to be explained by intrinsic accretion variability or random methodological uncertainties, which suggests that stellar parameters other than mass also affect the accretion rate.
\footnotetext[8]{The \citet{Har03} data are used to re-calculate accretion rates, as in \S 5.1.  The \citet{Muz05} accretion rates are used only if they are not duplicates from this work and are multuplied by a factor of 5 (see \S 5.3).  Duplicate measurements of $\dot{M}$ for the same star are averaged.}
A lack of completeness in our sampling of the true range in 
mass accretion rate for a given mass can bias the $\dot{M}-M^\alpha$ 
relationship by target selection and by sensitivity limits.

We specifically selected several targets (CIDA 1, GM Tau, and 2M0414+2811) based on our expectation of high $\dot{M}$.    
Most other stars were selected based in part on visual magnitude, which may tend to yield sources with higher $\dot{M}$.  If this selection bias is only present in the current data, which focuses on the low-mass range, then the $\alpha$ in the $\dot{M}-M^\alpha$ relationship will be underestimated.  
However, a similar bias could also be present in previous $\dot{M}$ measurements from samples of higher-mass stars.

The $\dot{M}$ from higher-mass accretors may be incomplete
because sensitivity to $\frac{\dot{M}}{M}$ improves to smaller masses for both UV excess and H$\alpha$ accretion measures. 
Figure 15 shows that accretion luminosity and the measured $\frac{L_{acc}}{L_{bol}}$ decreases with mass.  The improved sensitivity to low $\frac{L_{acc}}{L_{bol}}$ at smaller masses results because the percentage of total photospheric emission  that escapes at $<3700$ \AA\ is much less for mid-late M-dwarfs than for late-K dwarfs.  In
\citet{Val93} and \citet{Gul98}, the minimum value of 
$\frac{L_{acc}}{L_{bol}}$ is 0.008 and 0.018, respectively, with most
targets well above that limit.
This detection limit leads to $\dot{M}$ $\sim 5-10\times10^{-10}$ $M_\odot$ yr$^{-1}$
for a 2 Myr old 0.7 $M_\odot$ star with  $L=0.57$ $L_\odot$.
For the lower-mass accretors studied here, accretion from 2M0455+3028 is at our detection limit with  $\frac{L_{acc}}{L_{phot}}\sim0.001$.  For a 2 Myr old, 0.1 $M_\odot$ brown dwarf, the lowest detectable mass
accretion rate is $\sim 10^{-11}$ $M_\odot$ yr$^{-1}$.  The
$\frac{\dot{M}}{M}$ ratio is therefore about 5.5 times lower for a $0.1$
$M_\odot$ accretor than for a $0.7$ $M_\odot$ accretor.  Incompleteness at higher masses would cause us to overestimate $\alpha$ in the $\dot{M}-M^\alpha$ relationship.  On the other hand, inclusion of the five tentative accretors would increase $\alpha$ to 2.08.

\section{Conclusions}
We analyzed blue spectra of 18 accreting very low-mass stars and brown dwarfs obtained with LRIS on {\it Keck I}.  Ours are the first data on the Balmer continuum and high Balmer line emission in this mass range.   Most of this sample was selected based on the presence of accretion identified with other diagnostics.  Our observations are compared with archival {\it HST}/STIS spectra of the Balmer continuum from several higher-mass stellar accretors.  We obtain the following results:

1.  We detect excess hydrogen Balmer continuum emission from 16 of the 18 mid-late M-dwarfs in our sample.  11 of these objects also show hydrogen Paschen continuum emission.  This continuum is attributed to accretion for most of our objects.  Weak Balmer continuum emission onto earlier-type stars, including the K6 dwarf V836 Tau, can be attributed to accretion because non-accretors show no evidence for chromospheric Balmer continuum emission.  However, we cannot rule out a chromospheric origin for weak excess Balmer emission from mid-M dwarfs.

2.  Measurements of the Balmer and Paschen continua emission yield $\dot{M}=10^{-12}-10^{-8}$ for very-low mass stars and brown dwarfs with masses from 0.024--0.17 $M_\odot$.  Most of these accretion rates are smaller than those typically measured for accreting $\sim 0.3-1.0$ $M_\odot$ stars.  Analysis of the random and systematic errors indicate that these $\dot{M}$ may be underestimated by a factor of 1.2 (or 2.4 if the hydrogen continuum emission incident upon the star is reprocessed into photospheric emission), with a relative uncertainty of $\sim 0.6$ dex.  Comparing these $\dot{M}$ with $\dot{M}$ from other studies requires accounting for different assumptions regarding radiative transfer.

3. The Balmer jump tends to be larger and the Balmer continuum tends to be redder for stars with lower mass accretion rates.  These properties indicate lower opacity (with $\log n_e=13.5$ for 2M1207, compared with $\log n_e\sim14.3$ for stars with higher $\dot{M}$ in the plane-parallel slab models) and lower temperature (with $T\sim7000$ K for 2M1207, compared with $\sim 9500$ for stars with higher $\dot{M}$).

4.  The $\dot{M}$ calculated here are well correlated with but $\sim 4-7$ times larger than the $\dot{M}$ calculated from published models of the H$\alpha$ line profiles.   We suggest that this offset is methodological, although the difference is within the range of the combined errors for both methods.  The $\dot{M}$ measured here also correlates with H$\alpha$ 10\% width but introduces 0.9 dex uncertainty in $\dot{M}$ predicted from H$\alpha$ 10\% width alone.

5.   We provide empirical relationships between $L_{acc}$ and $L_{line}$, which are often well correlated.  In particular, estimating $L_{acc}$ from the luminosity in the high Balmer and many \ion{He}{1} lines introduces an uncertainty of only 0.2--0.35 dex.  On the other hand, estimating the $L_{acc}$ from line surface fluxes alone introduces an uncertainty of $\sim 0.7-1.1$ dex in $\dot{M}$ estimates.

6.  We find that $\log \dot{M}\propto M^{1.87}$ for Taurus stars with $A_V<2.5$, which is consistent with previous estimates.  This relationship may be biased if the higher sensitivity of ${\dot{M}}/M$ at lower $M$ causes the percentage of accretors that are detected to also vary with $M$.

\section{Acknowledgements}
We thank Jeff Valenti and Chris Johns-Krull for use of their
plane-parallel slab model codes that they developed in \citet{Val93}.
We thank Russel White for discussion of the proposal and Nuria Calvet for discussion of the line emission.  We thank the anonymous referee for helpful comments, which served to improve the structure and clarity of the paper.

Most of data presented herein were obtained at the W.M. Keck Observatory, which is operated as a scientific partnership among the California Institute of Technology, the University of California and the National Aeronautics and Space Administration. The Observatory was made possible by the generous financial support of the W.M. Keck Foundation.  
Some of the data presented in this paper were obtained from the Multimission Archive at the Space Telescope Science Institute (MAST). STScI is operated by the Association of Universities for Research in Astronomy, Inc., under NASA contract NAS5-26555. Support for MAST for non-HST data is provided by the NASA Office of Space Science via grant NAG5-7584 and by other grants and contracts.

\section{Appendix}

\subsection{Photospheric Emission}
The photospheric emission from M-dwarfs is dominated by strong TiO
 and
CaH bands at $\lambda>4800$ \AA\ that can be used to estimate spectral type \citep[e.g.][]{Kirk93}. 
Table 3 lists spectral types for our targets estimated from the TiO 7140 and
8465 \AA\ indices \citep{Sle06} and by visual comparison to existing
{\it Keck I}/LRIS spectra of M4.5-M9.5 M-dwarfs$^{9}$.  Most
spectral types are consistent with existing literature estimates.   We classify V927 Tau as an M5 star, compared with previous classifications of M3, M4.75, and M5.5. 2M0455+3028, S0439+2336, and MHO 6 have the same spectral type as V927 Tau and are
classified as M5. 
MHO 7 is a slightly later spectral type 
than V927 Tau and is classified as M5.5,
0.25 classes later than in \citet{Bri02}.  The photospheric emission from CIDA 14 is nearly
identical to MHO 7.  S0518+2327 is about the same as MHO 7 and
classified as M5.5, while KPNO Tau 11 slightly later than MHO 7 and
is classified as an M5.75.  MHO 5 is between MHO 7 and CFHT 7 and is
classified as M6.  We confirm spectral types from \citet{Luh04} for 2M0438+2611, 2M0439+2544, 2M0441+2534,
and 2M0444+2512, and from \citet{Giz02} for 2M1207-3932.  
We conservatively assign a 0.5 subclass uncertainty to all spectral types.
\footnotetext[9]{Obtained from
  http://www-int.stsci.edu/$\sim$inr/ultracool.html (Gizis et al. 2000ab, Kirkpatrick et al. 1999).}

The spectral types for GM Tau, CIDA 1, and 2M0414+2811 are difficult to
determine because of high veiling at $\lambda>7000$ \AA.  In order to make some estimate we assume that
the accretion continuum emission is constant in flux per \AA\ across
short wavelength intervals.  The depth of absorption
bands are then compared to spectral type standards to estimate the spectral
classification.  The photospheric emission of CIDA 1 is
consistent with M5, GM Tau with M5.5, and 2M0414+2811 with CFHT 7.  The
spectral types of CIDA 1 and GM Tau are 0.5 and 1 class earlier, respectively, than that measured by \citet{Whi03} and are both assigned an error of 1.0 subclasses.

\subsection{Extinction}
Extinction estimates are important to convert the measured flux in the accretion continuum into a luminosity. 
Many of our targets were selected for observation based on low extinction.  We adopt extinctions of $\sim 0.4$,
$0.1$, 0.0, and 0.0 to the non-accretors V927 Tau, MHO 7, CFHT
7, and the accretors 2M0444+2512 \citep{Ken95,Bri02,Gui07,Luh04}. 
We confirm these estimates with an accuracy of $\sim 1.0$ mag  by comparing  our flux-calibrated observations to existing {\it
  Keck I}/LRIS red spectra (6000--10000 \AA) of 
photospheric templates.  

The photospheric emission from the rest of our targets are then compared
to those four templates to calculate an extinction.  
For accretors, we also add a flat accretion
continuum at red wavelengths.
We use the extinction law from
\citet{Car89} and a total-to-selective extinction ($R_V$) of $3.1$.  Some
evidence suggests that $R_V>3.1$ in the Taurus Molecular Cloud because
grains in the molecular cloud may be larger than those in the average
interstellar medium \citep{Whittet01}.  If $R_V$ is large, the slope
of extinction versus wavelength is shallow and $A_V$ will be underestimated.
The spectrum of 2M0438+2611 is likely seen through an edge-on disk and suffers from gray extinction and scattering \citep{Luh07} that is not corrected for here.

We confirm our extinction estimates for sources with strong veiling by assuming that the Paschen continuum
between 4000--6000 \AA\ is flat (see \S 3.1), which is roughly consistent with observations
of the higher-mass accretors \citep[e.g.][]{Bas90,Har91}, with observations of accretors
with low $A_V$ \citep{Gul00},  and with shock models \citep{Cal98}.

Our extinction estimates are listed in Table 3.  Any error in the $A_V$ for V927 Tau, MHO 7, CFHT 7, and 2M0444+2512
 will propogate to the remaining data.
 Our estimates are consistent with
previous extinction estimates \citep{Bri02,Luh04}, with a few exceptions.  We estimate
$A_V\sim1.0$ and $\sim 0.7$ to 2M0455+3028 and 2M0439+2544, respectively, compared
with $A_J=0.0$ and $0.07$ from \citet{Luh04}.  We measure $A_V=2.0\pm0.7$ to GM
Tau, less than previous estimates of $\sim4.2$ mag \citep{Bri02,Gui07}.

2M1207-3932 is in the TW Hya Association and isolated from
molecular gas.  We assume $A_V=0$ to 2M1207-3932, and confirm that $A_V<1$ by comparison to a template LRIS M8-dwarf standard.

\subsection{Stellar Radius and Mass}
Estimates for the stellar mass and radius are required to calculate mass accretion rates.  
The spectral class is converted to temperature using the scale derived
by \citet{Luh03}.  We then use our spectral types, extinction estimates, synthetic R-band photometry (after correcting for
the measured veiling), and Baraffe et al. (1998) evolutionary tracks
to compute the stellar luminosity, mass, radius, and age (see Table 3).  The calculated ages for the higher-mass sample in Taurus are overestimated, which is consistent with known systematic problems in stellar evolutionary tracks \citep{Hil07}

A distance of 140 pc is assumed for Taurus stars \citep{Ken94}.  
\citet{Sle06} identified S0518+2327 as an accretor away from the main
Taurus molecular cloud and classify it as an intermediate-age star based
on the depth of \ion{Na}{1} gravity indices.  The distance to S0518+2327 is therefore more uncertain the most other Taurus stars.
We adopt a parallax distance of $54\pm3$ pc to 2M1207-3932 (Gizis et al. 2007; see also Biller \& Close 2007) and of 56 pc for TW Hya \citep{Wic98}.

Any binarity in our sample would reduce the stellar luminosity and estimated radius.   \citet{Kra06} found that GM Tau and MHO 5 are spatially unresolved to $\sim0\farcs05$ and are either single stars or spectroscopic binaries.  We assume our other targets are single stars with the exception of V927 Tau (see \S9.4.2).

\subsection{Notes on Individual Objects}
 
\subsubsection{2M1207-3932}
 2M1207-3932 is a M8-L dwarf binary \citep{Giz02,Cha04} in the 10 Myr old TW Hya Association \citep{Web99,Mam05}.  The primary has an IR excess and ongoing accretion.  The secondary is faint and undetected in our data.
We observed 2M1207-3932 on both November 24 and February 8.  The
February observation was deeper and obtained with the target at lower airmass.  The flux from the November integration is calibrated based on the strength of photospheric emission from 6000--8000 \AA\ measured in the February run.   The blue and UV continuum emission was fainter in November than on February 8.  
 Most of these parameters have a higher confidence for 2M1207-3932 than the Taurus members.

 \citet{Ste07} applied the \citet{Nat04} relationship for H$\alpha$ 10\% width to measure 
$\dot{M}\sim10^{-10.1}-10^{-9.8}$ \mdotyr\ from 2M1207-3932, which is two orders of magnitude larger than measured here.  This difference is not likely attributed to variability because the H$\alpha$ equivalent width during our February observation is the largest yet detected from 2M1207-3932.  During an epoch in 2003 with low $\dot{M}$ \citep{Ste07}, both the equivalent width and H$\alpha$ 10\% width were small.  The variable equivalent width in the \ion{Ca}{2} $\lambda8542$ line of $<0.3$ \AA\ \citep{Ste07} corresponds to $\dot{M}<2.1\times10^{-12}$ \mdotyr\ (Table 15), which confirms our low $\dot{M}$.

\subsubsection{V927 Tau}

V927 Tau is a binary separated by $\sim 0\farcs27$ \citep{Lei93,Whi01}.  
\citet{Muz03} measured a broad 10\% width of 290 \kms\ and only a small self-reversal in
the symmetric H$\alpha$ line profile from V927 Tau.  They suggested that this H$\alpha$ emission may be produced by chromospheric flaring since
most H$\alpha$ profiles produced by accretion are asymmetric.  At the time
no excess IR or sub-mm emission had been detected from either component
\citep{Sim95,Whi01,And05}.  However, \citet{McC06} measured K-band
magnitudes that were the same as measured by \citet{Whi01}, but L-band
emission that was brighter by 1 magnitude for both V927A and B.  Such variability may suggest that
the amount of warm dust in the disk is variable, and that perhaps accretion
onto V927 Tau may sometimes be present. 

We detect some excess Balmer continuum and \ion{He}{1}
$\lambda5876$ emission from V927 Tau relative to MHO 7.  
V927 Tau and its template MHO 7 are
slightly different spectral types, which may lead to some discrepancy
in this comparison.  Such emission could be
interpreted as either chromsopheric or accretion-related.  \citet{Har03} found weak excess Paschen continuum emission from the secondary, if the secondary is classified as an M3.5 star.  V927 Tau remains
ambiguous as to whether accretion is ongoing.

\subsubsection{2M0438+2611}
We obtain a short LRIS spectrum of 2M0438+2611, which \citet{Luh07} describe as a young accreting star seen through an edge-on disk with a gray extinction of 5.5 mag.  The gray extinction cannot be measured from the shape of the continuum emission, which is similar to that of 2M0444+2512 ($A_V=0$ mag).  If we assume the two stars have the same radius, then we find $A_V=2.8$ mag to 2M0438+2611. That the optical forbidden lines are bright \citep[see also][]{Luh04,Luh07} indicates that some of the outflow emission is at least slightly extended and not absorbed by the edge-on disk, consistent with other accretors seen through edge on disks.

We classify 2M0438+2611 as a moderate accretor based on the detection of optical forbidden lines and non-detection of the \ion{Ca}{2} IR triplet.   We use the H$\gamma$ line luminosity and Table 12 to calculate $L_{acc}\sim4\times10^{-5}$ $L_\odot$, or $\dot{M}\sim1.5\times10^{-11}$ $M_\odot$ yr$^{-1}$ (assuming same $M$ and $R$ are the same as 2M0444+2512).  The upper limit for the \ion{Ca}{2} $\lambda8542$ equivalent width of $<1.2$ \AA\ indicates $L_{acc}<1.6\times10^{-4}$ $L_{acc}$.   We do not use the $\dot{M}$ of 2M0438+2611 in any analysis.

\subsubsection{2M0414+2811}
 2M0414+2811 is about a magnitude brighter and is bluer
than UBV photometry obtained by \citet{Gro07}, which suggests a higher $\dot{M}$ during our observation. 

\subsubsection{MHO 5}
Emission in the [\ion{O}{1}] $\lambda6300,6363$ and \ion{He}{1} $\lambda5876$ lines are detected with equivalent width typical of accretors.  MHO 5 is therefore classified as an accretor, despite a low H$\alpha$ 10\% width of 154 \kms\ \citep{Muz03}.  Ongoing accretion indicates that the weak excess Balmer continuum emission is likely produced by accretion, although we cannot rule out a chromospheric origin.

\subsubsection{CIDA 14}

The excess Balmer continuum emission onto CIDA 14 is very weak, and only detected because the photospheric emission is very similar  to that from the non-accretor MHO 7.  We detect an equivalent width of 0.27 \AA\ in the [\ion{O}{1}] line from CIDA 14, which is slightly lower than the 0.4 \AA\ measured by \citet{Bri99} but larger than the $<0.1$ \AA\ upper limit from high-resolution spectra by \citet{Muz03}.  This low upper limit may be underestimated if the line is broad.  \citet{Muz03} measured an H$\alpha$ equivalent width of 289 \kms.  CIDA 14 is a likely accretor.

\subsubsection{KPNO 11}
Several \ion{He}{1} lines that are undetected from the non-accretors are detected from KPNO 11 (Table 5).  However, the strength of the \ion{He}{1} $\lambda5876$ \AA\ line and the upper limit on the \ion{Na}{1} D line luminosity are below the expected luminosity for the calculated $\dot{M}$.  These luminosities may suggest that the excess Balmer continuum emission and \ion{He}{1} lines are chromospheric.  The H$\alpha$ 10\% width of 182 \kms\ \citep{Moh05} is intermediate between accretors and non-accretors.

\subsubsection{GM Tau}
The flux ratio of Balmer line emission to Balmer continuum emission
from GM Tau is much smaller than measured for any other source here, which requires a large optical depth in the slab.

\subsubsection{S0518+2327}
S0518+2327 is an accretor located away from the main Taurus Molecular Cloud \citep{Sle06}.  The distance to this star is more uncertain the the other Taurus stars studied here.

\subsubsection{2M0436+2351}
The U-band atmospheric correction for 2M0436+2351 is somewhat uncertain.  As a result, no slope for the Balmer continuum is listed in Table 5.  The mass accretion rate for this star is also more uncertain than that of other stars in our sample.

\subsubsection{DF Tau}

 DF Tau is a $0\farcs1$ binary \citep{Sch06} that is unresolved in these spectra.  When calculating the stellar radius we assumed that the luminosity of the two stars are equal.  STIS observed DF Tau three times with the G430L grating, two of which are analyzed here.  The third observations shows much weaker photospheric and continuum emission with an accretion rate 10 times lower than the other observations \citep{Her06}.  We do not analyze this third spectrum here because it was obtained with a narrower slit and could have included the secondary but not the primary.

\subsubsection{DG Tau}
We measure $\dot{M}=4.1\times10^{-9}$ $M_\odot$ yr$^{-1}$, an order of magnitude less than that measured by \citet{Gul00}.  The spectrum of DG Tau during the \citet{Gul00} observation was heavily-veiled and an order of magnitude brigther than that detected here.  The different $\dot{M}$ result from real variability.
Strong emission is detected in forbidden lines and \ion{Fe}{1} lines despite the lower $\dot{M}$.

\clearpage

\begin{table}
\caption{Observations with {\it Keck I}/LRIS}
\label{tab:obsx.tab}
{\footnotesize 
\begin{tabular}{cccccccccc}
\hline
Star     & MJD$^a$   & $t_{exp}$ & U$^b$ & B$^b$   & V$^b$ & R$^b$  & $\sigma^c$ (mag) & Abbreviation$^d$\\
\hline
\hline
CIDA 1   & 63.318 &   720  & 18.6 &  18.3 &  17.2 &  15.3       & 0.6 & CIDA 1\\
2MASS J04141188+2811535  & 63.329 & 1200  & 18.2 &  18.2 &  17.6 &  16.4 & 0.6 &   2M0414+2811\\
MHO-Tau 5    & 63.382  & 1620  &  19.2 &  18.3 &  16.9 &  15.0 & 0.4 & MHO 5\\
MHO-Tau 6    & 63.405 & 1260  &  18.5 &  17.8 &  16.6 &  15.1 & 0.4 & MHO 6\\
GM Tau   & 63.422 &  1050 & 18.7 &  18.7 &  17.8 &  16.3 & 0.4 & GM Tau\\
2MASS J04390396+2544264   & 63.438  & 2700  &  22.0 &  21.5 &  20.1 &  17.6 & 0.4 & 2M0439+2544\\
2MASS J04442713+2512164   & 63.475 & 2100  &  21.0 &  20.4 &  19.5 &  17.3 & 0.4 & 2M0444+2512\\
\end{tabular}}
\end{table}

\begin{table}
\caption{Archival {\it HST}/STIS G430L Observations}
\label{tab:obsx.tab}
{\footnotesize \begin{tabular}{ccccccc}
\hline
Star     & MJD    & $t_{exp}$ (s) & Slit & PI  & Ref. & Abbreviation \\
\hline
\hline
DF Tau  & 51439.84 & 144 & $52\times0.2$  & Linsky & 2 & DF Tau (Sep. 99)\\
TW Hya  & 51671.88 & 260 & $52\times0.2$  & Linsky & 1  & TW Hya (May 00)\\
TW Hya  & 52474.58 & 120 &  $52\times0.2$ & Johns-Krull & -- & TW Hya (Jul. 02) \\
DF Tau  & 52512.49 & 120 &  $52\times0.2$ & Johns-Krull & -- & DF Tau (Aug. 02)\\
RU Lupi & 51737.65 & 120 & $52\times0.1$ & Walter & 2 & --\\
V819 Tau$^a$ & 51786.87 & 360& $52\times0.1$ & Walter & 2 & --\\
DG Tau  & 51839.52 & 240 & $52\times0.1$ & Walter & 2 & --\\
V836 Tau & 51954.36& 360 & $52\times0.1$ & Walter & 2 & --\\
LkCa 7A$^a$ & 53039.71 & 162 & $52\times0.2$ & Johns-Krull & -- & LkCa 7\\
\hline
\multicolumn{3}{l}{1: \citet{Her04}} &
\multicolumn{3}{l}{2: \citet{Her06}}\\
\multicolumn{7}{l}{$^a$We extract and use only the primary spectrum and do not discuss the faint secondary}\\
\end{tabular}}
\end{table}

\begin{table}
\caption{Adopted Stellar Properties}
\label{tab:obsx.tab}
\begin{tabular}{ccccccccccccc}
\hline
Star    & SpT   & $\log$ Age$^a$ & $A_V$ & $T$ & $L$ & $R$ & $M$ & Ref.\\
 & & $\log$ yr & mag & K & $L_\odot$ & $R_\odot$ & $M_\odot$ & \\
\hline
\hline
CIDA 1  & M5    & 6.0* & 3.0  & 3125 & 0.21   & 1.57  & 0.15 & 1\\
GM Tau  & M5.5  & 6.4 & 2.0  & 3091 & 0.070 & 0.93 & 0.15 & 2\\
2M0414+2811  & M6.5  & 6.5 & 1.0  & 2935 & 0.014  & 0.46  & 0.065 & 3\\
S0518+2327   & M5.5  & 6.9 & 1.0  & 3091 & 0.016  & 0.44  & 0.10 & 4\\
MHO 6   & M5    & 6.1 & 1.0  & 3125 & 0.14   & 1.31  & 0.17 & 5\\
S0439+2336   & M5    & 6.3 & 0.0  & 3125 & 0.089  & 1.02  & 0.17 & 4\\
\end{tabular}
\end{table}

\begin{table}
\caption{Veiling Measurements of M-dwarfs from {\it Keck} spectra$^{a}$}
\label{tab:obsx.tab}
{\scriptsize
\begin{tabular}{cccccccccc}
\hline
Star   & Template&    4227 &    4955 &    5450 &    6190 &    6880 &   7100 & 7800 & 8600\\
\hline
\hline
CIDA 1 &V927 Tau&    $>2$ &    0.98 &    0.75 &    0.94 &    0.42 &    0.63 &  0.38 &    0.56\\
       &        &   0.85  &   1.4 &   2.1 &   3.6 &   6.1 &   8.0 & 10.9 &  25.0\\
\hline
GM Tau & MHO 7  &    $>2$ &    1.84 &    1.55 &    3.63 &    0.88 &    0.79 &   0.27 &    0.26\\
       &        &    0.8 &   1.2 &   1.6 &   2.7 &   3.7 &   3.5 &  2.9 &   4.7\\
\hline
2M0414+2811 &CFHT 7  &    $>2$ &    $>3$  &    $>3$ &    $>3$ &    1.52 &    1.59 &   0.76 &    0.16\\
       &        &    1.6   &   2.3   &   2.5   &   $3.0$ &   4.2 &   4.3 &  4.7 &   2.9\\
\hline
S0518+2327  & MHO 7  &    $>2$ &    1.62 &    0.79 &    0.87 &    0.38 &    0.38 &  0.14 &    0.40\\
\end{tabular}}
\end{table}

\begin{table}
\caption{Veiling Measurements for higher-mass accretors observed with {\it HST}$^a$}
\label{tab:obsx.tab}
{\scriptsize \begin{tabular}{cccccccccc}
Star    & Template & 4000 & 4600 & 5400\\
\hline
DF Tau & LkCa 7   & $>5$   & 1.8  & 1.3\\
(Sep. 99) &          &   310   & 310      & 275\\
\hline
DF Tau & LkCa 7   &  $>5$  & 0.73 & 0.6\\
 (Aug. 02)   &          & 150    & 126  & 114\\
\hline
RU Lup  & V819 Tau & $>5$ & 1.7  &  2.3 \\
        &          & 840  & 590  &  710 \\
\hline
DG Tau  & V819 Tau &3.0 & 1.4  &   1.7 \\
        &          & 37   & 59    &  64\\
\hline
TW Hya & V819 Tau & 2.2   & 0.58 &  0.36  \\
(May 00)       &          &  510   & 450  & 330\\
\hline
TW Hya & V819 Tau & 1.4  & 0.69 &  0.72 \\
(Jul. 02) &          &  360  & 460  & 470\\
\hline
V836 Tau& V819 Tau & 0.23 & $<0.1$ & $<0.08$\\
        &          &  3.1 & $<4.9$ & $<4.0$ \\
\hline
\hline
\multicolumn{5}{l}{$^a$Format same as in Table 4}\\
\end{tabular}}
\end{table}

\begin{table}
\caption{Observed properties of the accretion continuum}
\label{tab:obsx.tab}
\begin{tabular}{cccccccccc}
\hline
Star    & $J_{obs}^a$ & $J_{mod}^b$ & $S_B^c$ & $F_{4955}/F_{4227}^d$  \\
\hline
\hline
\multicolumn{5}{c}{Accretors in {\it Keck I} sample}\\
\hline
CIDA 1  &  3.4   &  $5.6\pm0.7$   & $4.1\pm2.0$ & 0.85\\
GM Tau  &  3.5   &  $5.9\pm0.7^e$ & $6.1\pm1.5$ & 0.9\\
2M0414+2811  &  2.7   &  $3.10\pm0.17$ & $8.1\pm1.0$ & 1.14\\
S0518+2327   &  2.1   &   $2.35\pm0.15$& $2.4\pm2.0$ & 0.83 \\
MHO 6   &  1.10  &  $3.9\pm0.8$   & $7.7\pm1.5$ & $<0.5$\\
S0439+2336   &  0.76  &   $3.1\pm1.5$  & $6.2\pm1.1$ & $<1.2$\\
\end{tabular}
\end{table}

\begin{table}
\caption{Balmer Line Equivalent Widths and Fluxes}
\label{tab:obsx.tab}
{\scriptsize
\begin{tabular}{ccccccccccccc}
\hline
Star & H$\alpha$ & H$\beta$ & H$\gamma$ & H$\delta$ &  H$6$ & H$7$ & H8 & H9 & H10 & H11 & H12 \\
\hline
\hline
CIDA 1 &    232 &    134 &     80 &     64 &     41 &     49 &     37 &     27 &     18 &     17 &      10\\
      &    31.2 &     3.0 &     1.08 &     0.65 &     0.38 &     0.38 &     0.29 &     0.24 &     0.21 &     0.20 &     0.14\\
\hline
GM Tau &    281 &     68 &     39 &     30 &      8.8 &     26 &     13 &     11 &     12 &     11 &      9.1\\
      &    14.6 &     1.11 &     0.41 &  0.24 &     0.12 &     0.21 &     0.13 &     0.11 &     0.15 &     0.14 &     0.12\\
\hline
2M0414+2811 &    280 &    103 &     72 &     58 &     30 &     32 &     23 &     17 &     10.6 &      8.7 &      5.6\\
      &    15 &     2.5 &     1.3 &     0.97 &     0.57 &     0.53 &     0.39 &     0.32 &     0.26 &     0.22 &     0.15\\
\hline
S0518+2327 &     50 &   67 &     60 &     54 &     28 &     27 &  19   &    9.8 &   5.3 &      4.7 &      1.9\\
\hline
\hline
\end{tabular}}
\end{table}

\begin{table}
\label{tab:obsx.tab}
{\scriptsize
\begin{tabular}{ccccccccccccc}
\multicolumn{9}{l}{Table 7 (continued): {\it HST}/STIS Sample}\\
\hline
Star & H$\alpha$ & H$\beta$ & H$\gamma$ & H$\delta$ &  H$6$ & H$7$ & H8 & H9 & H10 & H11 & H12 \\
\hline
\hline
DF Tau&      -- &     29 &     27 &     26 &     22 &     17 &     13 &      8.4 &      5.2 &      3.4 & 2.9   \\
 (Sep. 99)       &     --  &   150 &   109 &    90 &    68 &    54 &    42 &    30 &    19 &    13 &   11.5\\
\hline
DF Tau &      -- &     26 &     20 &     17.7 &     16&     15 &     10.9 &      7.8 &      6.1 &      3.9 & 4.2  \\
(Aug. 02) &     --  &    80 &    44 &    32 &    24 &    24 &    16 &    13.6 &    11.3 &     7.6 &   7.9\\
\hline
RU Lupi &      -- &     37 &     16 &     14.5 &      7.3 & 8.7 &      6.6 &     9.3 &     4.6B &    4.4 &  3.0   \\
      &     --    &   360   &   160 &   113    &    68    &   74 &    57   &     80 &     51B   &   48 &   33\\
\end{tabular}}
\end{table}

\begin{table}
\caption{He Line Equivalent Widths and Fluxes$^a$}
\label{tab:obsx.tab}
{\scriptsize
\begin{tabular}{ccccccccc}
\hline
Star  &    $\lambda4026$  & $\lambda4470$    &  $\lambda4922$   & $\lambda5016$  & $\lambda5876$   & $\lambda6678$  & $\lambda7065$
&   \ion{He}{2} $\lambda4686$\\ 
\hline
\hline
\multicolumn{9}{c}{{\it Keck I}/LRIS sample of low-mass stars}\\
\hline
CIDA 1 &      4.3B &      5.3 &      8.3 &     10.4 &      9.6 &      1.9 &      2.0 &      1.3\\
      &   0.035B &   0.086 &   0.22 &   0.26 &   0.49 &   0.22 &   0.31 &   0.23\\
\hline
GM Tau &    2.5    &   5.0 &     13.9 &     16.1 &      4.6 &      1.1 &      0.5 &      0.2\\
      &    0.019B &   $0.055$ID    &   0.23 &   0.26 &   0.12 &  0.052 &   0.030 &  0.015\\
\hline
2M0414+2811 &   2.8 B  &      4.2 &      8.2 &      9.6 &      7.3 &      2.4 &      1.5 &      1.1\\
      &   0.038 B &   0.085 &   0.20 &   0.24 &   0.23 &   0.11 &   0.084 &  0.062\\
\hline
\hline
\end{tabular}}
\end{table}

\begin{table}
\label{tab:obsx.tab}
{\scriptsize
\begin{tabular}{ccccccccccccc}
\multicolumn{9}{l}{Table 8 (continued)}\\
\hline
Star  &    $\lambda4026$  & $\lambda4470$    &  $\lambda4922$   & $\lambda5016$  & $\lambda5876$   & $\lambda6678$  & $\lambda7065$
&   \ion{He}{2} $\lambda4686$\\ 
\hline
\hline
\multicolumn{9}{c}{{\it HST}/STIS sample of higher mass stars}\\
\hline
DF Tau   &     1.9 &      2.8 &      2.0 &      1.3 &     -- &   -- &  -- &      0.7\\
Sep. 99  &     6.1 &  12.4 &  10.8 &   6.6 &  -- &   -- &   -- &   3.3\\
\hline
DF Tau    &  1.7 &      1.3 &      2.4 &      2.6 &   -- &    -- &   -- &      0.4\\
 Aug. 02  &  2.8 &   3.5 &   8.3 &   7.9 &   -- &   -- &   -- &   1.13\\
\hline
RU Lupi&   3.3B &    3.3 &      9.5 &      9.5 &   -- &    -- &   -- &  $<0.35$*\\
      &  25B &  33 &  90 &  92 &   -- &    --&   -- &  $<3.2$*\\
\hline
DG Tau & 1.3   & 1.7  & 3.2 & 4.4 & -- & -- & -- & $<0.5$\\
       & 0.65  & 1.5  & 3.9 & 5.0 & -- & -- & -- & $<0.5$\\
\hline
TW Hya &      2.2 &      2.8 &      0.9 &      1.0 &     --   &   -- &  -- &      1.1\\
 May 00     &  16.8 &  29.8 &  10.9 &  11.8 &   -- &   -- &   -- &  14.1\\
\hline
TW Hya &  2.3 &      2.9 &   1.5 &  1.2 &   -- &   -- &    -- &      1.0\\
 Jul. 02     &  14.9 &  28 &  17 &  12.8&   --  &   -- &   -- &  11.4\\
\hline
V836 Tau &    1.1 &      0.6 &      0.7 &      0.3 &     -- &     -- &  -- &      0.5\\
         &  0.18 &   0.26 &   0.40 &   0.19 &   -- &  -- &   -- &   0.27\\
\hline
V819 Tau & $<2.5$ & $<1.5$ &$<1.4$ & $<0.7$  &-- &   -- &    -- &  $<1.0$\\
         & $<1.0$ & $<1.5$ & $<2.0$ & $<1.0$ & -- &   -- &    -- &  $<1.5$\\
\hline
LkCa 7   & $<2.8$ & $<1.5$ & $<1.2$ & $<0.4$ &-- &   -- &    -- &  $<1.2$\\
         & $<2.0$ & $<2.5$ & $<3.0$ & $<1.0$ &-- &   -- &    -- &  $<3.0$\\
\hline
\hline
\multicolumn{9}{l}{$^a$Equivalent width and line flux format as in Table 4.}\\
\multicolumn{9}{l}{B=blend, *=marginal detection, ID=questionable line
identification}\\
\end{tabular}}
\end{table}

\begin{table}
\caption{Ca and Na Line Equivalent Widths and Fluxes}
\label{tab:obsx.tab}
{\scriptsize
\begin{tabular}{ccccccccc}
\hline
Star & \ion{Ca}{2} K & \ion{Ca}{2} H$^c$& \ion{Ca}{2} $\lambda8498$ & \ion{Ca}{2} $\lambda8542$ & \ion{Ca}{2} $\lambda8662$ & \ion{Na}{1} $\lambda5893$ & \ion{O}{1} $\lambda7773$ & \ion{O}{1} $\lambda8446$\\
\hline
\hline
CIDA 1&   175 &   160 &    14.7 &    18.9 &    14.4 &    10.4 &     1.7 &     2.5\\
      &    1.49 &    1.32 &    7.8 &   10.4 &    9.1 &    0.48 &    0.54 &    1.36\\
\hline
GM Tau&   106 &   135 &    28 &    29 &    21 &    15.4 &     1.14 &     3.4\\
      &    1.35 &    1.15 &    4.4 &    4.5 &    4.1 &    0.41 &    0.114 &    0.55\\
\hline
2M0414+2811&   116 &   142 &    39 &    40 &    30 &    11 &     1.24 &     2.0\\
      &    1.9 &    2.3 &    4.9 &    5.4 &    5.1 &    0.35 &    0.108 &    0.28\\
\hline
S0518+2327 &    20 &    59 &     1.73 &     2.1 &     1.94 &     4.35 &     2.30 &     1.23\\
\end{tabular}}
\end{table}


\begin{table}
\caption{Wind Line Equivalent Widths and Fluxes}
\label{tab:obsx.tab}
{\scriptsize
\begin{tabular}{ccccccccc}
\hline
Star & \ion{O}{1} $\lambda6300$ & \ion{O}{1} $\lambda6363$ & \ion{O}{1} $\lambda5577$ & \ion{S}{2} $\lambda6716$ & \ion{S}{2} $\lambda6731$  & \ion{Fe}{2} $\lambda7155$ & \ion{S}{2} $\lambda4069$ & \ion{S}{2} $\lambda4076$ \\
\hline
\hline
CIDA 1 &     1.17 &     $<$0.32 &     0.48 &     $<$0.28 &     0.21 &     $<$0.30* &     $<$3.5B &     $<$2.7B\\
      &   0.084 &   $<$0.029 &   0.0096 &   $<$0.030 &   0.023 &   $<$0.042* &   $<$0.034B &   $<$0.028B\\
\hline
GM Tau &     5.4 &     2.0 &     1.9 &     0.34 &     0.64 &      $<$0.27* &      $<$7.5B &      $<$5.0B\\
      &   0.19 &   0.077 &   0.052 &   0.015 &   0.030 &    $<$0.015* &    $<$0.069B &    $<$0.040B\\
\hline
2M0414+2811 &     3.2 &     1.17 &     1.01 &     0.42 &     0.17 &     0.15 &     $<$2.8B &     $<$2.8B\\
      &   0.12 &   0.047 &   0.033 &   0.020 &   0.0078 &   0.0081 &   $<$0.047B &   $<$0.049B\\
\hline
S0518+2327 &     1.02 &     0.44B &     1.30 &      $<$0.21* &      $<$0.17 &     0.25 &      $<$0.63 &      $<$1.47B\\
\end{tabular}}
\end{table}

\begin{table}
{\scriptsize
\begin{tabular}{ccccccccc}
\multicolumn{9}{l}{Tables 9 and 10 (continued for {\it HST} Sample)}\\
\hline
Star & \ion{Ca}{2} K & \ion{Ca}{2} H$^c$& [\ion{O}{1}] $\lambda5577$ & [\ion{S}{1}] $\lambda4069+4076^d$\\
\hline
DF Tau &      10.4 &     30  & 0.7 & $<1.5$\\
Sep. 99      &     31 & 92   & 6.6 & $<4.7$\\
\hline
DF Tau &      28 &   38 & 0.9 & 3.0\\
 Aug. 02   & 42 &    56 & 4.1 & 4.7\\
\hline
RU Lupi &     50 & 39  &$<0.3$ & 9.2\\
      &    500 &   330 & $<3$  & 72 \\
\hline
DG Tau &  61  & 44 & 0.39  & 19$^e$\\
       &  33  & 22 & 0.67  & 9.2$^e$ \\
\hline
TW Hya &       9.6 &     32.0  &$<0.15$& $<0.8$\\
 May 00     &   65 &   230    & $<2.5$&  $<6.0$\\
\hline
TW Hya &   11.5 &    45  & $<0.15$& $<0.7$ \\
Jul. 02  &  65  &   270  & $<2.0$ & $<4.0$  \\
\hline
V836 Tau &      12.9 &     19.5 & $<0.26$* &  4.1\\
      &      1.37 &     2.6 & $<0.24$*     &  0.79\\
\hline
V819 Tau &  4.1  &  4.2 & $<0.5$ & $<1.7$\\
         &  0.95 &  1.2 & $<1.2$ & $<0.8$\\
\hline
LkCa 7   & 7.9 &  7.0  & $<0.5$ & $<2.5$\\
         & 3.8 &  4.2  & $<2.0$ & $<2.0$\\
\hline
\hline
\multicolumn{9}{l}{$^a$Equivalent width and line flux format as in
  Table 4.}\\
\multicolumn{9}{l}{B=blend, *=marginal detection, ID=questionable line
identification}\\
\multicolumn{9}{l}{$^c$Blended with H$\epsilon$}\\
\multicolumn{9}{l}{$^d$Lines are marginally resolved and are coadded here}\\
\multicolumn{9}{l}{$^e$Emission extended by $0\farcs3$ in slit and included in extraction}\\
\end{tabular}}
\end{table}



\begin{table}
\caption{Model Accretion Parameters}
\label{tab:obsx.tab}
{\footnotesize
\begin{tabular}{ccccccccccc}
  \hline
Star   & Template & $T$ & $n_e$ & $L$ & $\delta$  & $r_{3600}$ & $r_{4000}$  & $L_{acc}$ & $\dot{M}_{acc}$ \\
& & K & $10^{14}$ cm$^{-3}$ & $10^7$ cm & \% & & & $10^{-4}$ $L_\odot$ & $10^{-10}$ $M_\odot$ yr$^{-1}$\\
\hline
\hline
\multicolumn{10}{c}{Low-mass accretors in {\it Keck I} sample}\\
\hline
CIDA 1 & V927 Tau$^a$& 10500  & 0.8  &  10   & 1.01  & 70   &  4& 280  & 120 \\
GM Tau & MHO 7   & 9000   & 0.7  & 20   & 0.53   & 110  & 4.9  &  81  & 20 \\
2M0414+2811 & CFHT 7  & 7400   & 2.5  & 2   & 3.3    & 200  & 20   &  32 & 9.0\\
S0518+2327  & MHO 7   & 10500  & 2.5  & 1   & 0.59   & 68  & 6.7   &  28 & 5.0 \\
MHO 6  & V927 Tau& 7800   & 1.5  & 8    & 0.31   & 3.4  & 0.29 &  14 & 4.4 \\
\end{tabular}}
\end{table}

\begin{table}
\caption{Error Estimates}
\label{tab:obsx.tab}
\begin{tabular}{cccccccccc}
\hline
Parameter        & Uncertainty &  Dependent Variables   &$log (\sigma \dot{M})^a$ & Systematic\\
\hline
\hline
$A_V$            & 0.5 mag       & $L_{acc}$              &  0.4    & ? \\
SpT              & 0.5 classes   & M, R                  &  0.14   & No? \\
Pre-main sequence track & 0.03--0.4 dex   & M           &  $-0.03$ to $-0.4$ & Yes\\
Mass calibration &  0.2 dex    & $M$  & $0$ to $-0.2$ & Yes\\
Distance         & 15 pc       & $L_{acc}^2$, $R$       &  0.13   & No? \\
Exclusion of lines & $0.25 L_{acc}$   &  $L_{acc}$    &  $+0.03$ to +$0.25$  & Yes  \\
Flux Calibration & 0.18 dex   & $\propto L_{acc}$,$R^{1/2}$ & 0.26 & Sometimes\\
Bolometric Correction& 0.3 dex    & $L_{acc}$              & $+0.3$ & Yes\\
Truncation Radius    &  0.3 dex    &  $R_{in}$            & 0.08 & ?\\
Photospheric Template&  0--0.3 dex  & $L_{acc}$ &  $0$ to $+0.3$ & No?\\
Single/binary    & 0.15 dex  & $R$               & $-0.3$ & Yes\\
Re-radiation of emission  & 0.3 dex    & $L_{acc}$       & $+0.3$ & Yes\\
\hline
\multicolumn{5}{l}{$^a$$+$ or $-$ indicatesthat  the uncertainty is only in one direction.} \\
\end{tabular}
\end{table}


\begin{table}
\caption{Estimated $\dot{M}$ from V-band excess$^a$}
\label{tab:obsx.tab}
{\footnotesize \begin{tabular}{cccccccccc}
\hline
Star    &   V    & M     & R   &  $r_V$ & $A_V$ & $f_{5500}$& $L_{acc}$ & $\dot{M}$ \\  
Star    &   mag   & $M_\odot$ & R &   & mag & $10^{-16}$ erg cm$^{-2}$ s$^{-1}$ AA$^{-1}$& $10^{-4}$ $L_\odot$ &  $10^{-10}$ $M_\odot$ yr$^{-1}$ \\  
\hline
\hline
GM Tau  & 17.577 & 0.15  & 0.93  & $1.7\pm0.9$   & 2.0   & 13  & 79 & 19\\
KPNO 12 & 23.228 & 0.014 & 0.14  & $5^{+3}_{-2}$  &  1.6  &  0.067 & 0.41 &0.16\\
KPNO 4  & 24.722 & 0.01  & 0.42  & $1.6\pm0.8$  &  4.0  &  0.11 & 0.67 & 1.1 \\
KPNO 6  & 22.292 & 0.02  & 0.68  & $1.1^{+0.9}_{-0.7}$ &  0.7  &  0.044 & 0.27 & 0.36\\
\hline
\multicolumn{9}{l}{$^a$Magnitudes, Excess, and stellar properties (except for GM Tau) from \citet{Kra06}.}\\
\end{tabular}}
\end{table}

\begin{table}
\caption{Comparing $\dot{M}$ measured from UV Excess and H$\alpha$}
\label{tab:obsx.tab}
\begin{tabular}{ccccccc}
\hline
      &    \multicolumn{3}{c}{H$\alpha$ Profile$^a$} &  \multicolumn{2}{c}{UV Excess$^b$} & \\
Star   &  H$\alpha$ 10\% width & H$\alpha$ EW & $\log \dot{M}$ & H$\alpha$ EW & $\log \dot{M}^c$ & Ref\\
     &  (\kms)& (\AA) & ($\log$ $M_\odot$ yr$^{-1}$) & (\AA) & ($\log$ $M_\odot$ yr$^{-1}$) & \\
\hline
\hline
MHO 5       & 154   &  -60     &-10.8   &  -43 &-10.7 & 1\\
MHO 6        & 309  &  -25     &-10.3   & -55  & -9.3 & 1\\
CIDA 14      & 289  &  -34     &-10.3   & -10  & -10.9 & 1\\
2M0414+2811  & 362  &  -250    &-10       & -288 & -8.9 & 2\\
2M0439+2544  & 215  &  -102    &-11.3     & -161 & -11.1 & 2\\
\end{tabular}
\end{table}

\begin{table}
\caption{Line Luminosity-Accretion Luminosity Relationships for selected lines}
\label{tab:obsx.tab}
\begin{tabular}{cccc|cc}
\hline
Diagnostic    & $\lambda$  (\AA) & a$^a$      &   b$^a$ & \#$^b$ & $\sigma^c$ \\
\hline
\hline
\ion{He}{1}   & 5016     & $3.3\pm0.3$ & $1.04\pm0.05$ & 18 & 0.30 \\
\ion{He}{1}   & 5876     & $5.3\pm0.7$ & $1.46\pm0.12^d$ & 12 & 0.34 \\
\ion{He}{2}   & 4686     & $3.7\pm0.5$ & $1.04\pm0.08$ & 16 & 0.38  \\
\ion{Na}{1} D & 5893     & $3.3\pm0.7$ & $1.09\pm0.11$ & 11 & 0.41  \\
\ion{Ca}{2}   & 8542     & $2.5\pm0.5$ & $1.02\pm0.11$ & 20 & 0.44 \\
\ion{O}{1}    & 6300     & $2.8\pm0.8$ & $0.96\pm0.16$ & 26 & 0.67 \\
H$\alpha$  & 6563     & $2.0\pm0.4$ & $1.20\pm0.11$ & 35 & 0.57  \\
H$\beta$   & 4861     & $2.6\pm0.2$ & $1.22\pm0.05$ & 19 & 0.26  \\
H$\gamma$    & 4340     & $3.0\pm0.2$ & $1.24\pm0.04$ & 19 & 0.21  \\
\ion{Ca}{2} K & 3932     & $2.0\pm0.3$ & $0.96\pm0.05$ & 19 & 0.33 \\
H-9          & 3771     & $3.4\pm0.2$ & $1.17\pm0.03$ & 19 & 0.18\\
\hline
\multicolumn{6}{l}{$^a\log L_{line}=a + b \log L_{acc}$, with $L_{acc}$ and $L_{line}$ in units of $L_\odot$} \\
\multicolumn{6}{l}{$^b$Number of accretors with detections.}\\
\multicolumn{6}{l}{$^c$Standard deviation of detected line fluxes from linear fit.}\\
\multicolumn{6}{l}{$^d$Significant deviation from 1.0, use with caution outside our $\dot{M}$ range.}\\
\end{tabular}
\end{table}

\pagebreak
\pagebreak
\clearpage

\begin{figure}
\plotone{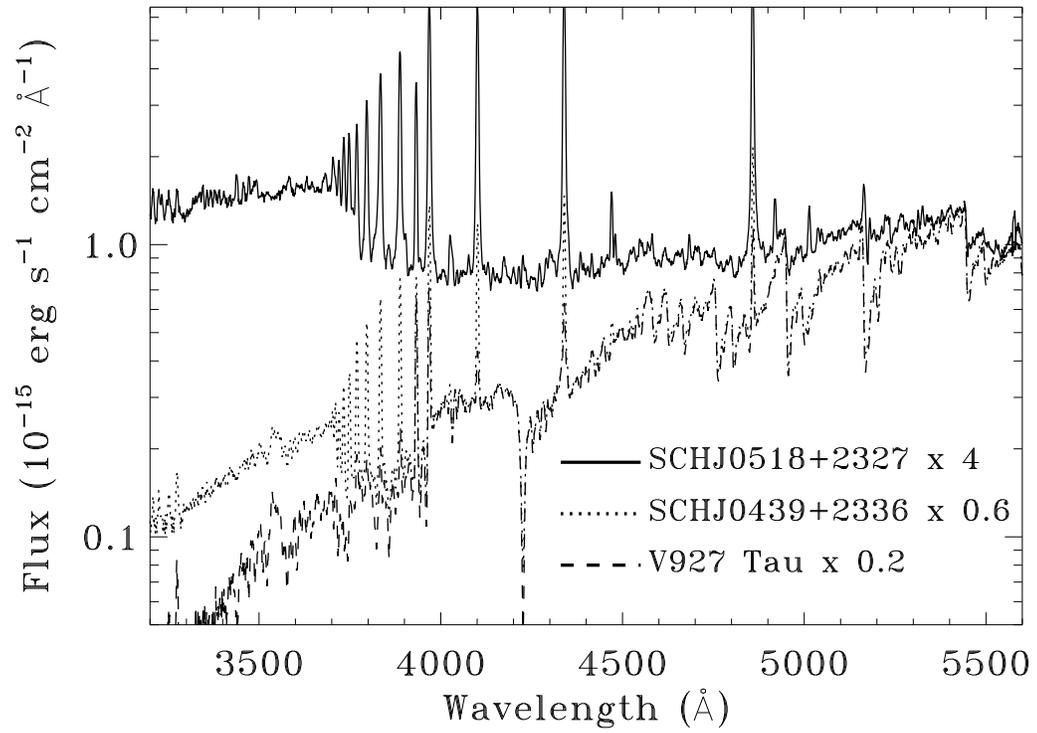}
\caption{Three spectra at 3300--5600 \AA.  The spectrum of
  V927 Tau is dominated by photospheric emission, with strong TiO bands
  and \ion{Ca}{1} $\lambda4227$ absorption line.  Moderate accretors, such
  as S0439+2336, are dominated by photospheric emission longward of the Balmer
  jump at 3700 \AA\ but excess Balmer continuum emission at shorter
  wavelengths.  The TiO bands at $<5600$ \AA\ are only weakly detected from
  the strong accretor S0518+2327, which shows a heavily veiled spectrum both
  shortward and longward of the Balmer jump.}
    \end{figure}

\begin{figure}
\epsscale{0.9}
\plotone{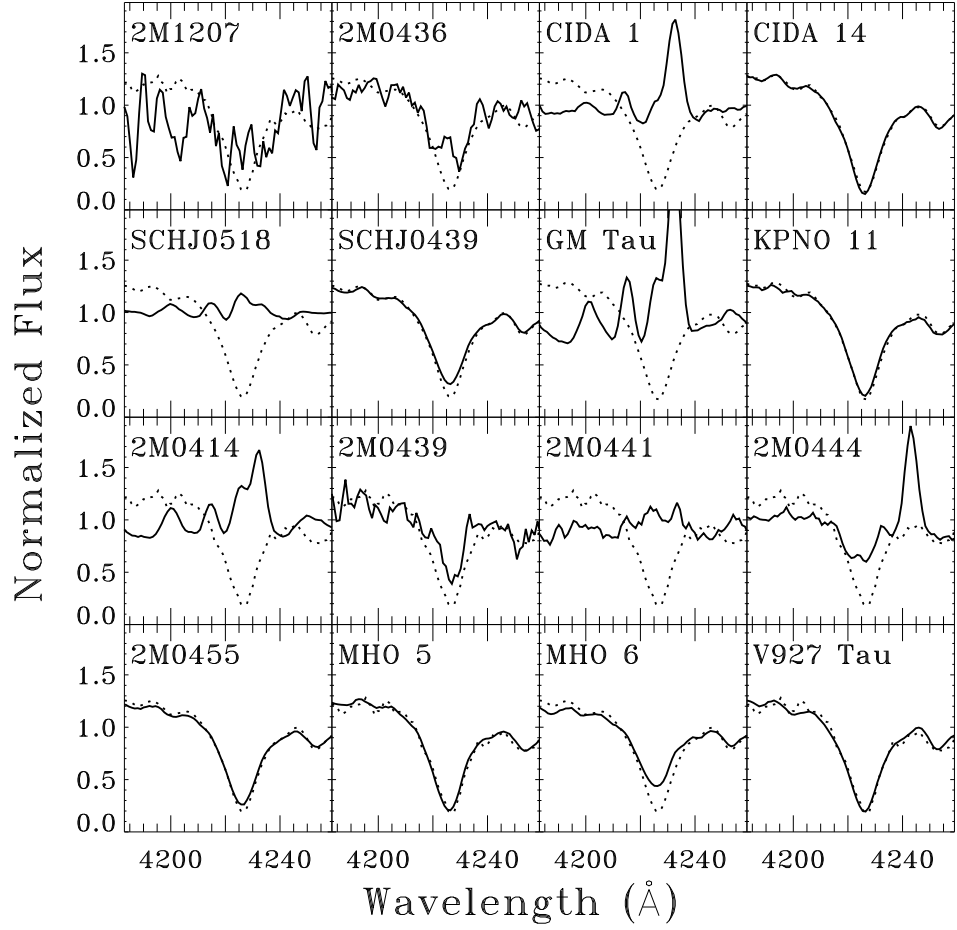}
\caption{The \ion{Ca}{1} $\lambda4227$ absorption line in our LRIS
  spectra (solid lines) compared with a photospheric template (dotted lines, template listed
  in Table 5).  When present, accretion fills in some of the absorption.  The veiling is calculated from the depth of the absorption line.  Emission lines in the vicinity of the \ion{Ca}{1} $\lambda4227$ line from several strong accretors (e.g., CIDA 1, GM Tau) limit our ability to
  measure high veilings and may lead to overestimated small
  veilings.}
\end{figure}

\pagebreak
\begin{figure}
\plotone{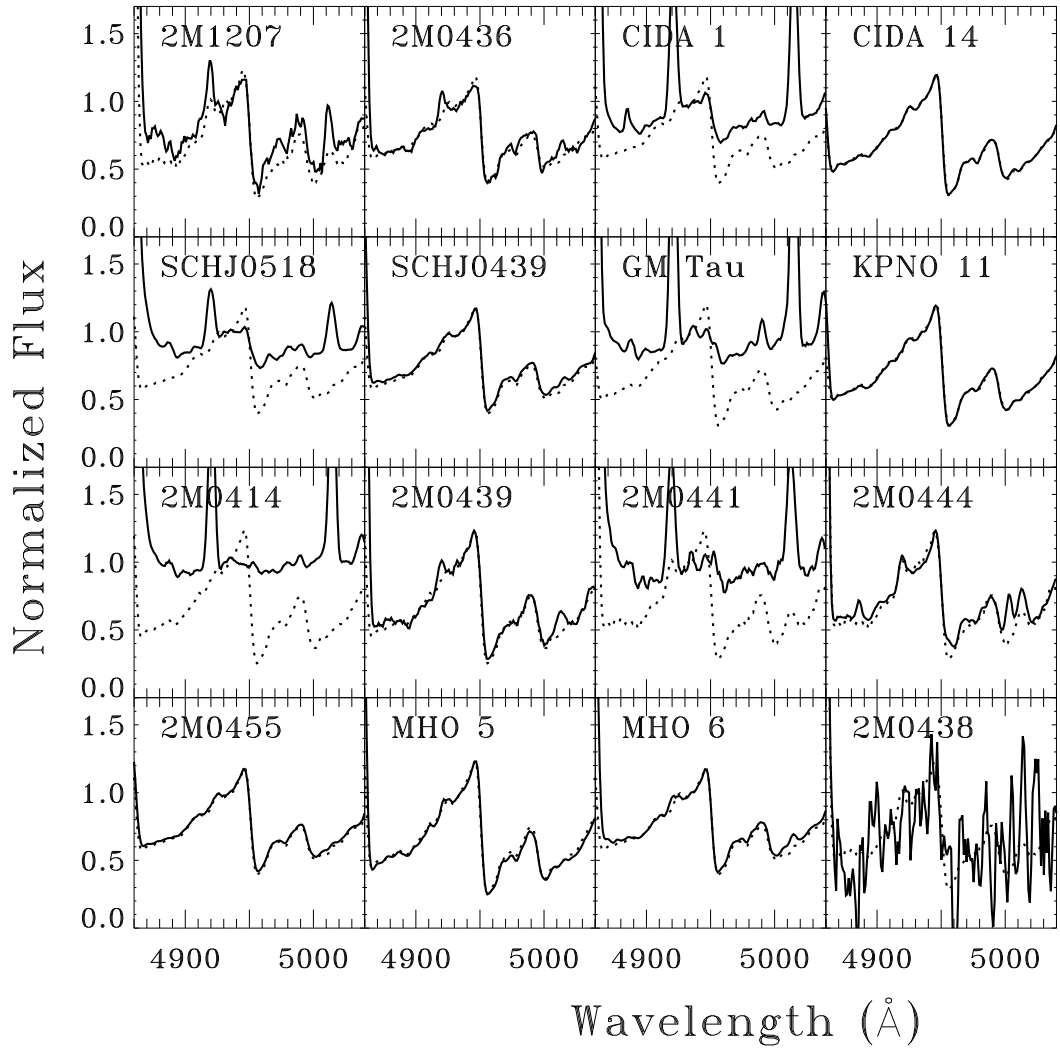}
\caption{A TiO
  absorption band at 4955 \AA\ used for veiling measurements. }
\end{figure}

\begin{figure}
\plotone{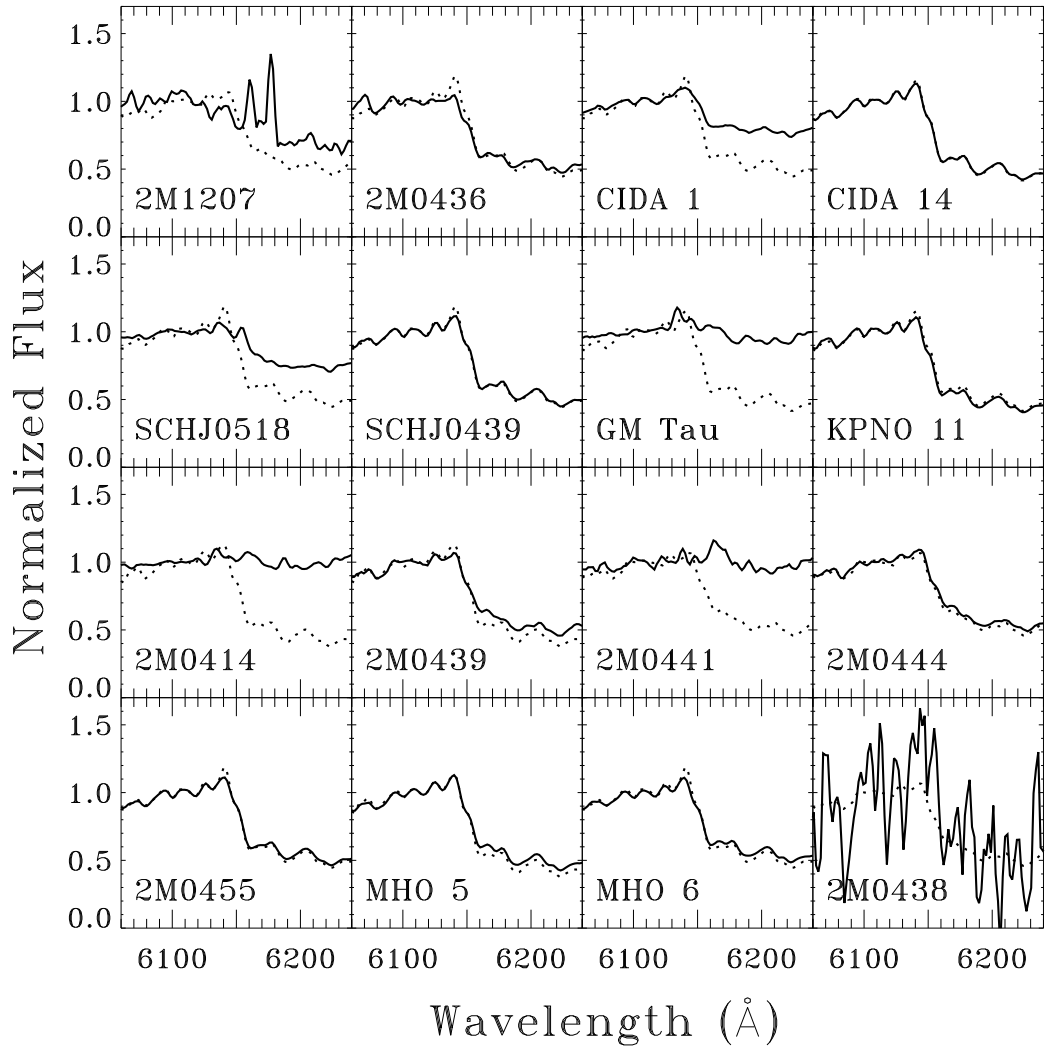}
\caption{A TiO
  absorption band at 6150 \AA\ used for veiling measurements. }
\end{figure}
\pagebreak

\begin{figure}
\plottwo{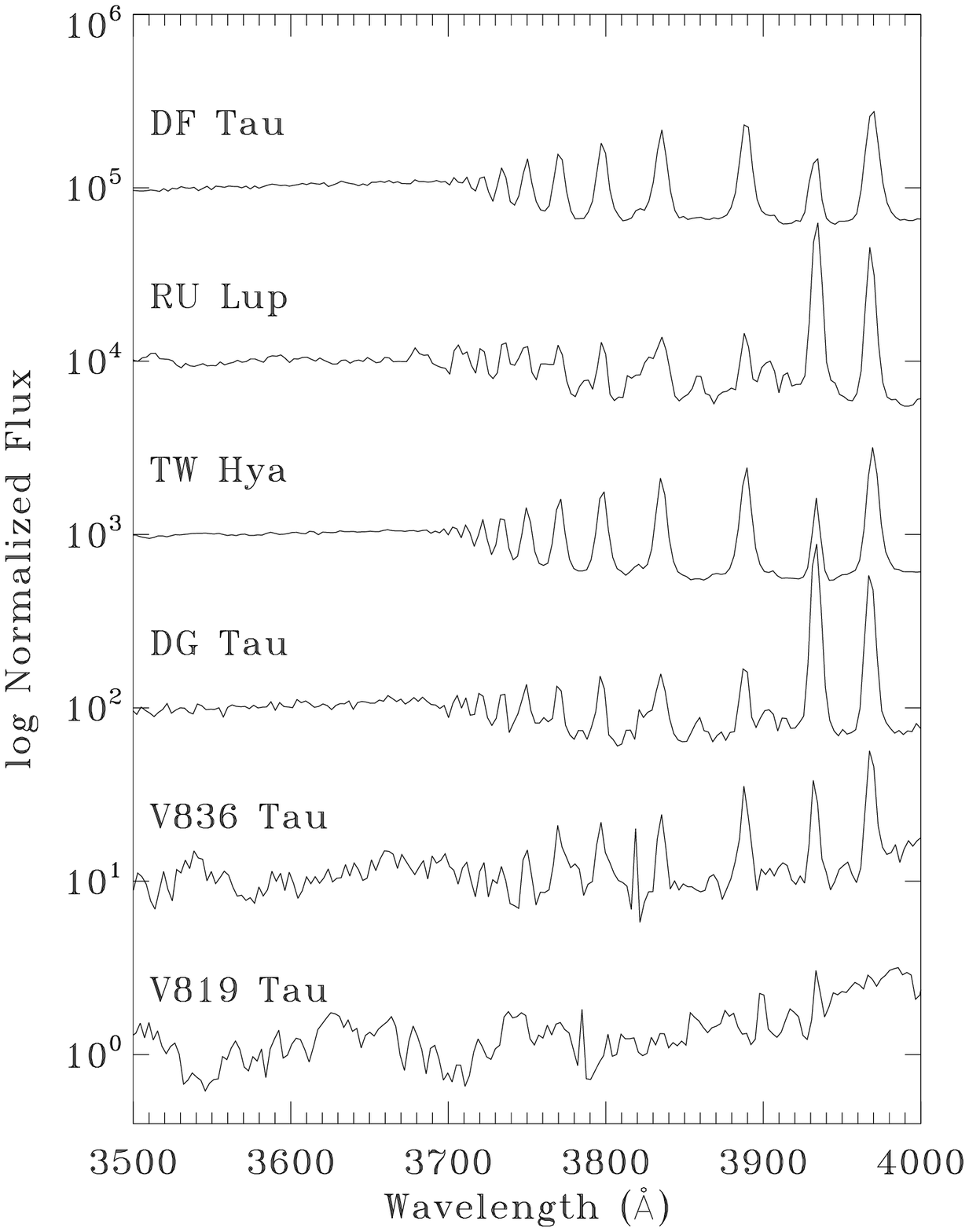}{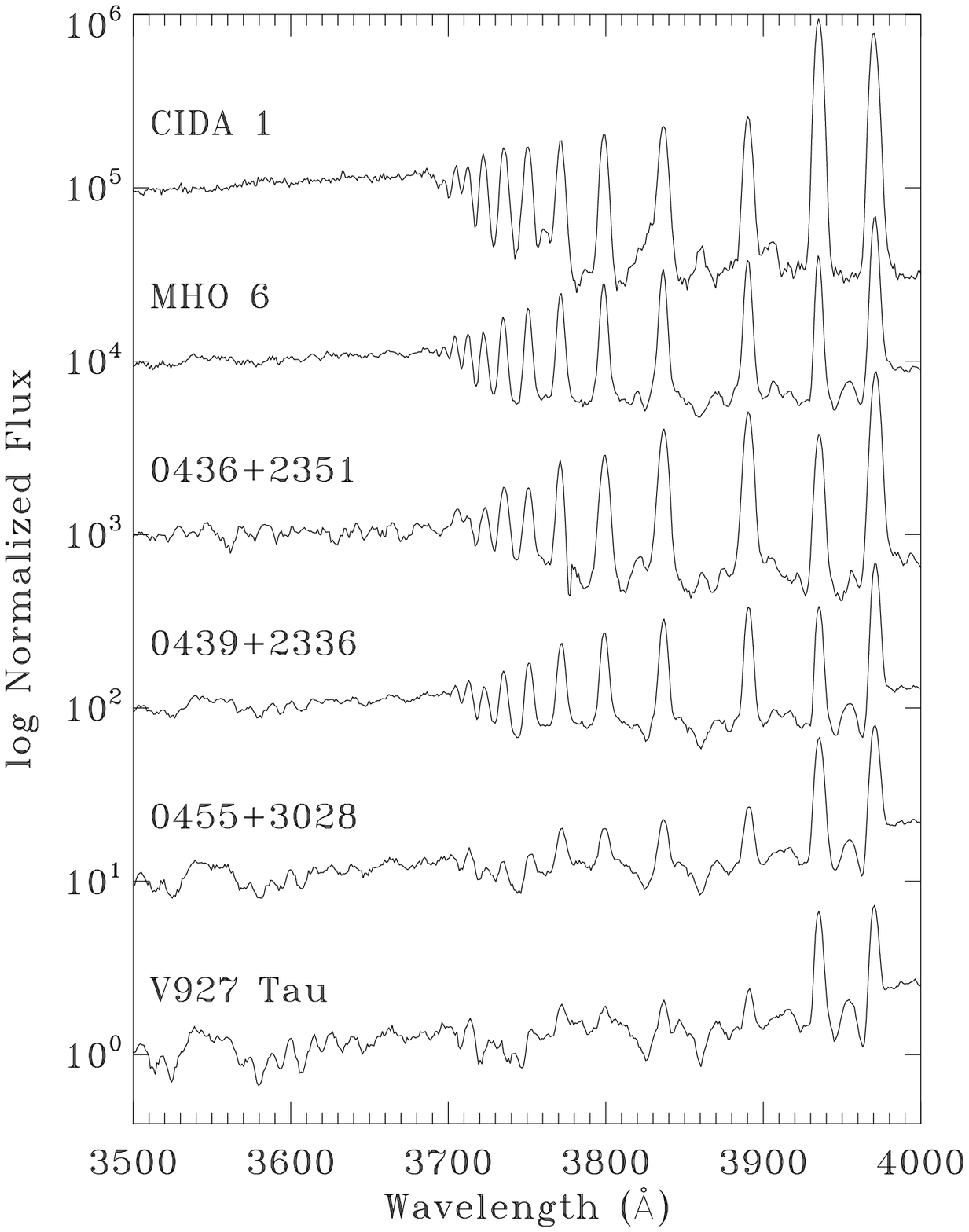}
\plottwo{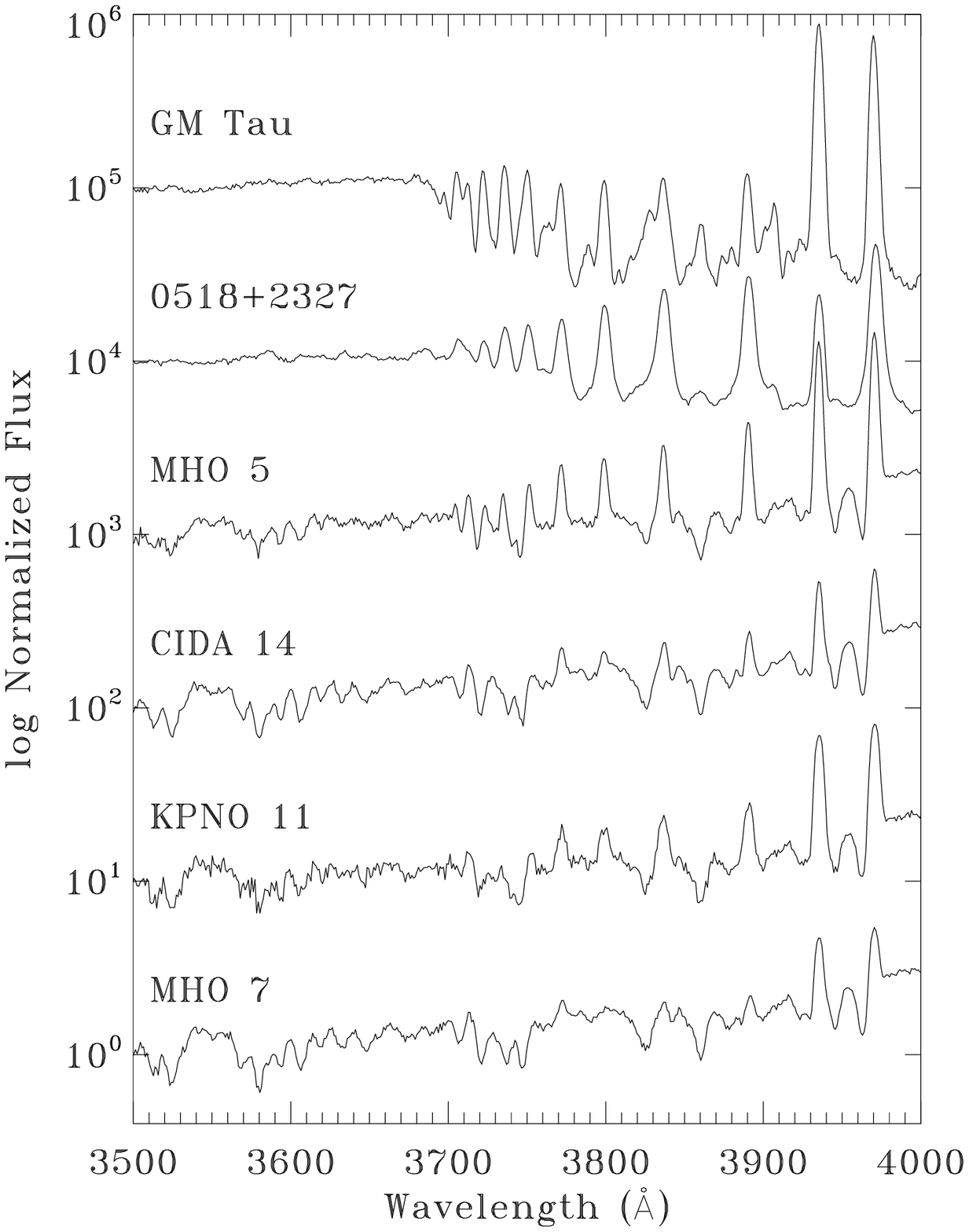}{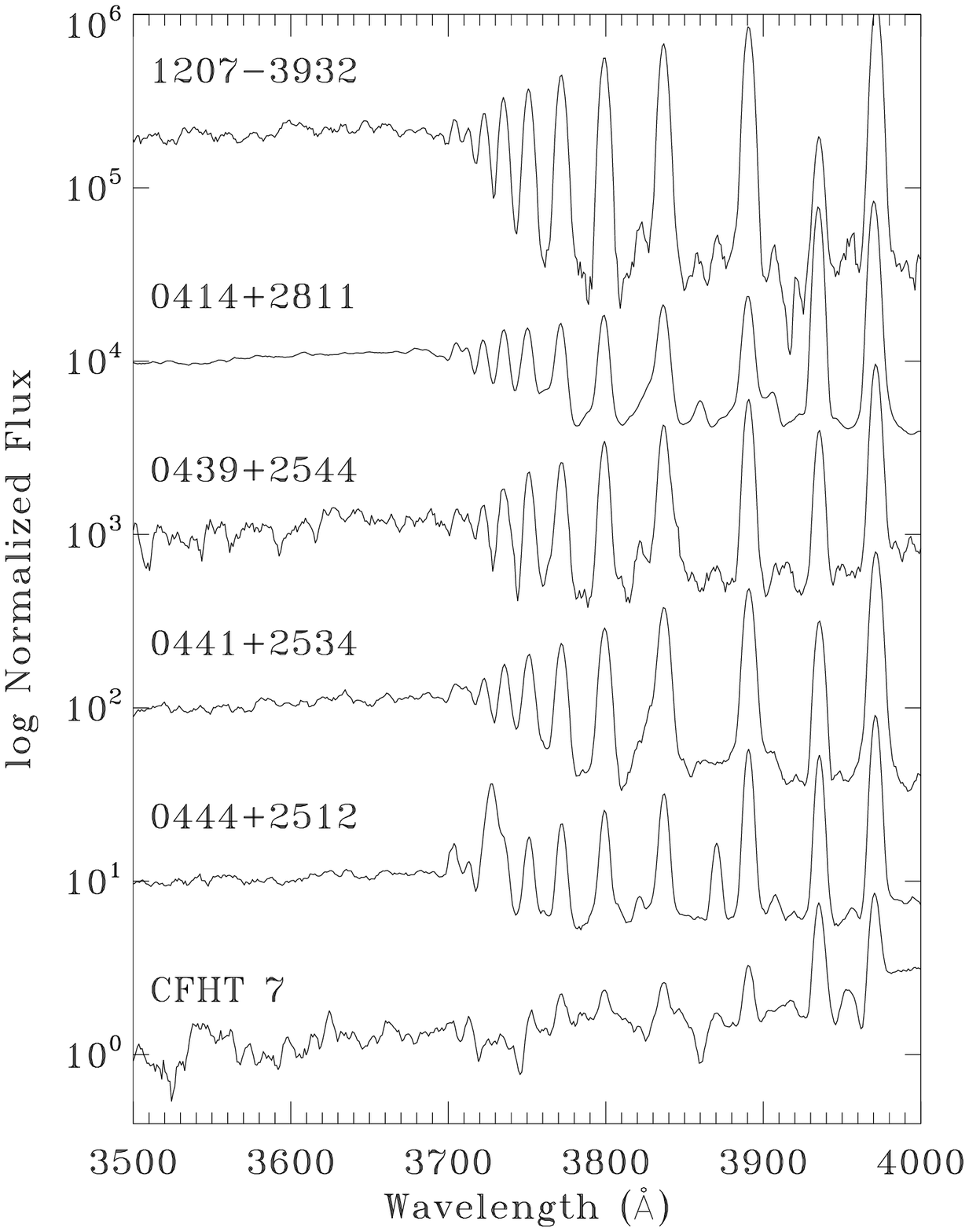}
\caption{The spectral region from 3500--4000 \AA\ for late-K/early M stars (upper left), M5 stars (upper right), M5.5-M6 stars (lower left), and M6.5-M8 stars (lower right). The Balmer continuum, Balmer series, and \ion{Ca}{2} H \& K are apparent in emission.}
\end{figure}


\clearpage

\begin{figure}
\epsscale{0.8}
\plotone{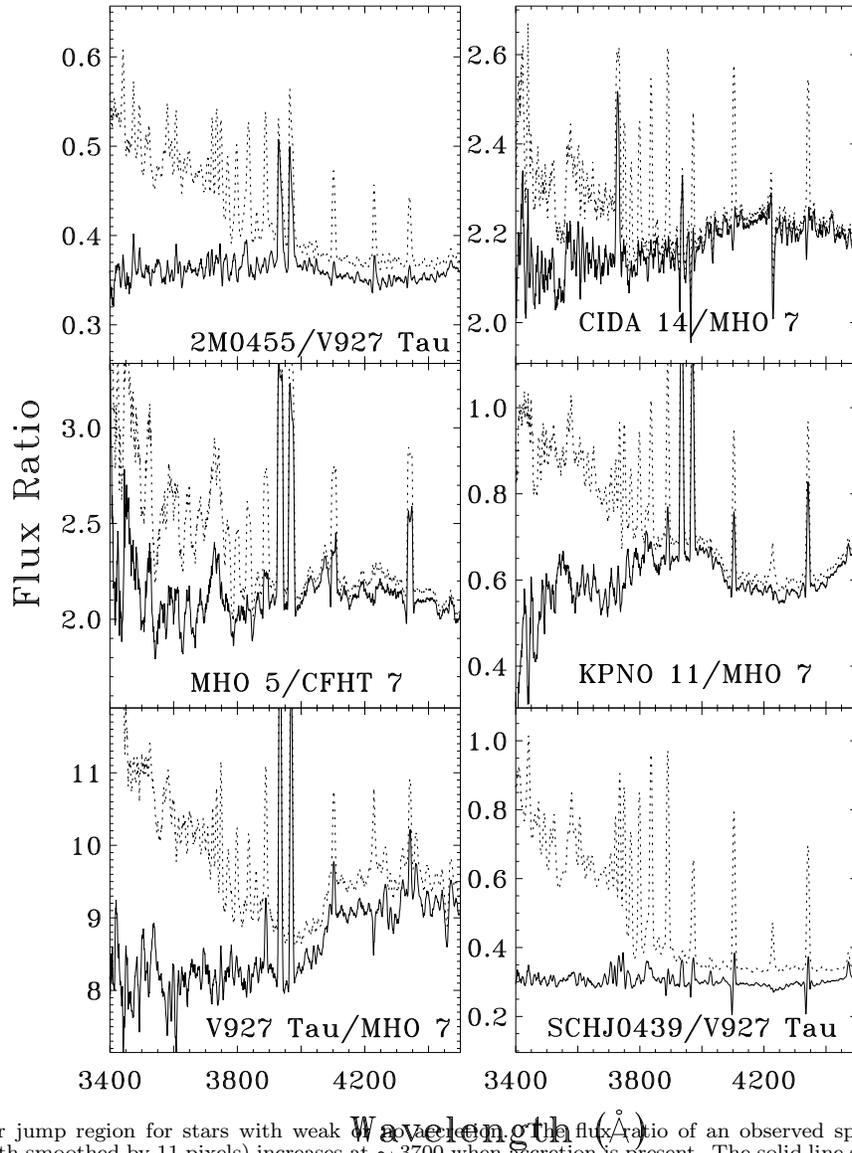}
\vspace{-20mm}
\caption{The Balmer jump region for stars with weak
  or no accretion.  The flux ratio of an observed spectrum to a photospheric
  template (dotted line, both smoothed by 11 pixels) increases at $\sim
  3700$ when accretion is present.  The solid line shows the same ratio, after subtracting excess Balmer line and continuum emission from the spectrum.  For a perfect model and photospheric template, the residuals should be a line equal to the flux ratio with noise introduced from both the spectrum and photospheric template.  Some excess Balmer continuum emission is seen from each source
  plotted here. 
The excess Balmer continuum emission from S0439+2336 is strong enough to be attributed to accretion.  Fainter Balmer excess emission may be produced by 
  either chromospheric activity or accretion.}
\end{figure}

\begin{figure}
\epsscale{1.2}
\plottwo{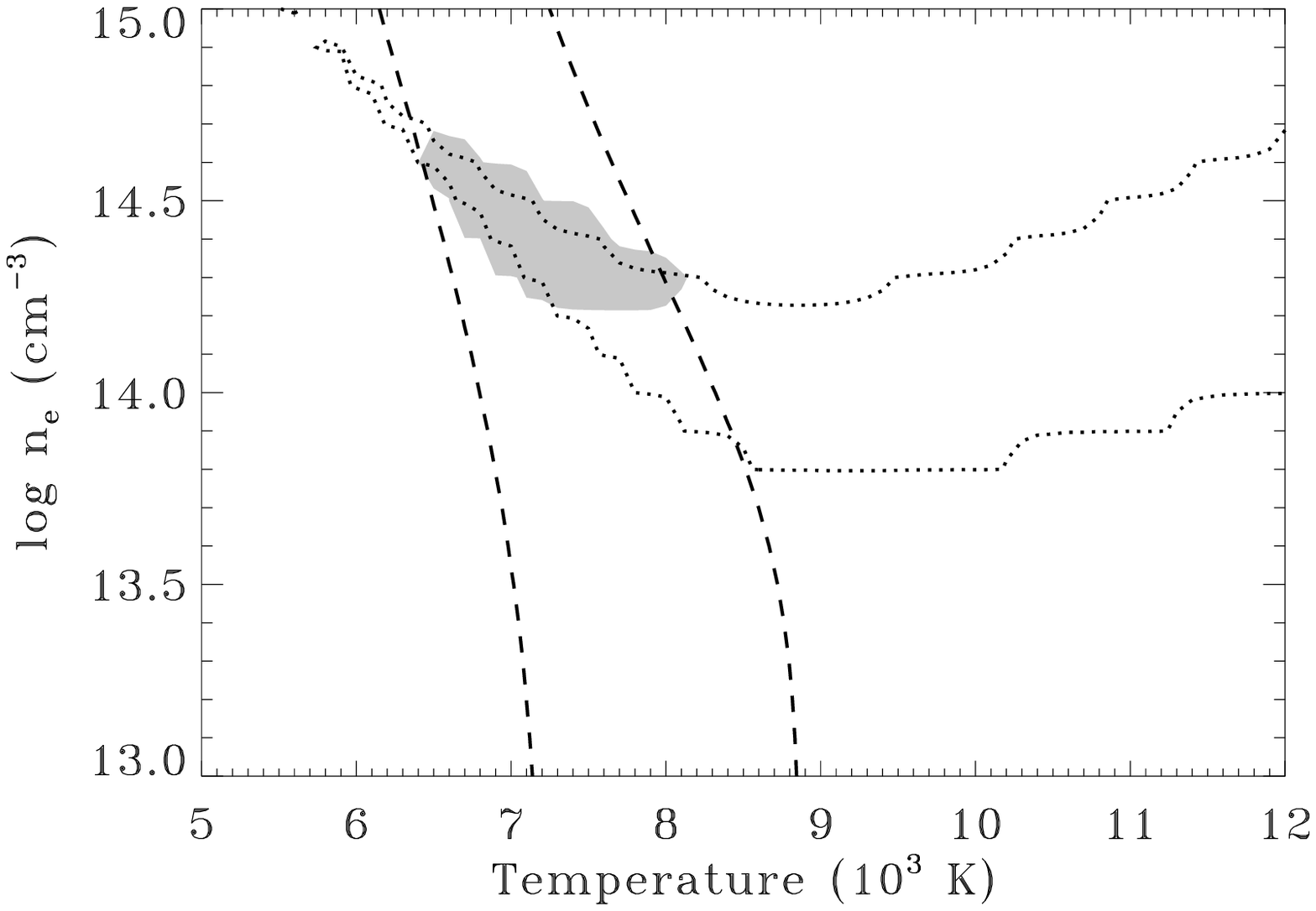}{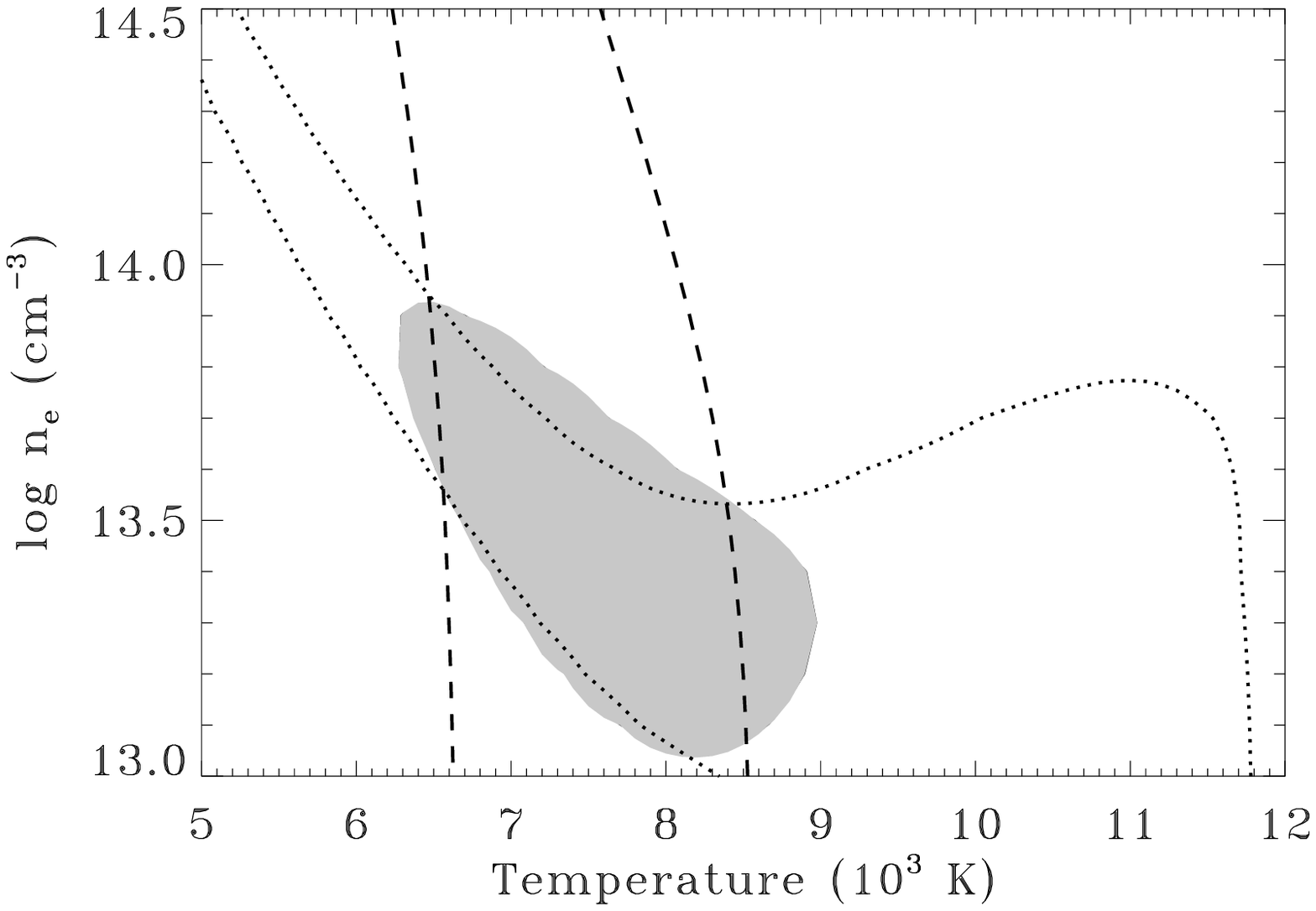}
\caption{Contours of acceptable $T$ and $n_e$ for isothermal hydrogen slab model fits to the spectra of 2M0414+2811 (left) and 2M1207-3932 (right).  The vertical contours that determine $T$ are constrained by the slope of the Balmer continuum.  The horizontal contours that determine $n_e$ are constrained by the size of the Balmer jump.  For 2M0414+2811 the acceptable values of $T$ and $n_e$ together differ from their intersection because they are only acceptable for different values of $L$.}
\end{figure}

\clearpage
\pagebreak
\pagebreak
\epsscale{1.}

\begin{figure}
\plottwo{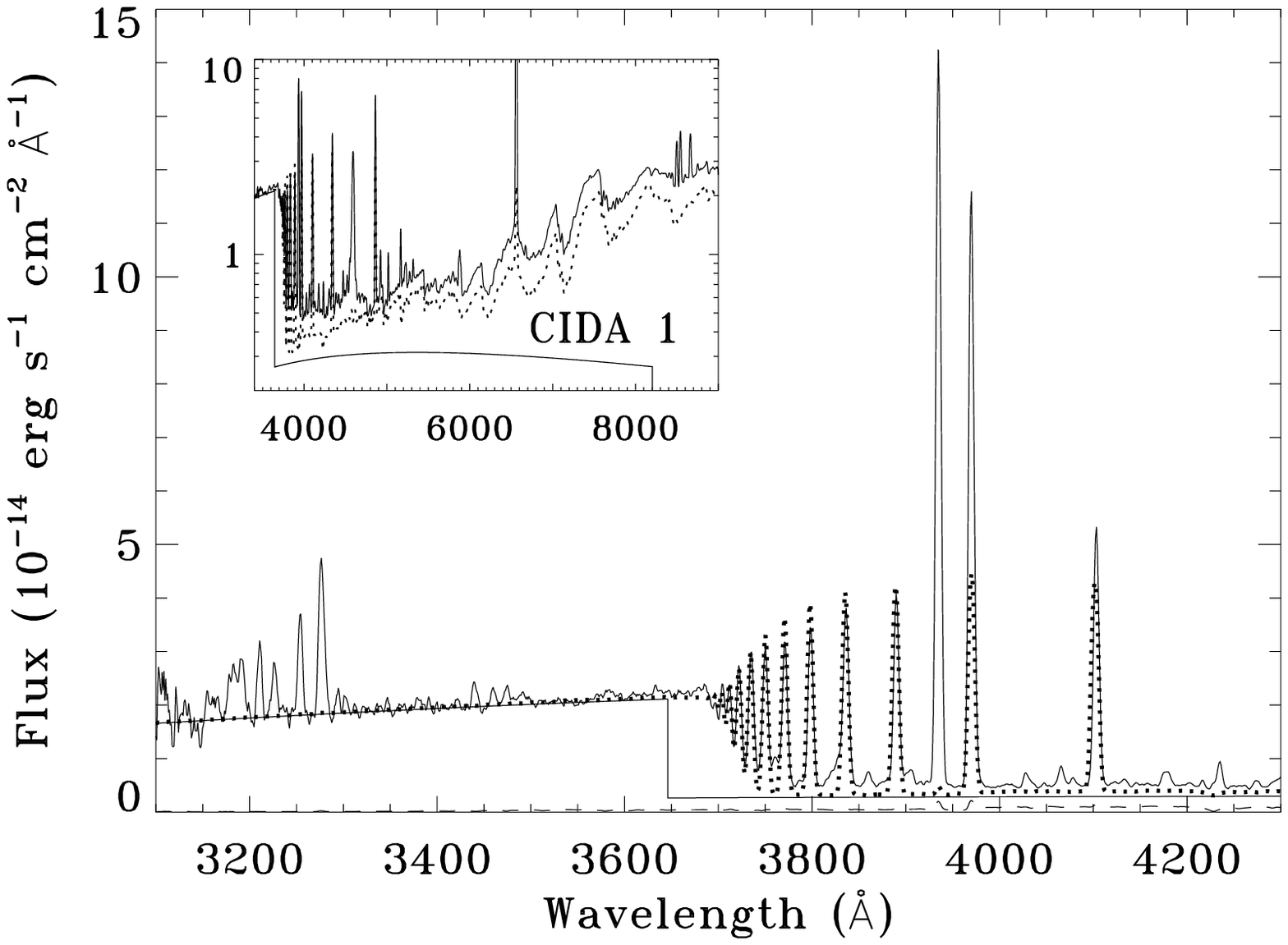}{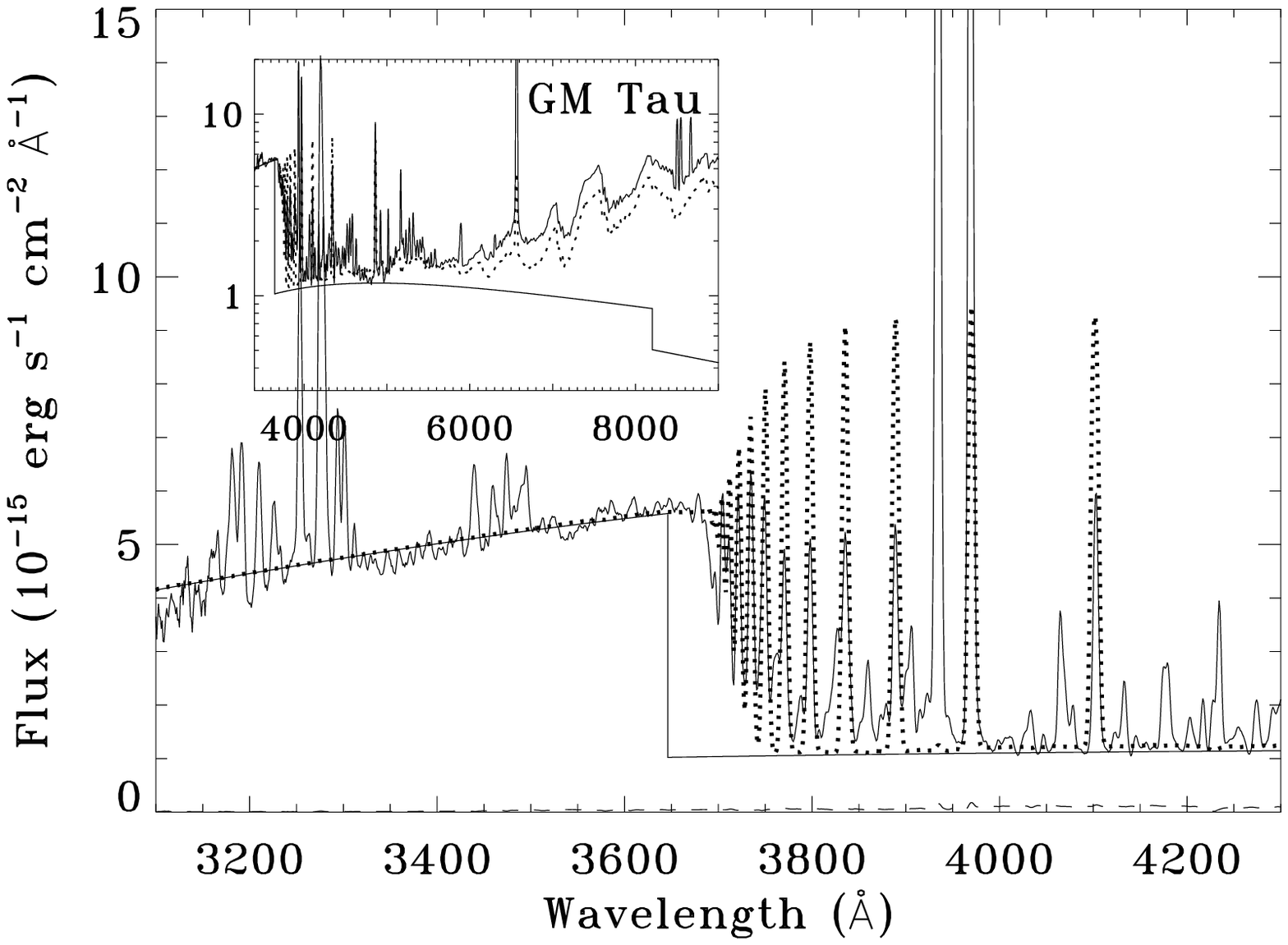}
\plottwo{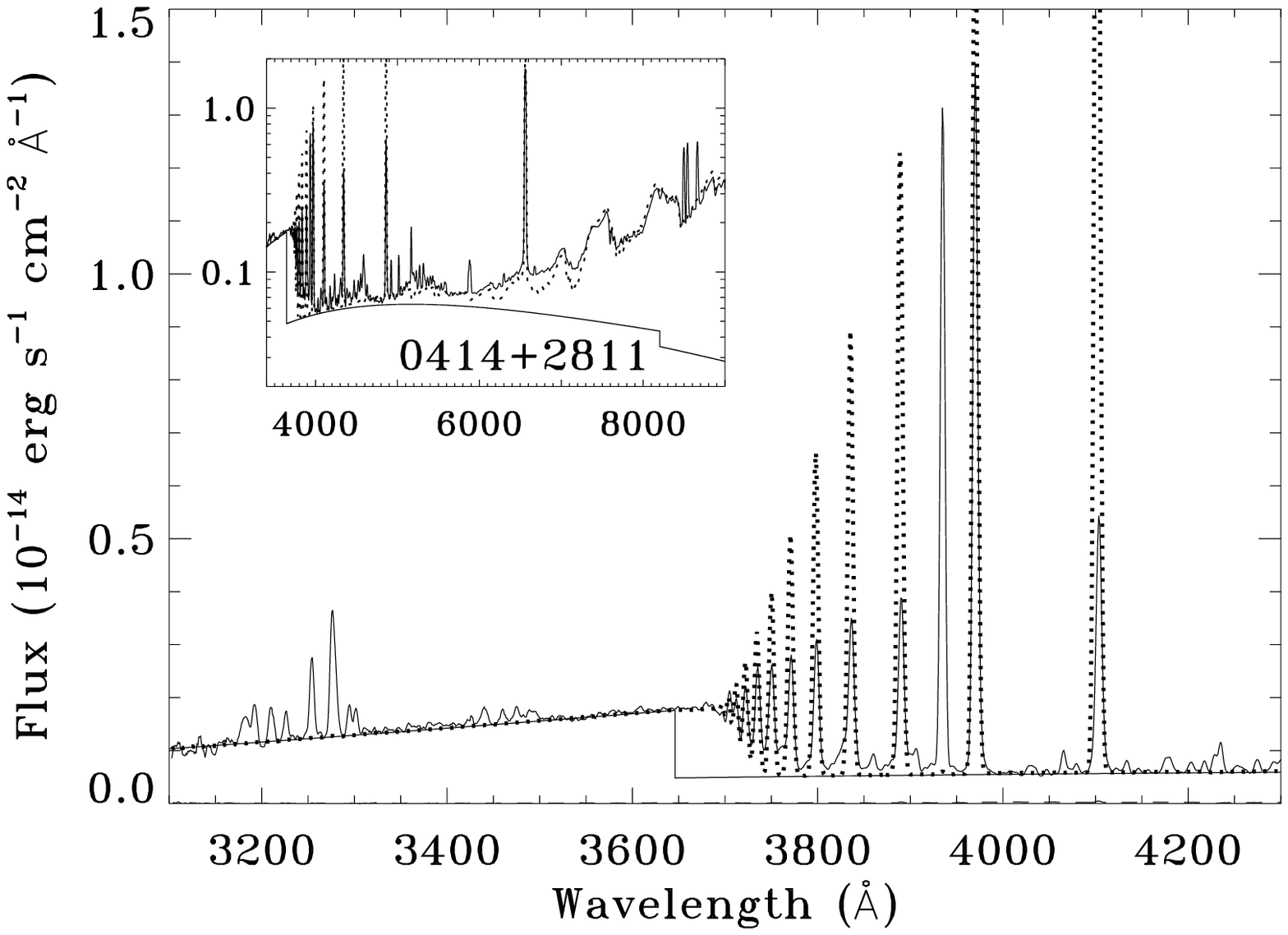}{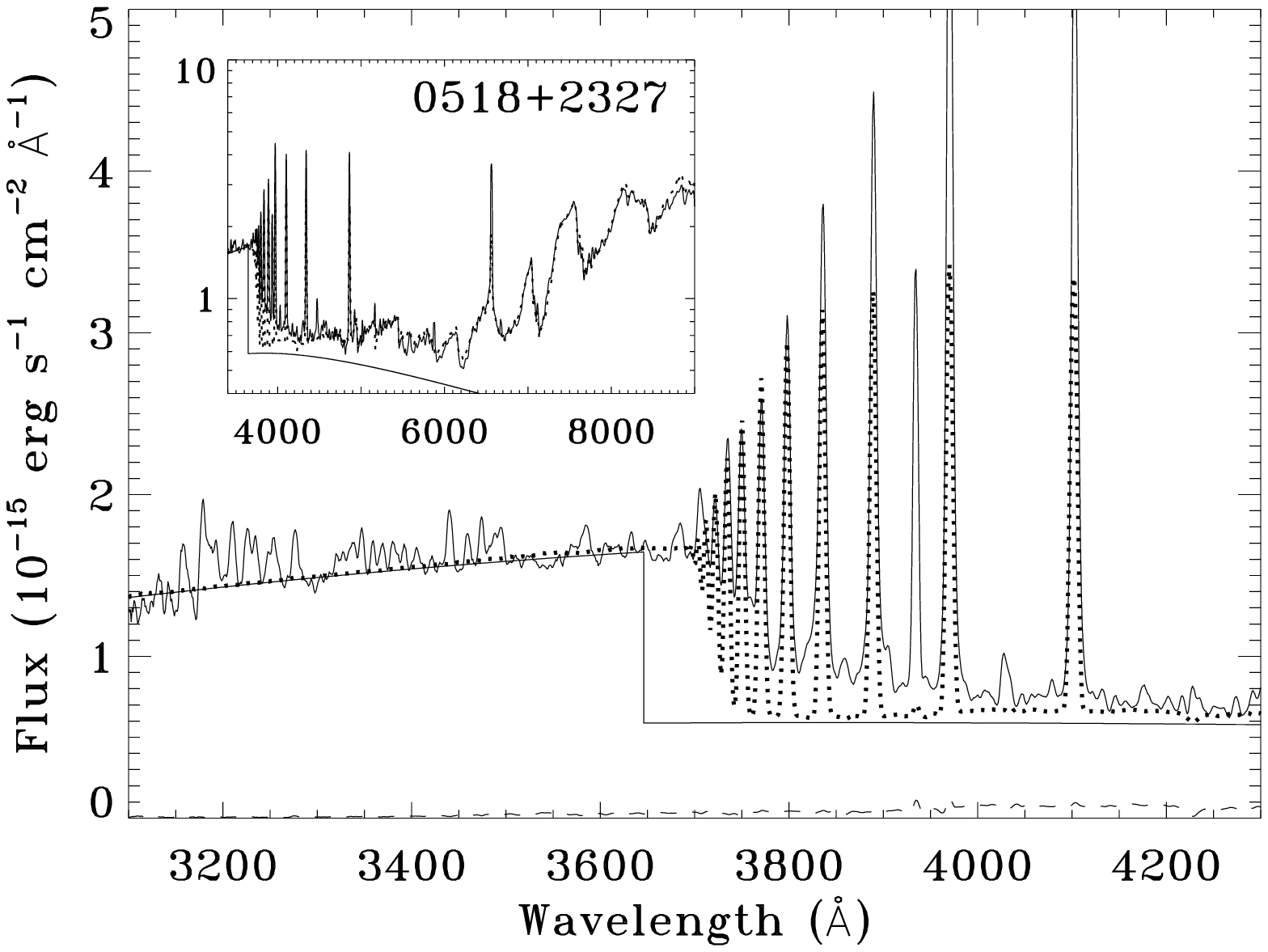}
\end{figure}

\begin{figure}
\plottwo{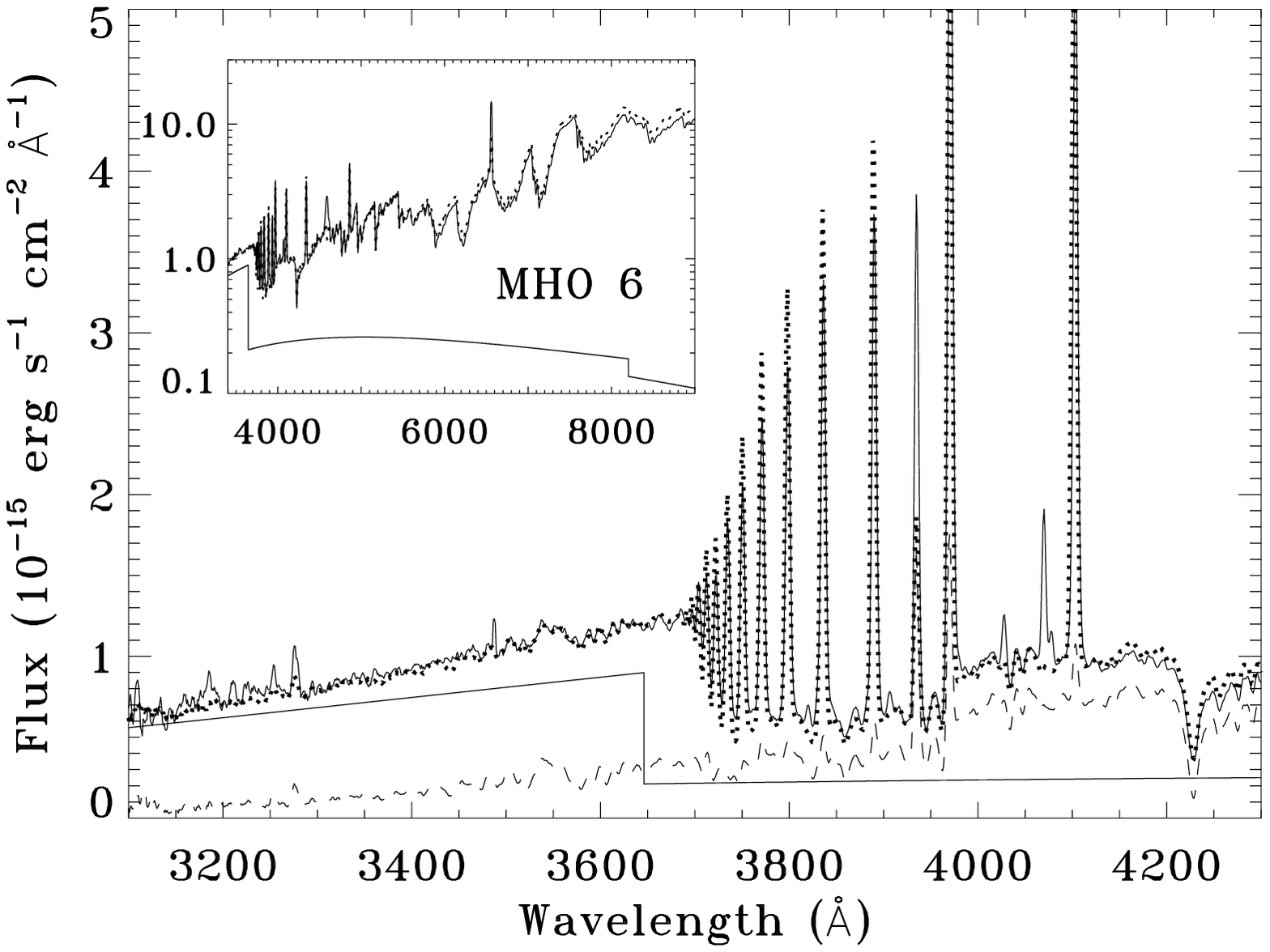}{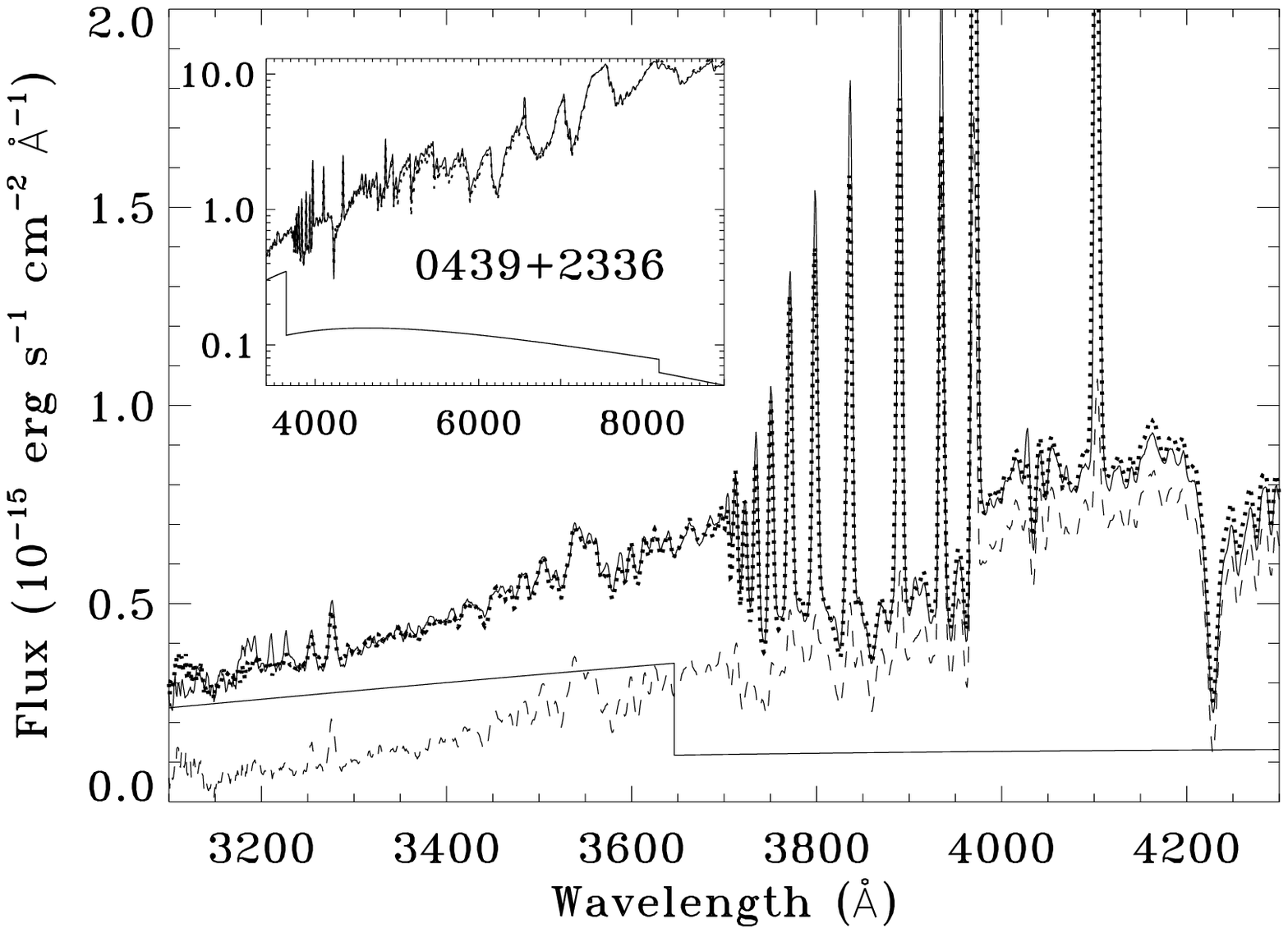}
\end{figure}

\begin{figure}
\plottwo{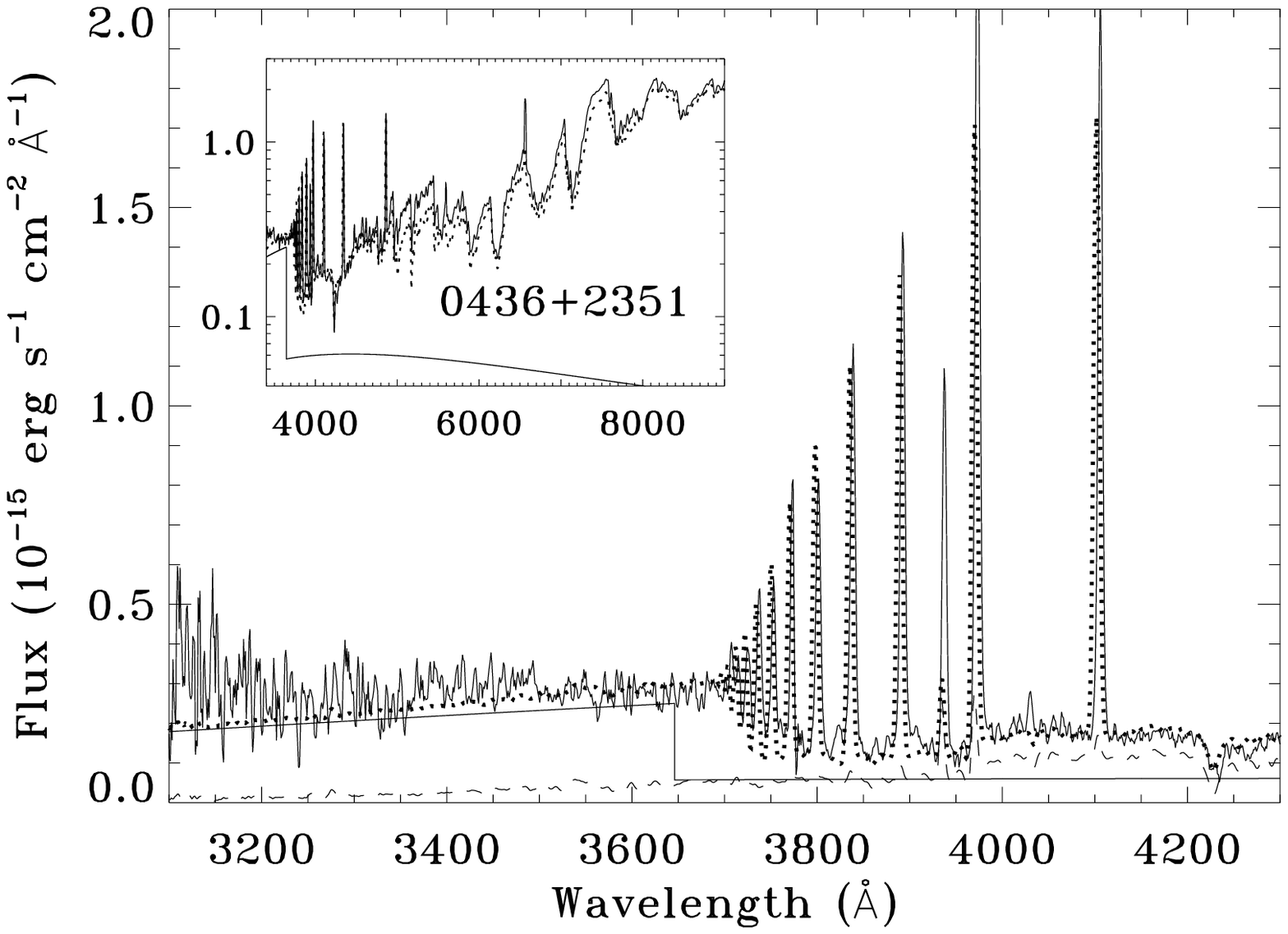}{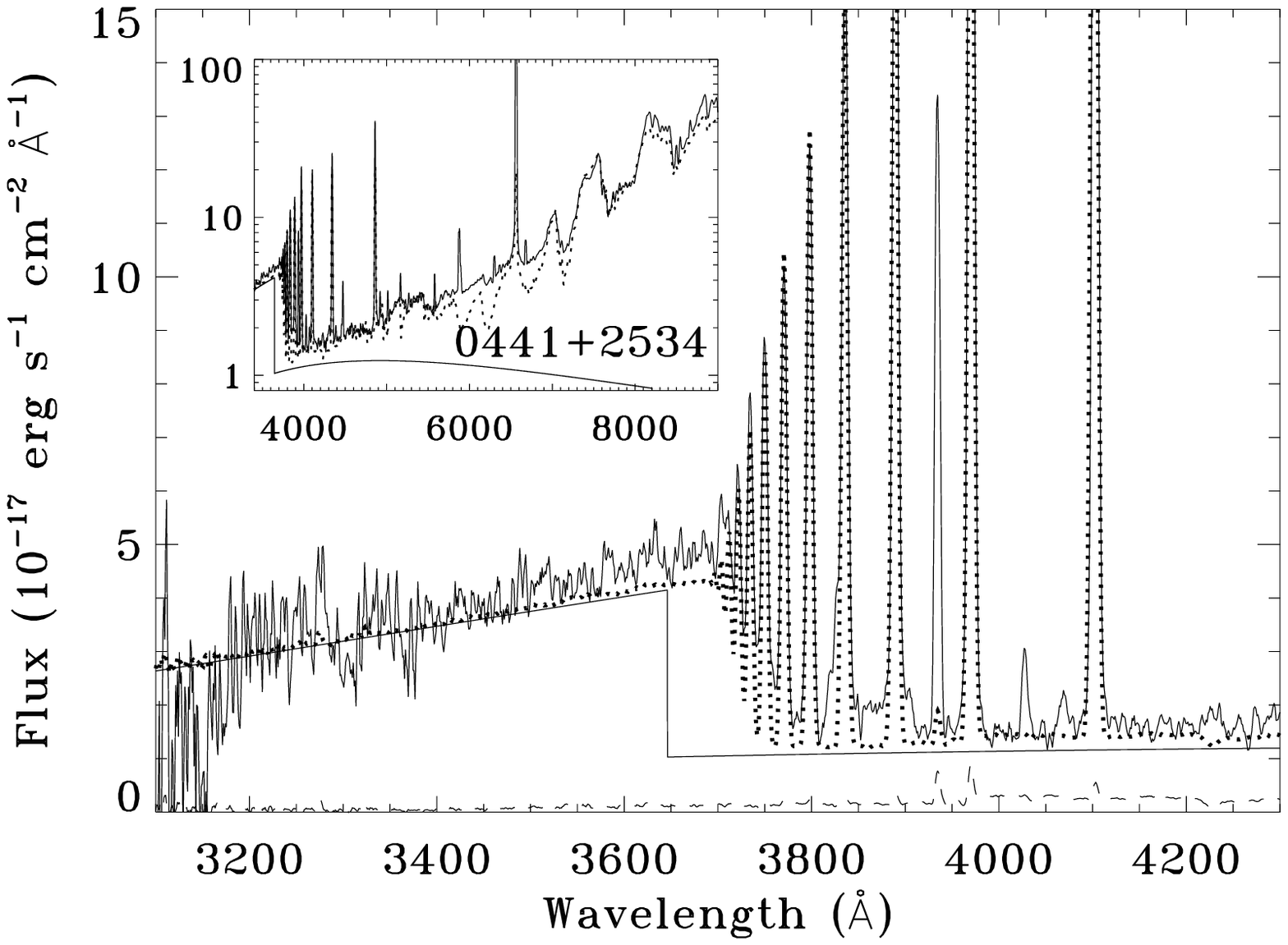}
\end{figure}

\begin{figure}
\plottwo{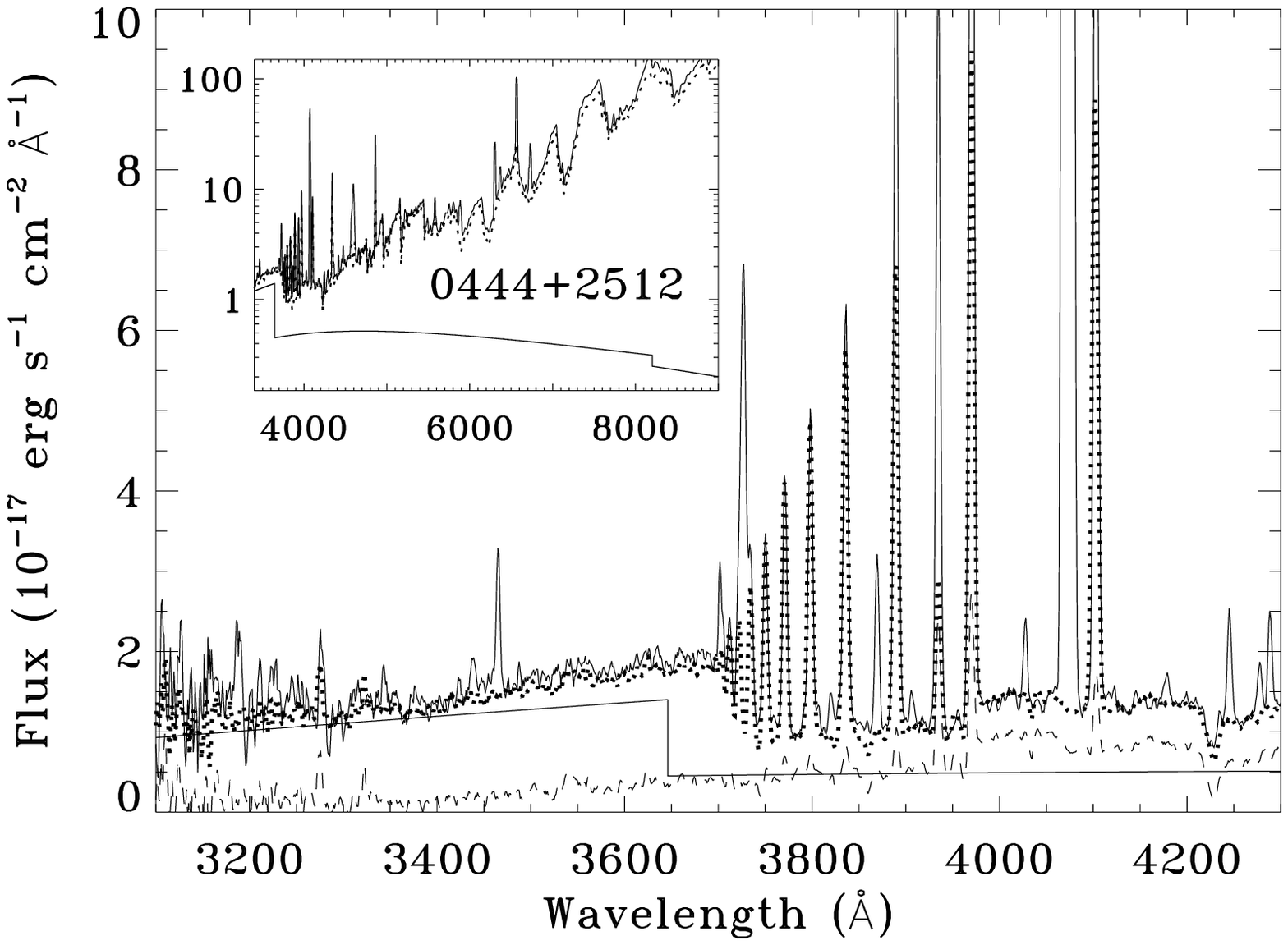}{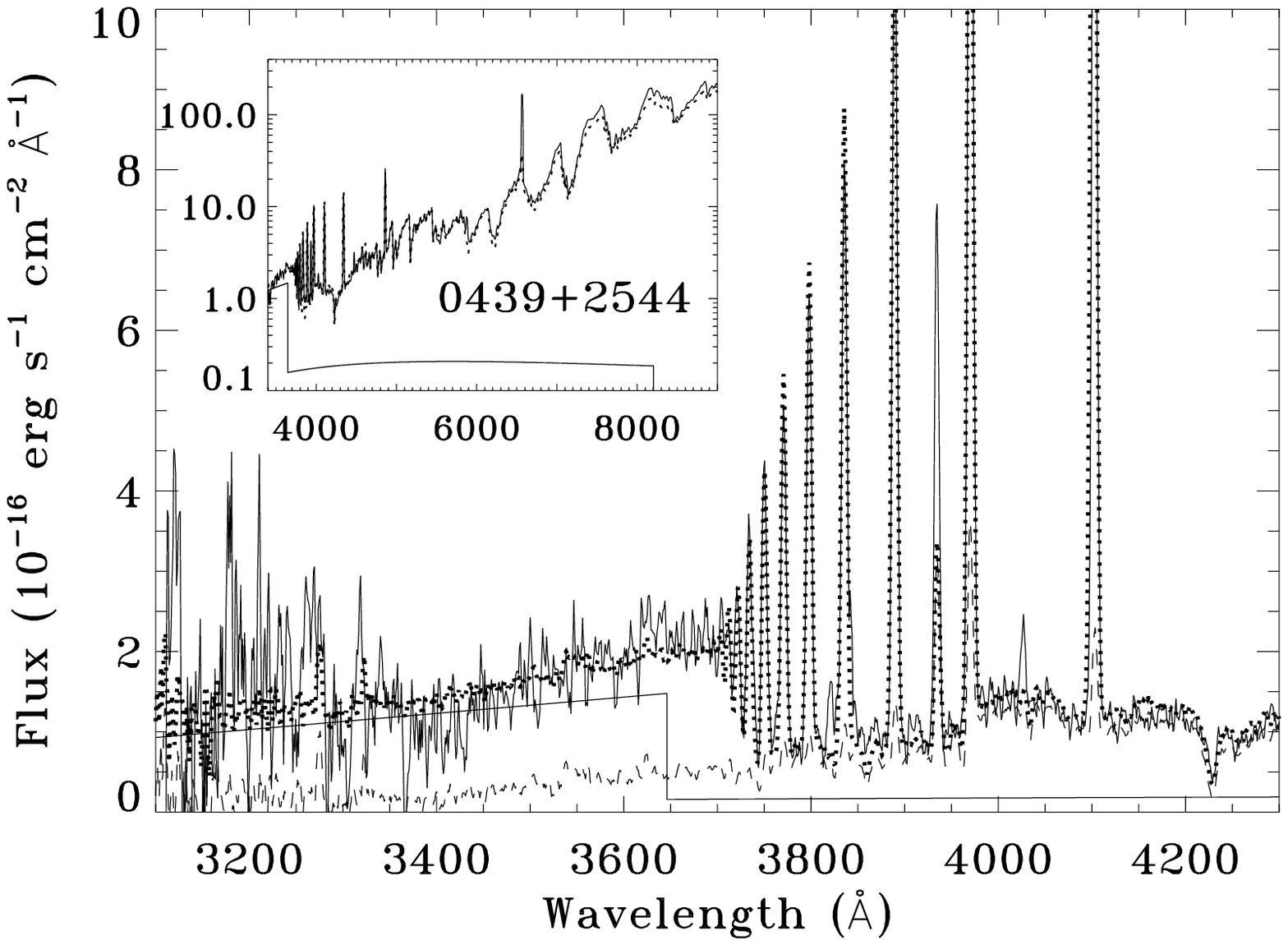}
\end{figure}

\begin{figure}
\plottwo{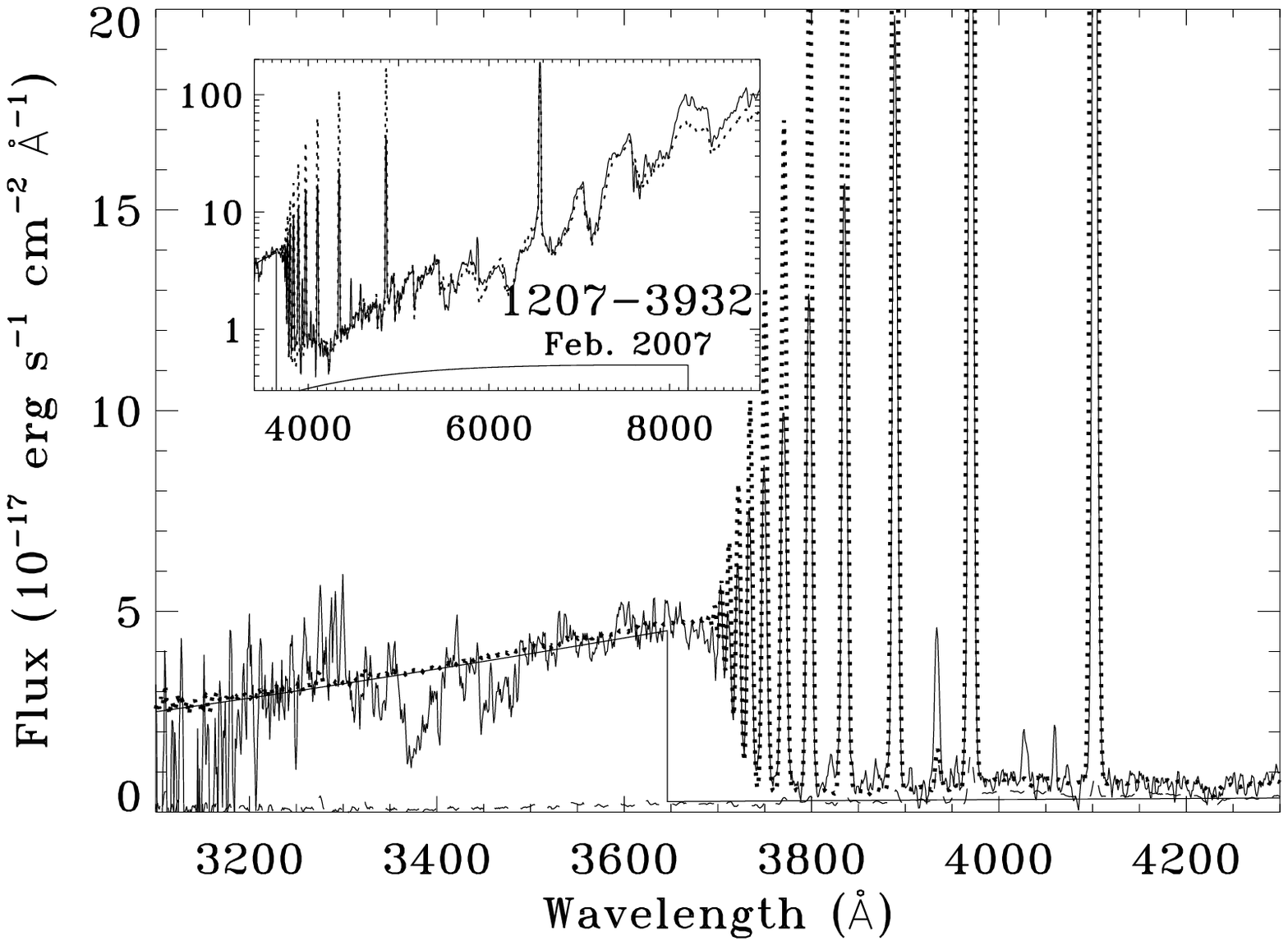}{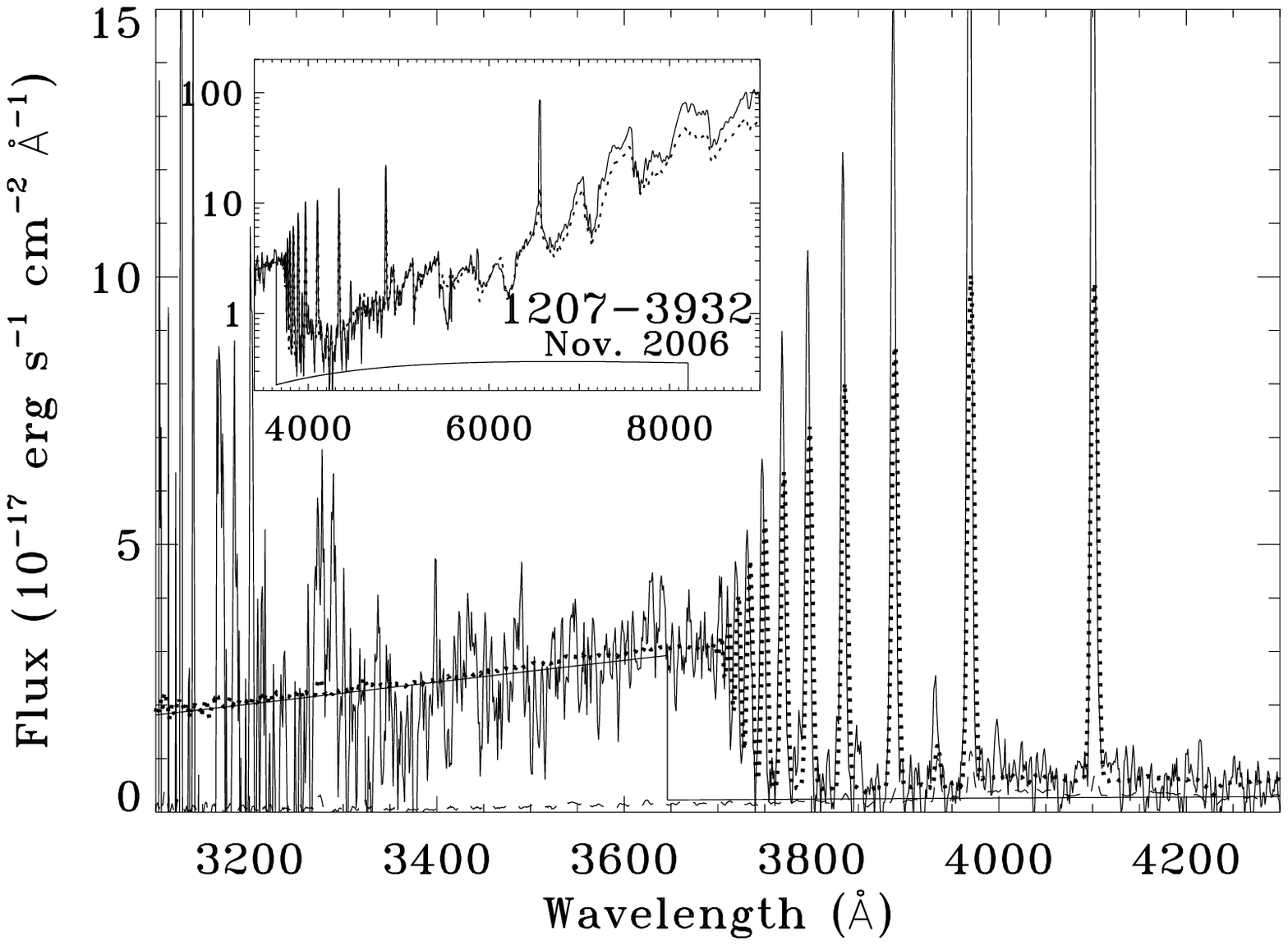}
\end{figure}

\begin{figure}
\plottwo{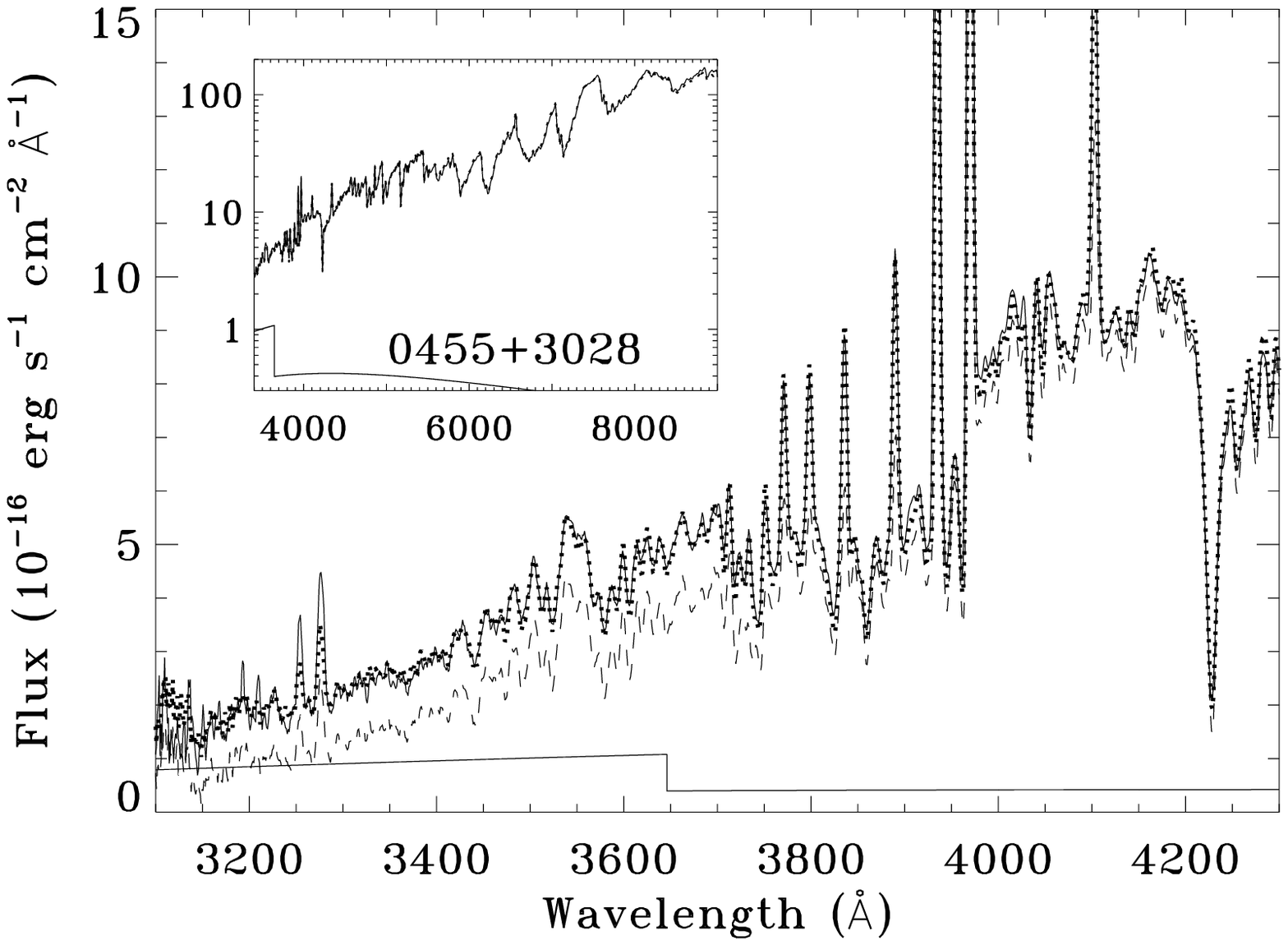}{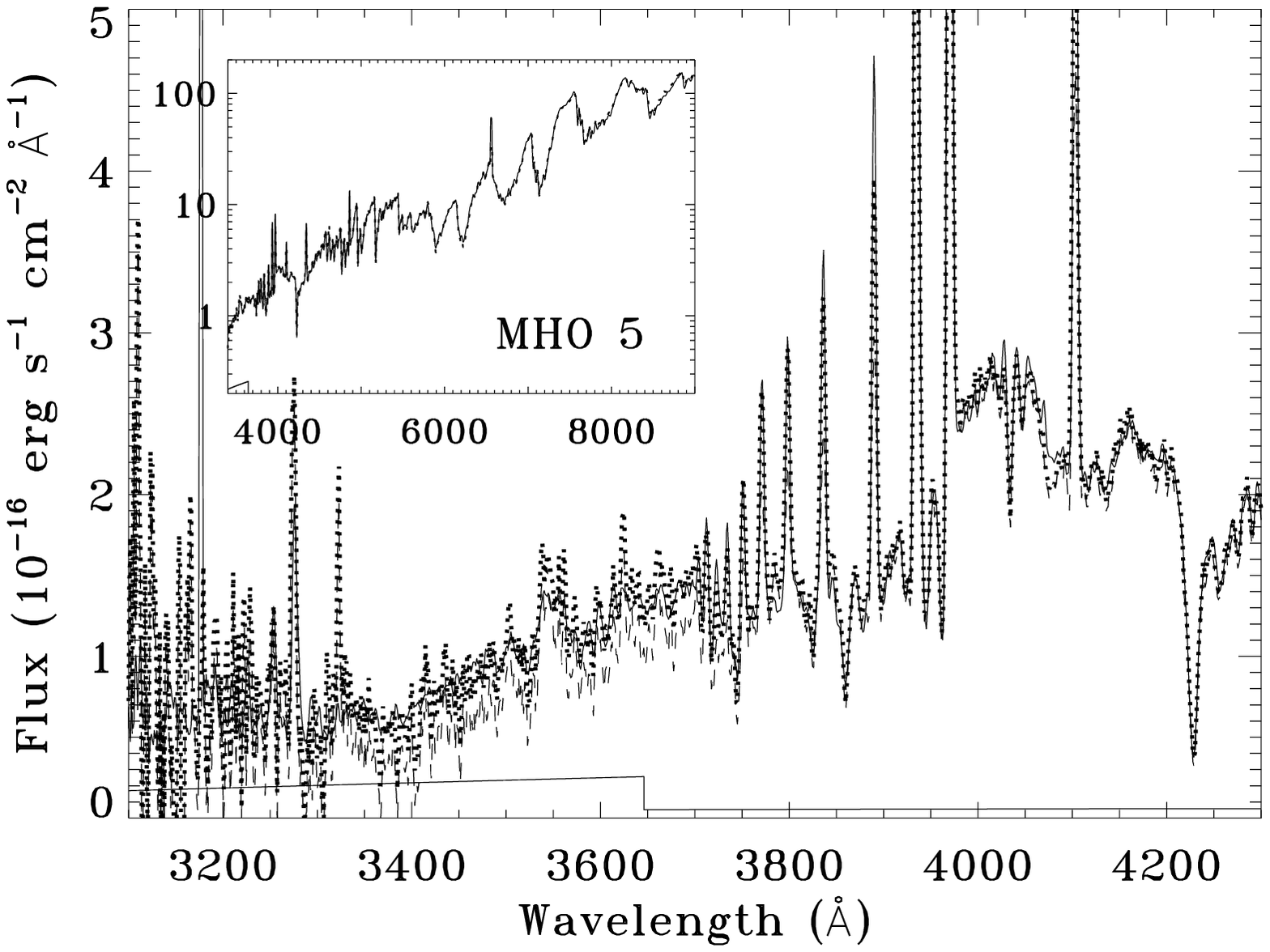}
\end{figure}

\begin{figure}
\plottwo{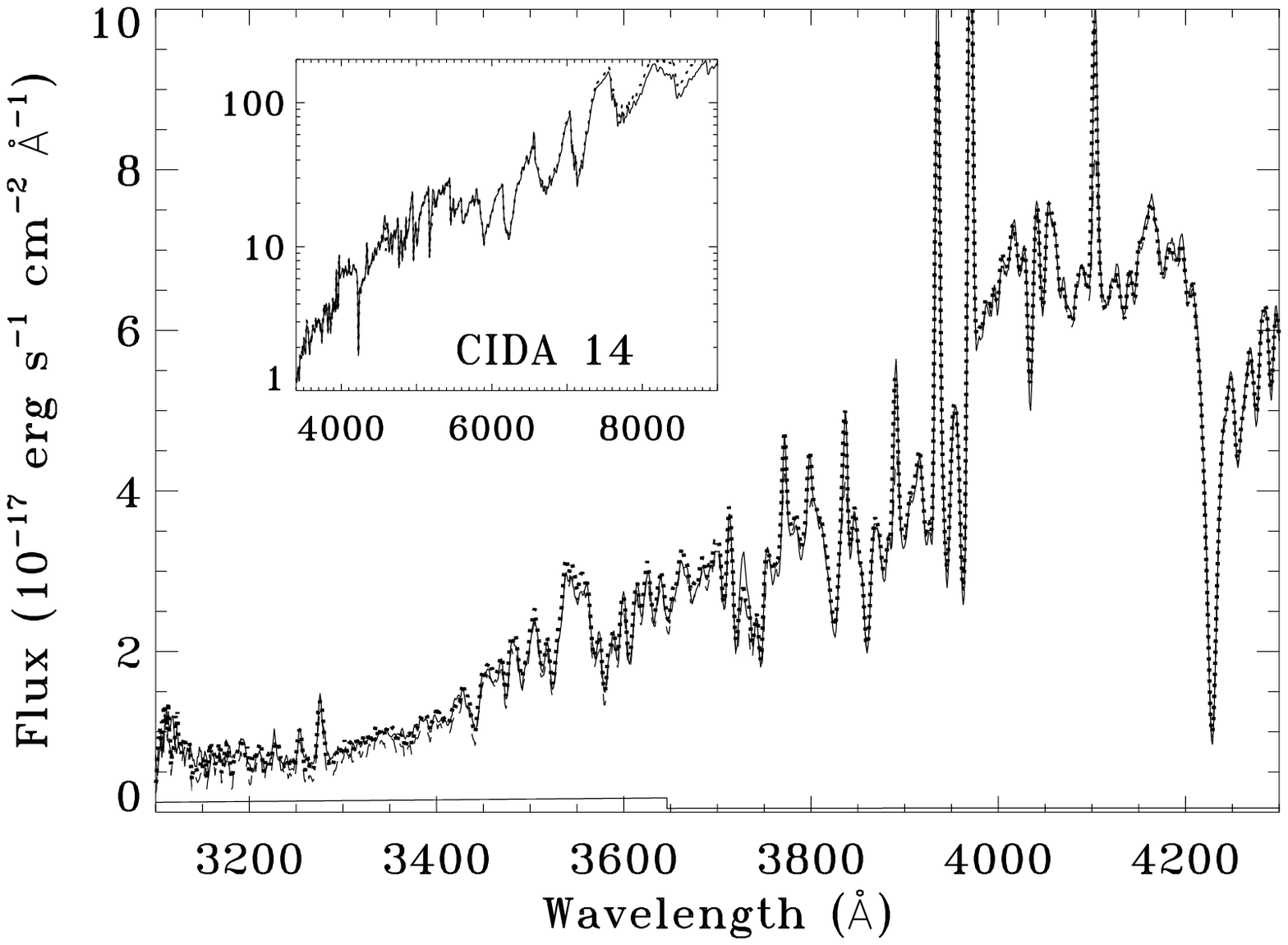}{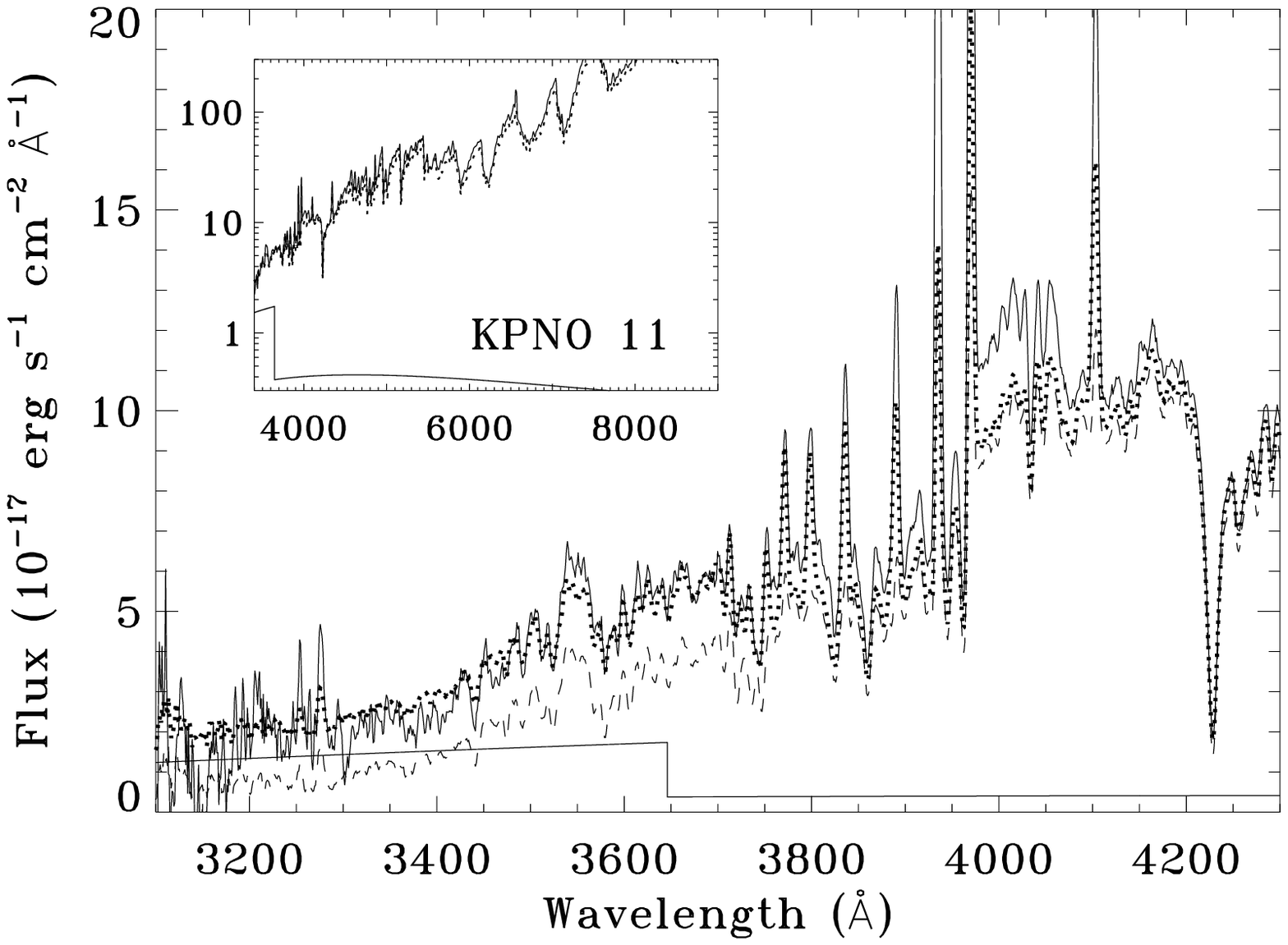}
\end{figure}

\begin{figure}
\plottwo{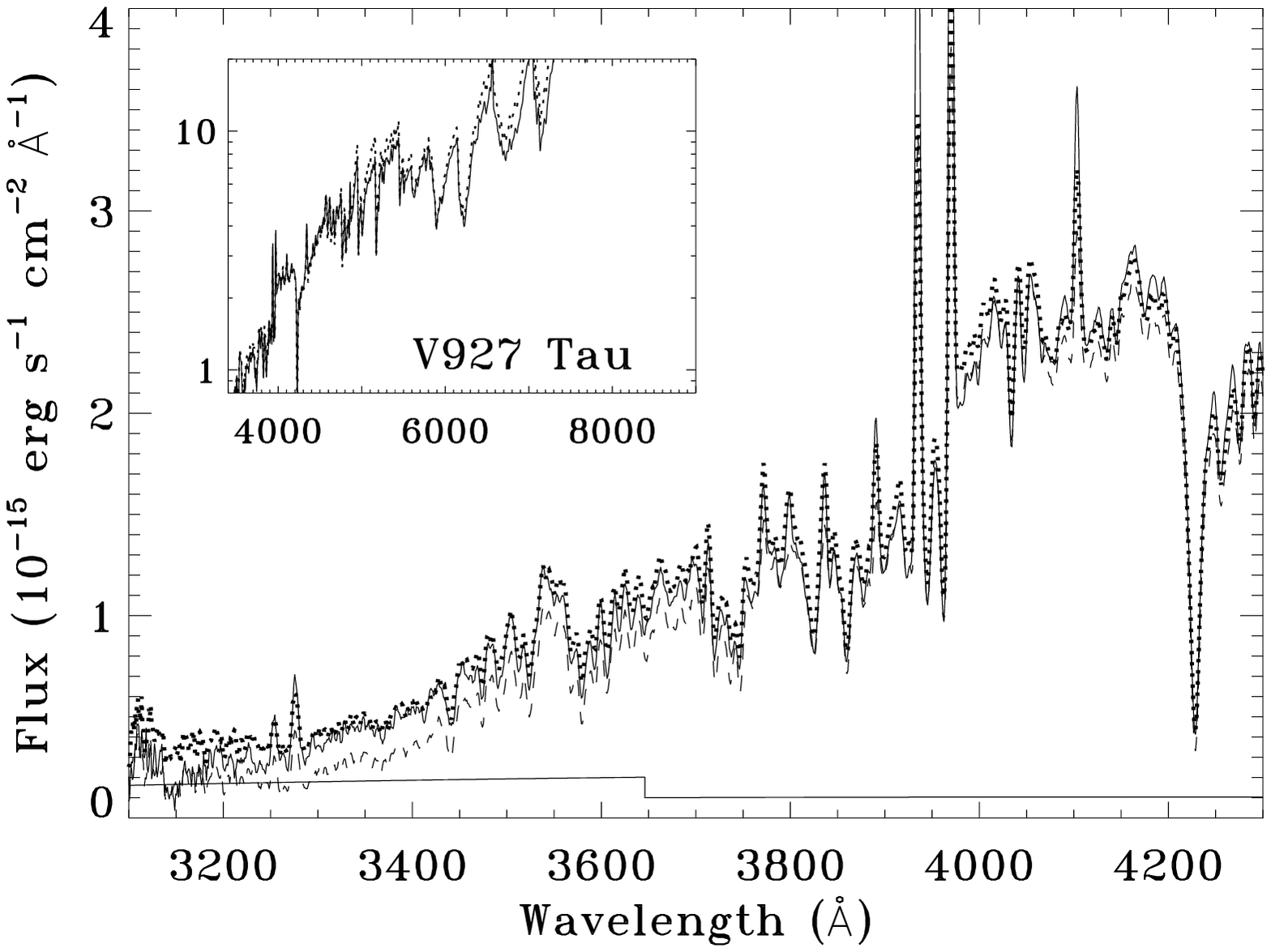}{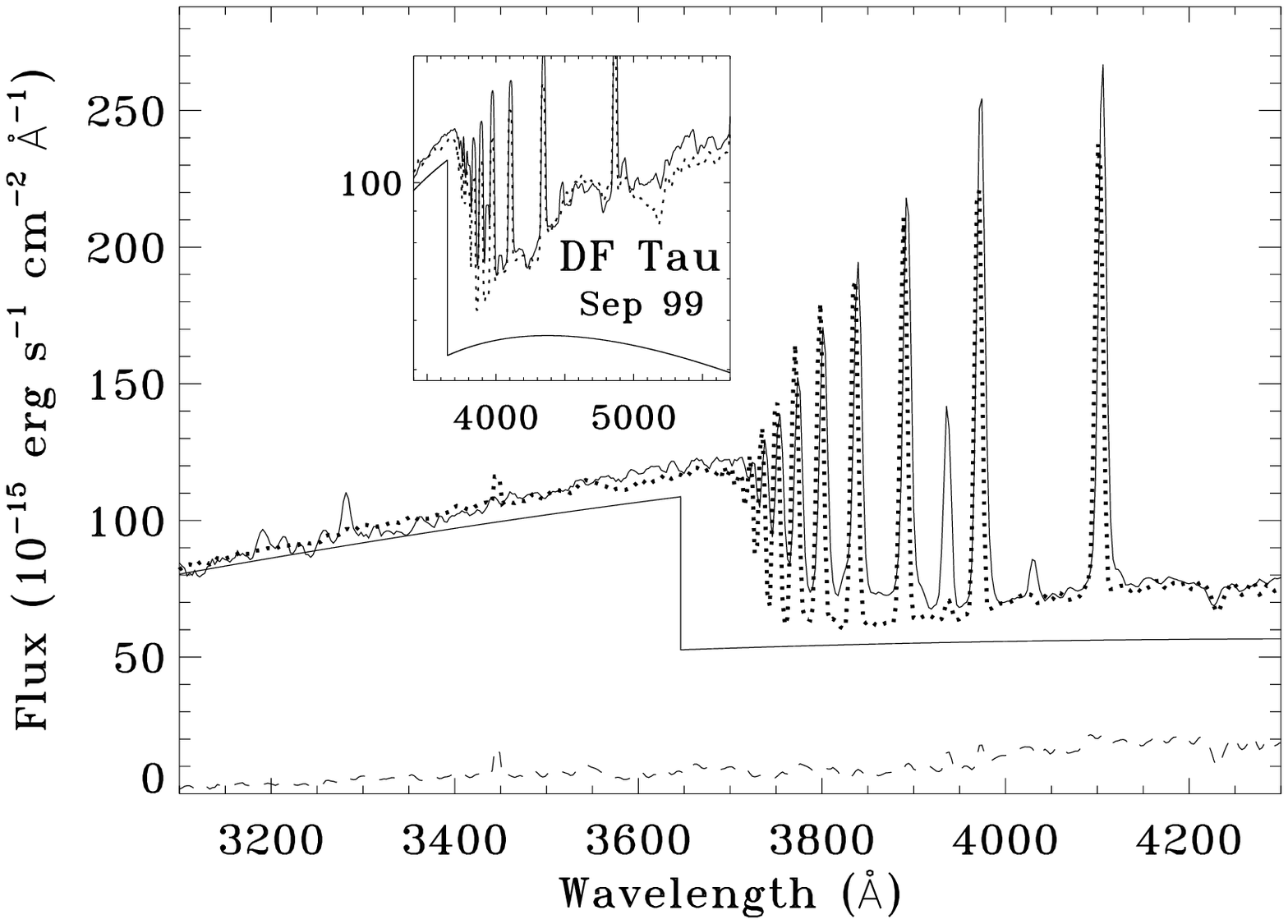}
\end{figure}

\begin{figure}
\plottwo{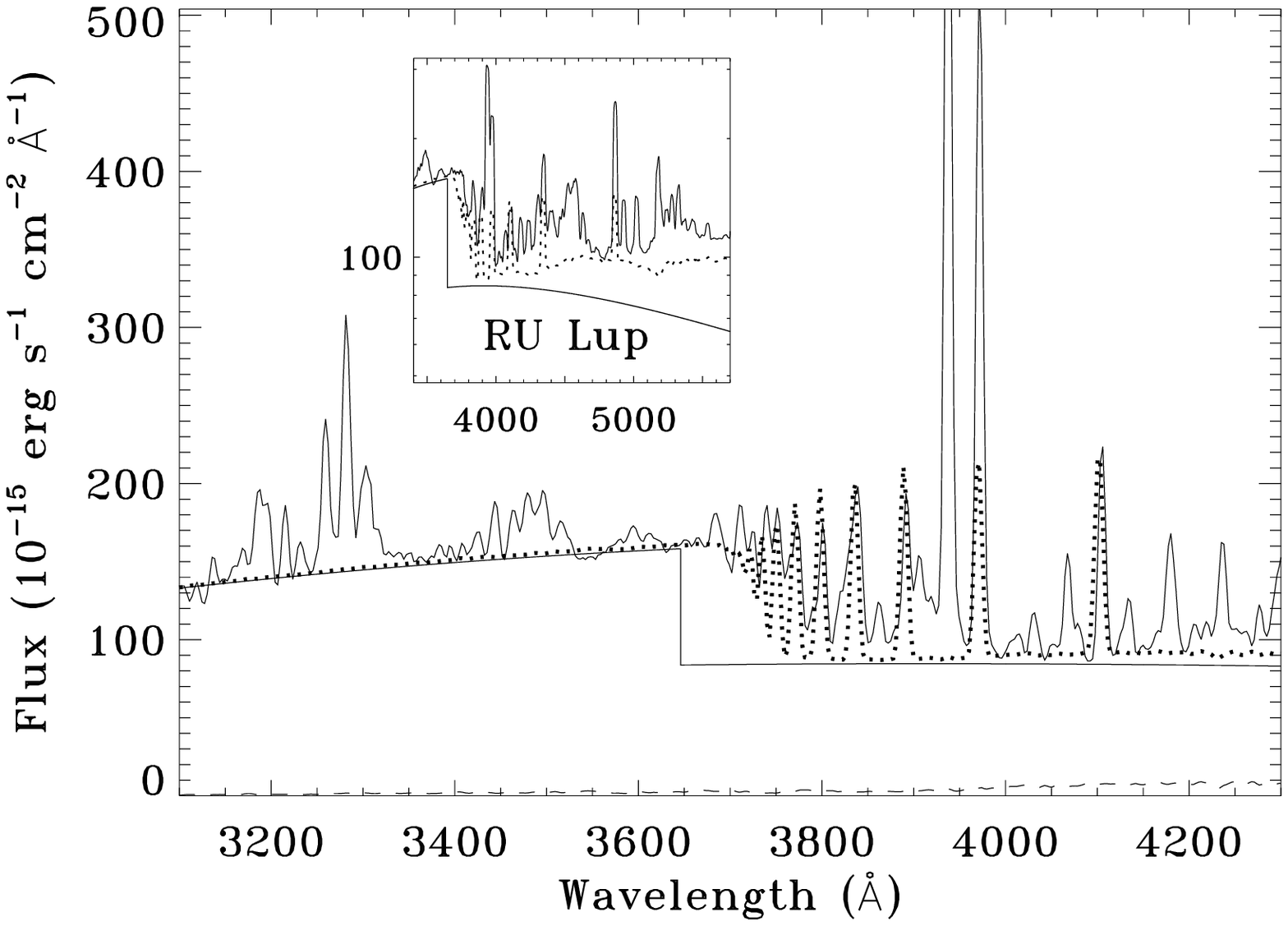}{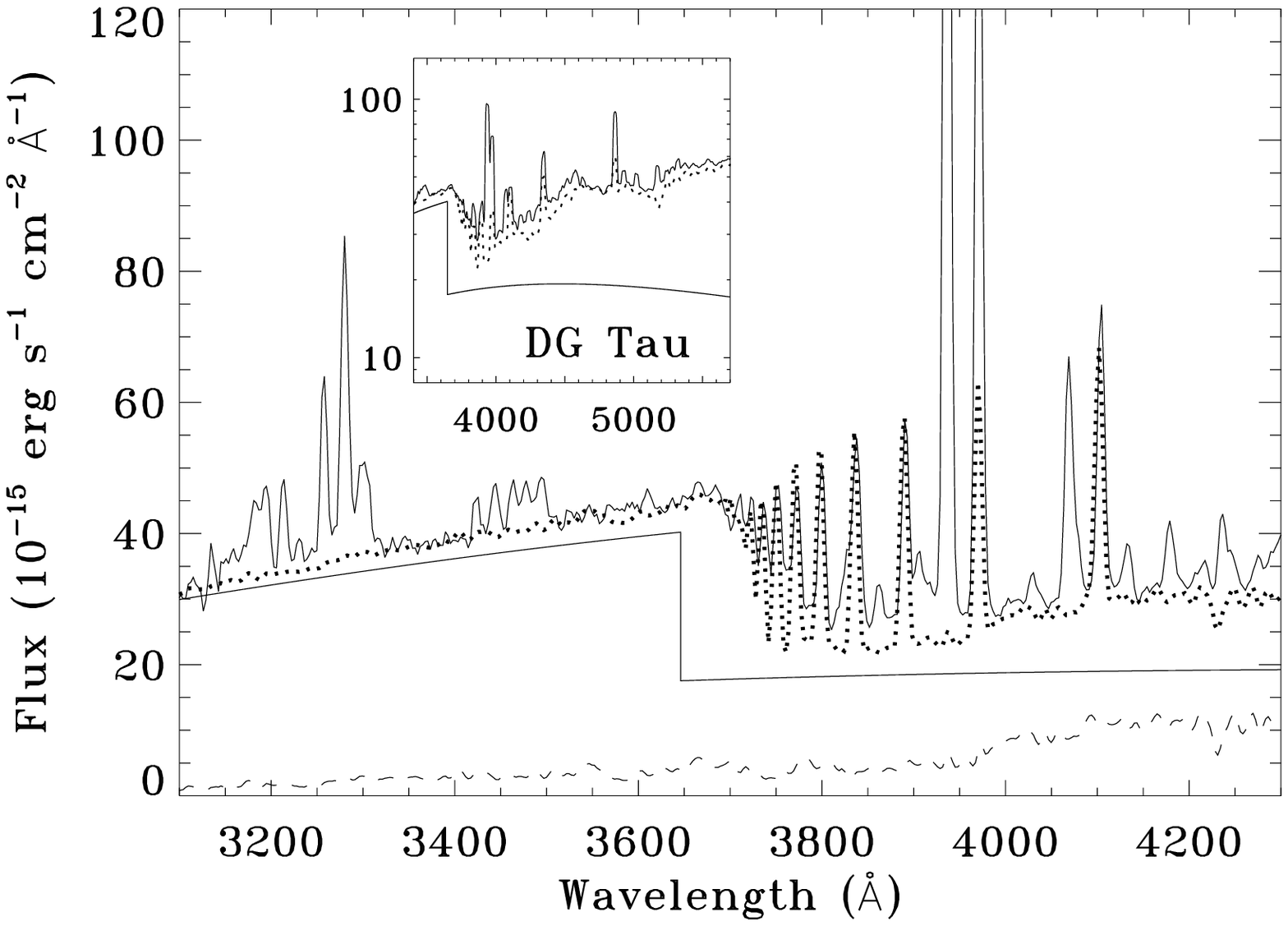}
\end{figure}

\begin{figure}
\plottwo{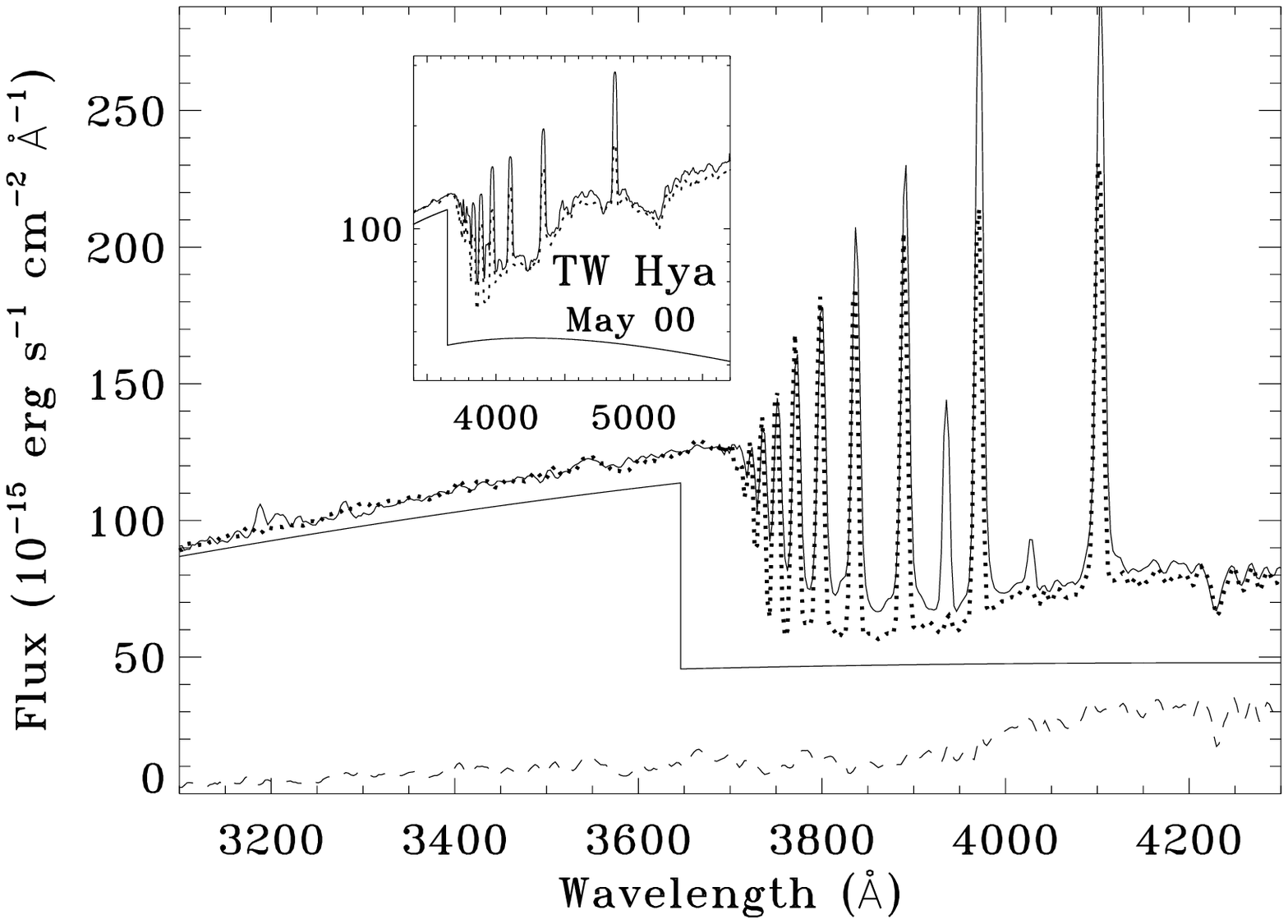}{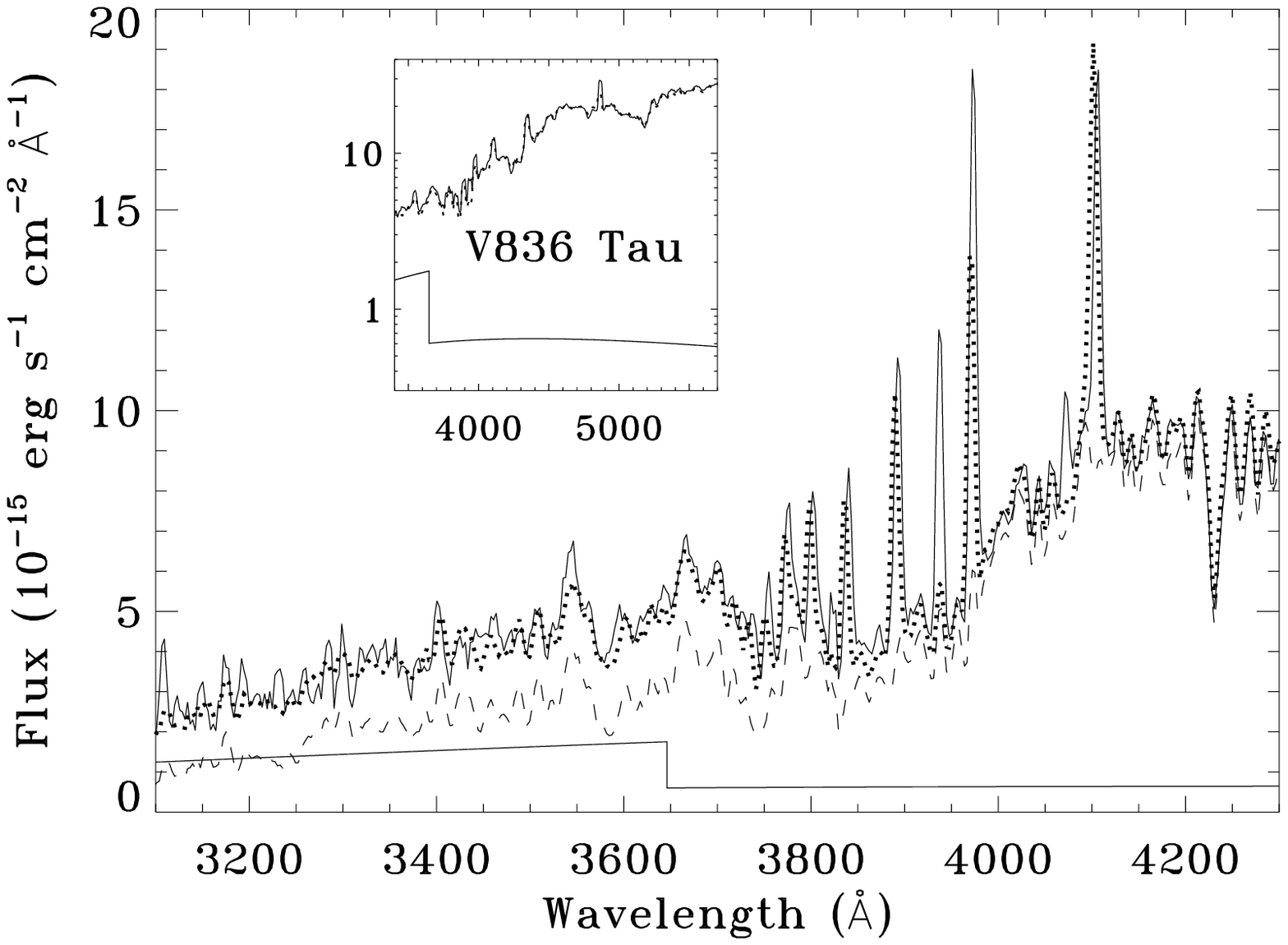}
\caption{The observed emission
  (solid spectrum), corrected for reddening 
  (Table 1), is fit with the combination of a photospheric template
  (dashed line, listed in Table 5 and not shown in the smaller panel) and a synthetic accretion spectrum
  from a slab model (light solid line, Balmer lines are not shown for clarity).
  The total fit is shown as a dashed line that overlaps the observed
  spectrum.}
\end{figure}

\clearpage
\clearpage
\pagebreak

\begin{figure}
\plottwo{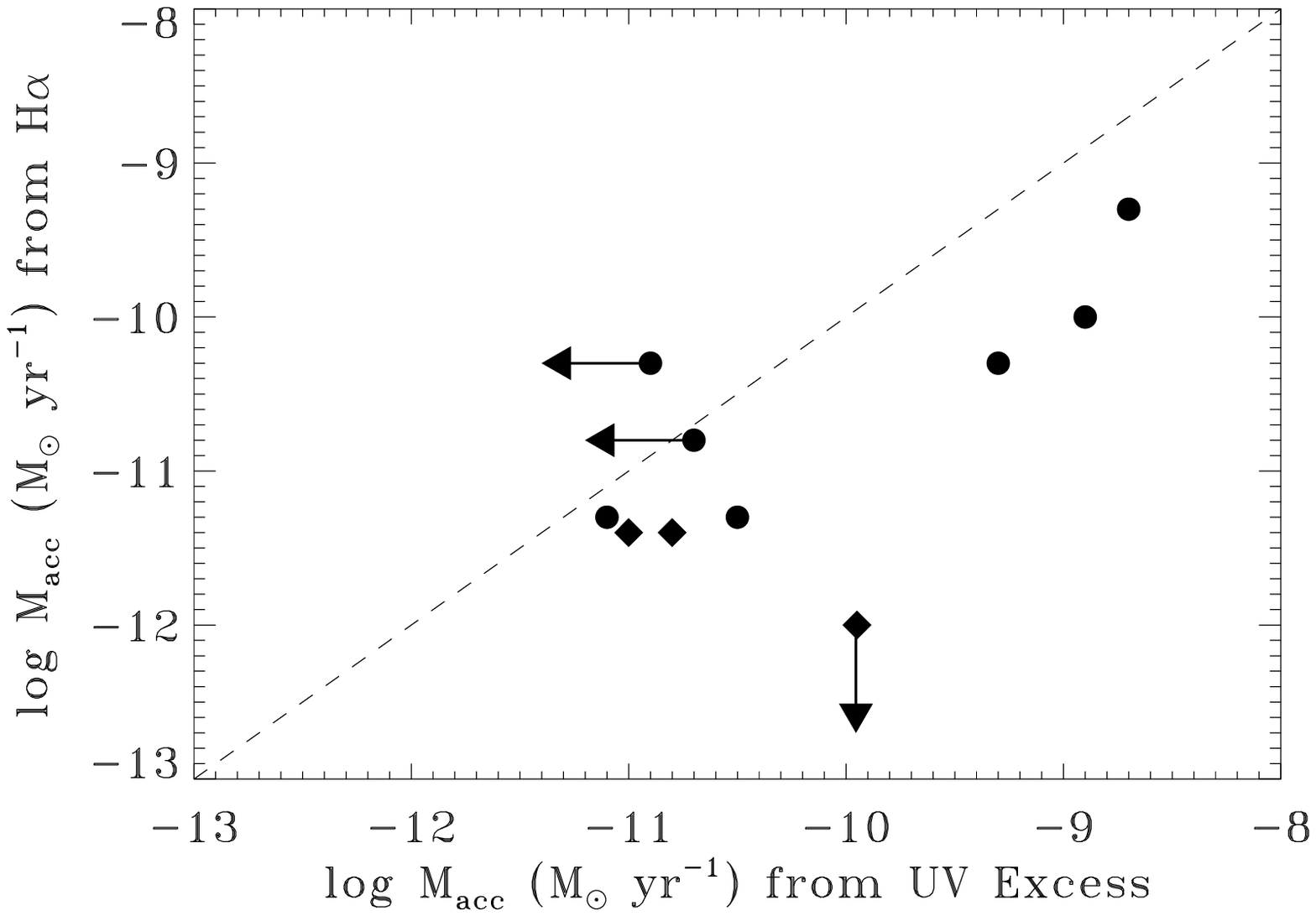}{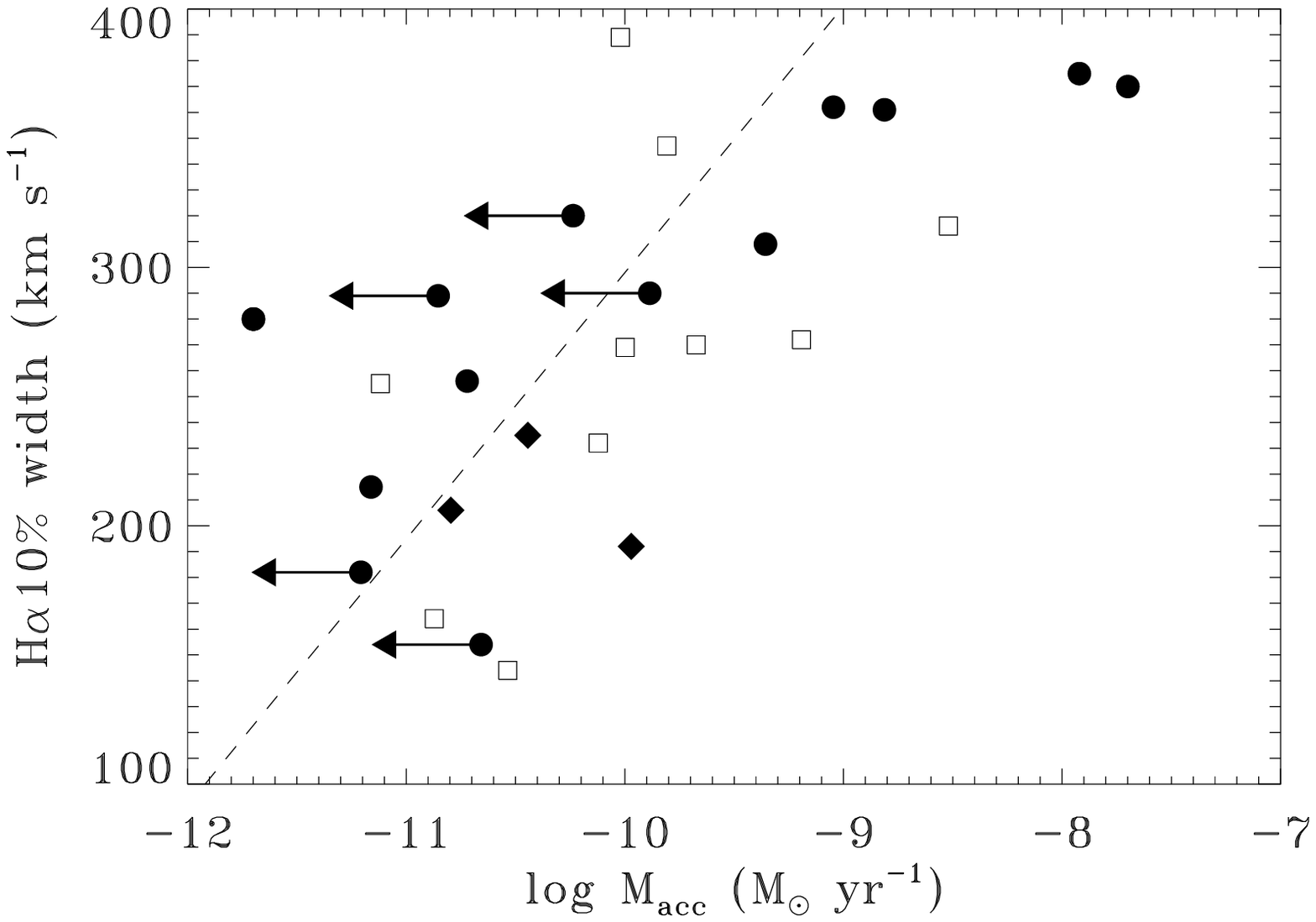}
\caption{UV excess measure of $\dot{M}$ versus non-simultaneous measurements of $\dot{M}$ from models of the H$\alpha$ line profiles (left, from Muzerolle et al. 2000, 2003, 2005) and versus non-simultaneous measurements of the H$\alpha$ 10\% width (left).  The circles are accretors with $\dot{M}$ measured from UV excess and the diamonds are accretors with $\dot{M}$ measured from the V-band excess (Table 14) reported by Kraus et al.~(2006).  The dashed line on the left panel denotes where two accretion rate measures would be equal.   The dashed line on the right panel denotes the $\dot{M}$--H$\alpha$ 10\% width relationship calculated by Natta et al.~(2004).}
\end{figure}

\epsscale{1.1}
\begin{figure}
\plottwo{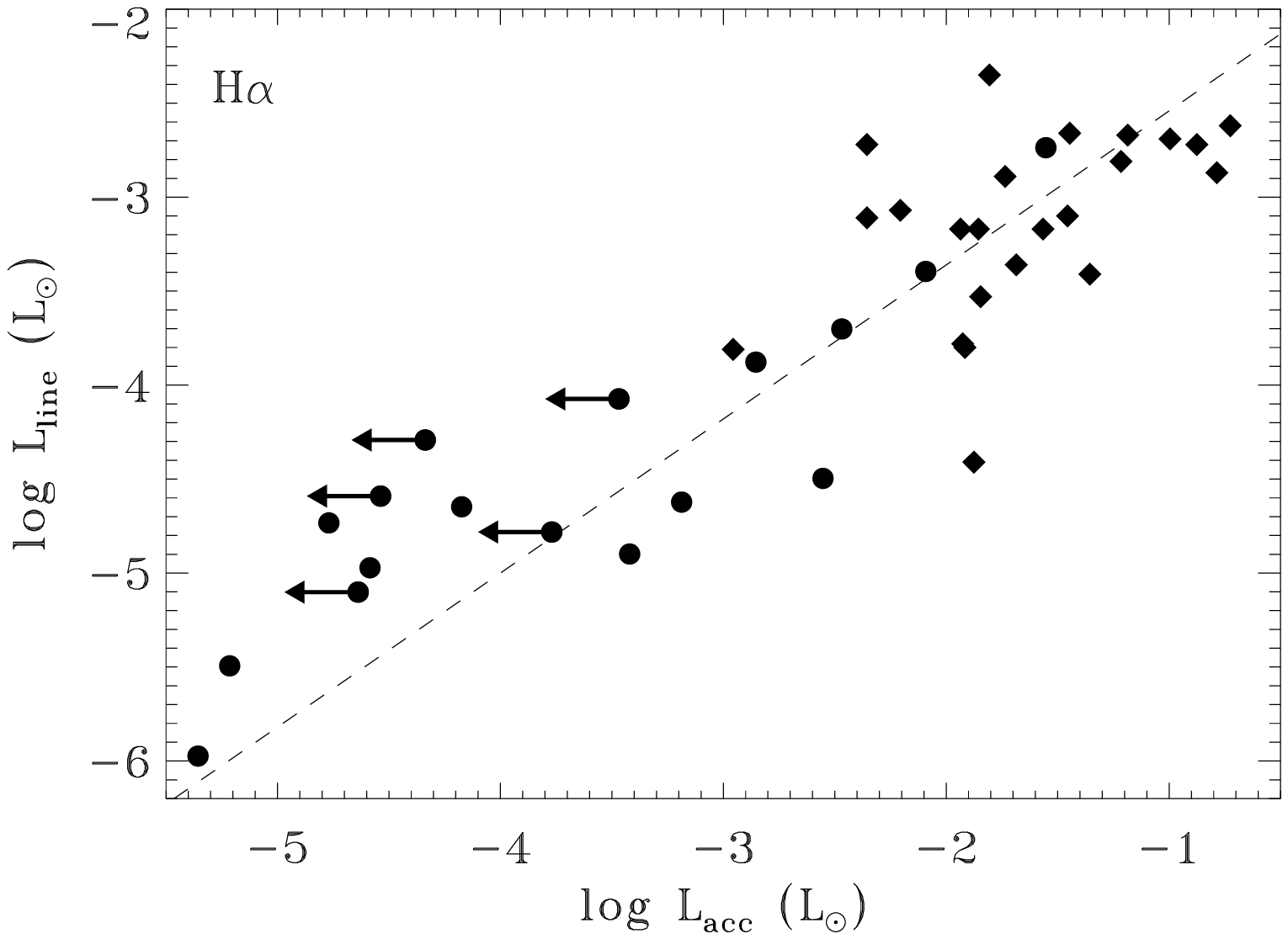}{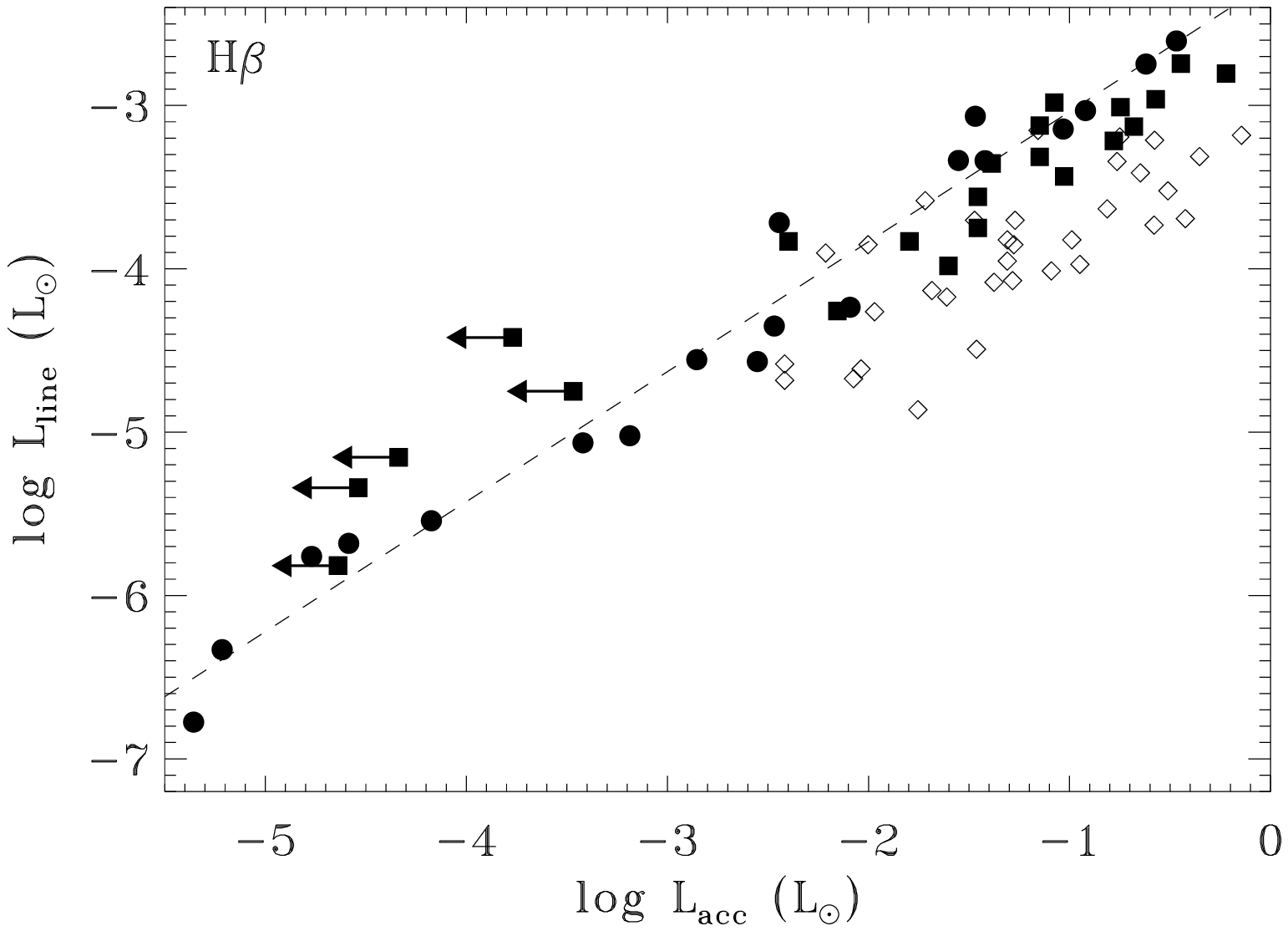}
\end{figure}

\begin{figure}
\plottwo{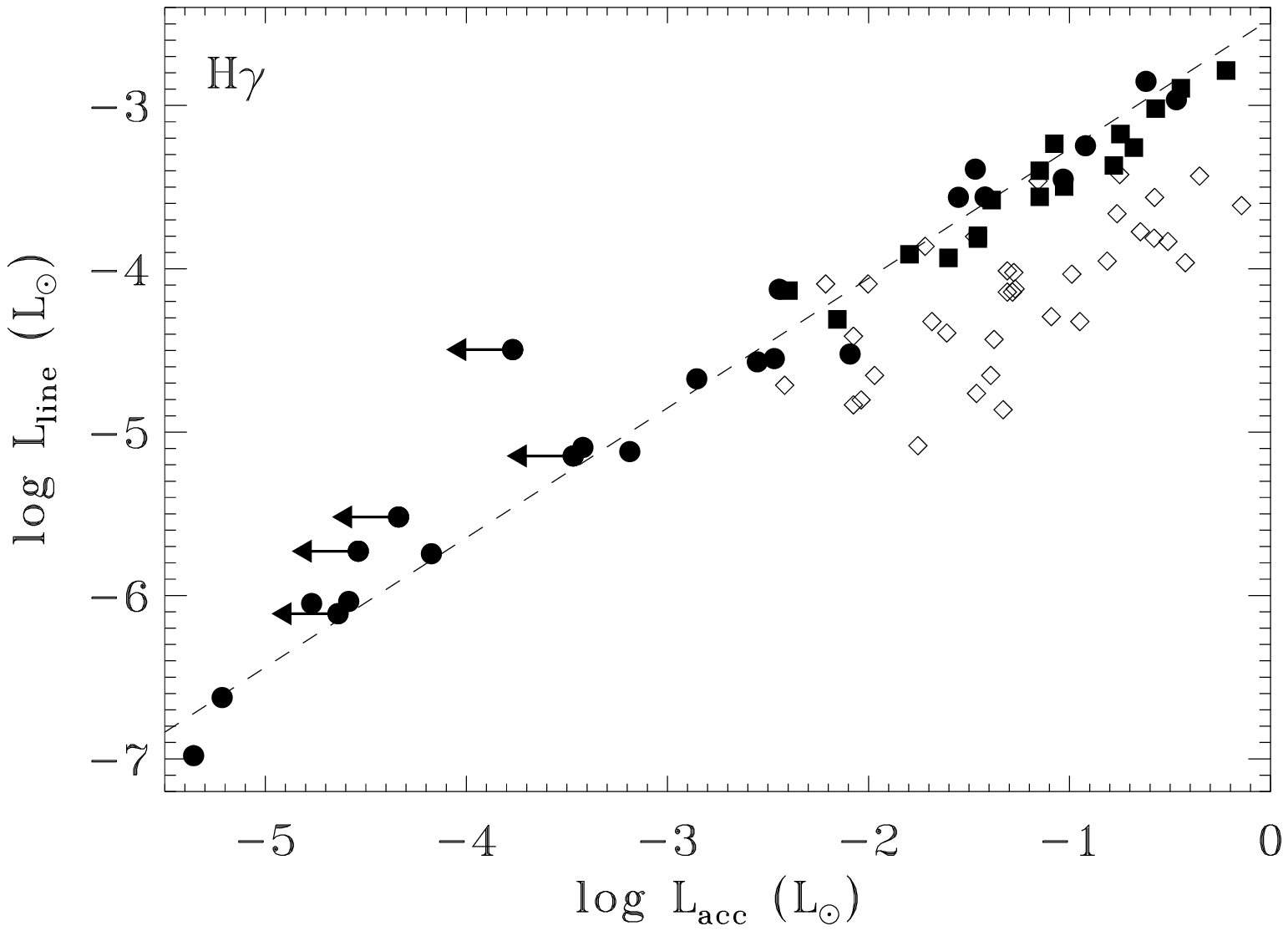}{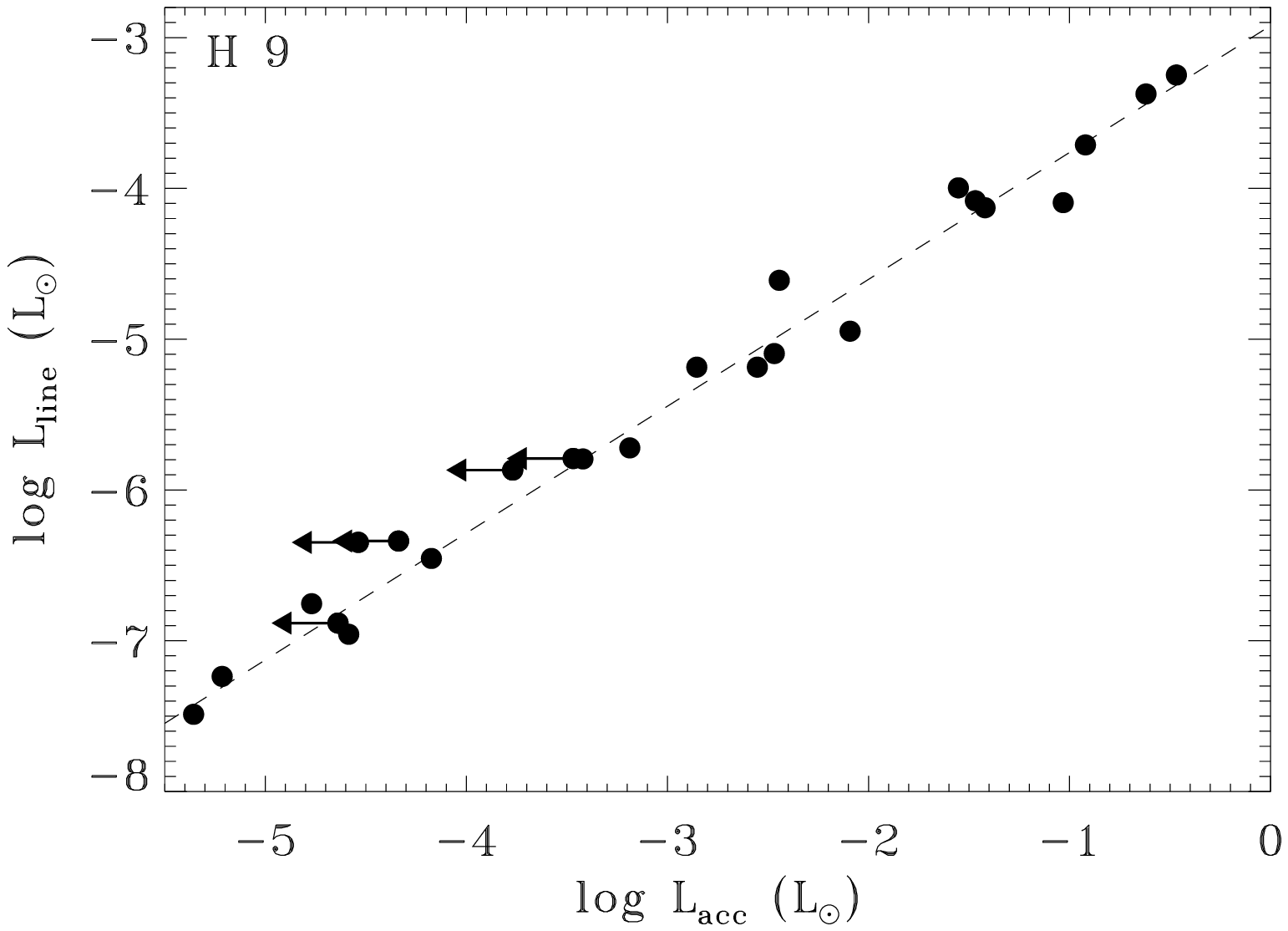}
\end{figure}

\begin{figure}[!t]
\plottwo{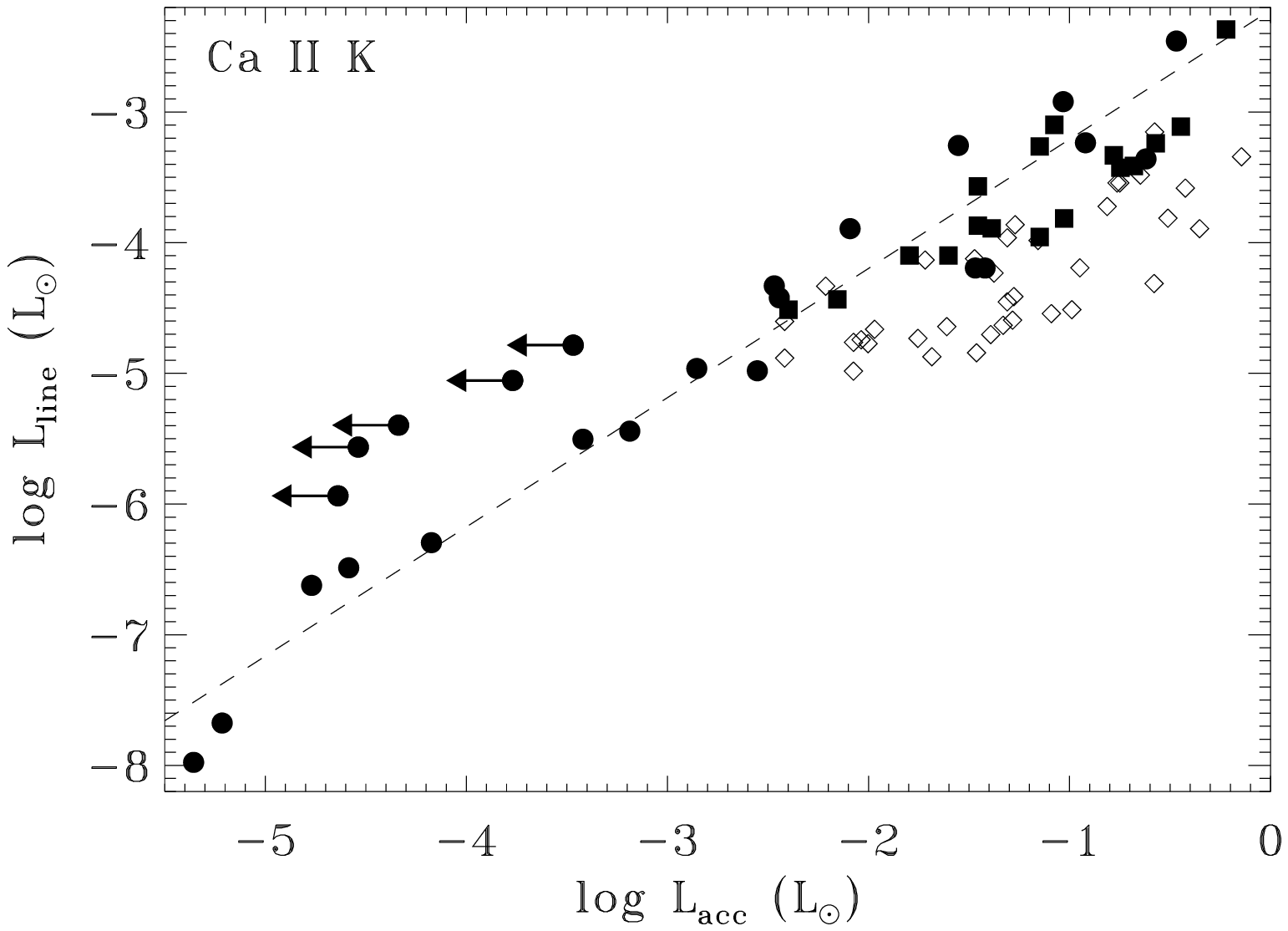}{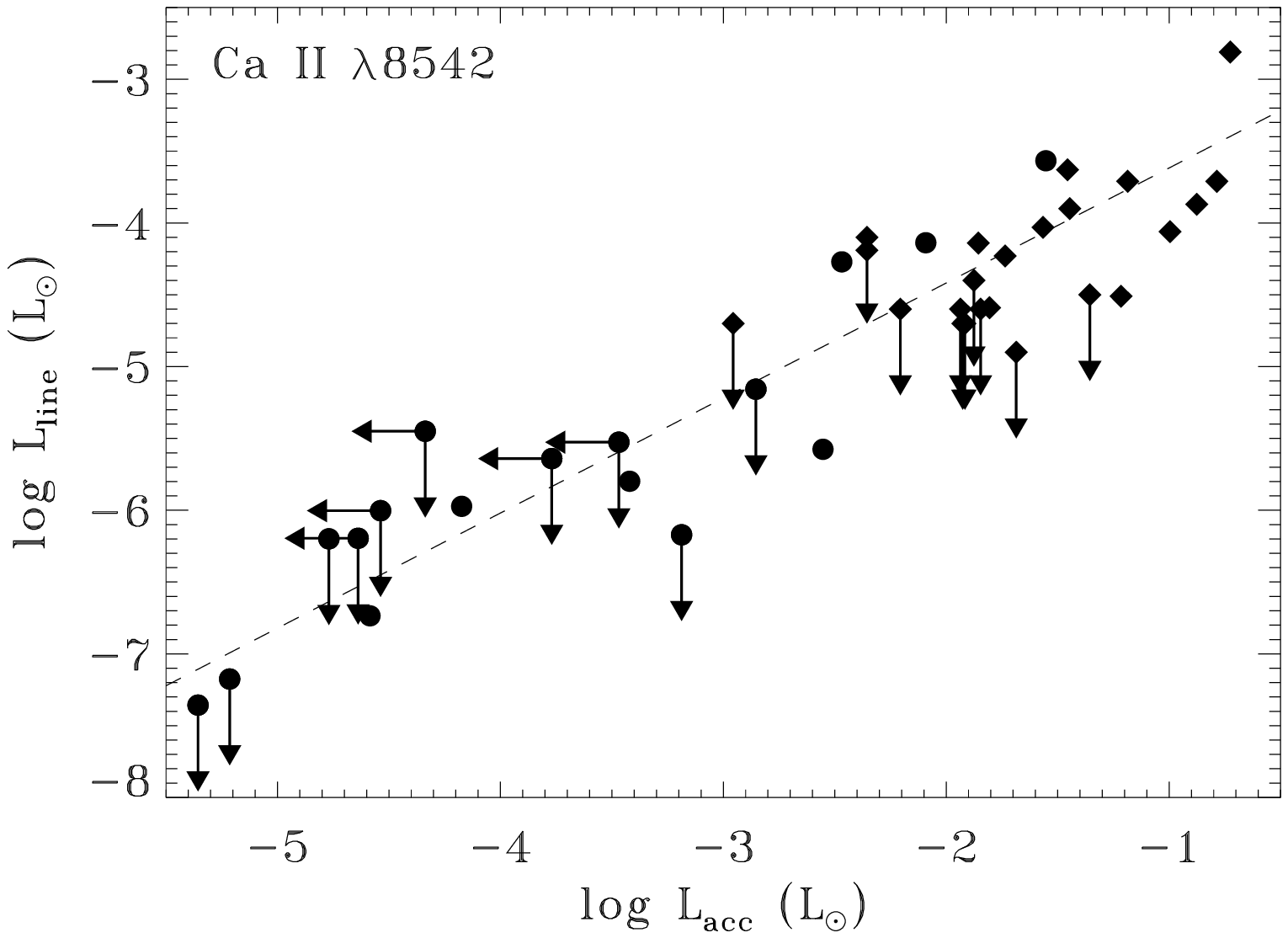}
\end{figure}


\begin{figure}[!h]
\plottwo{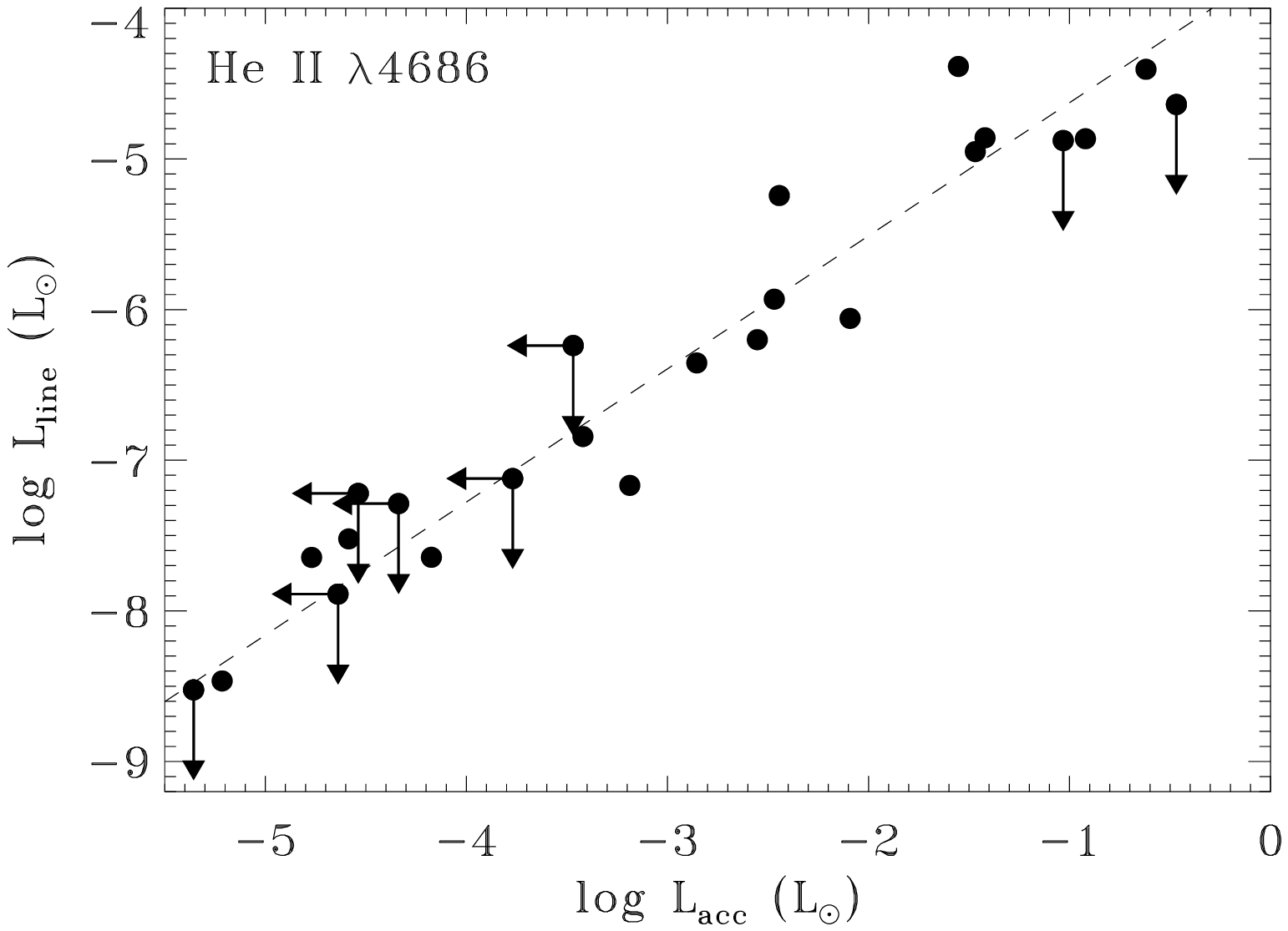}{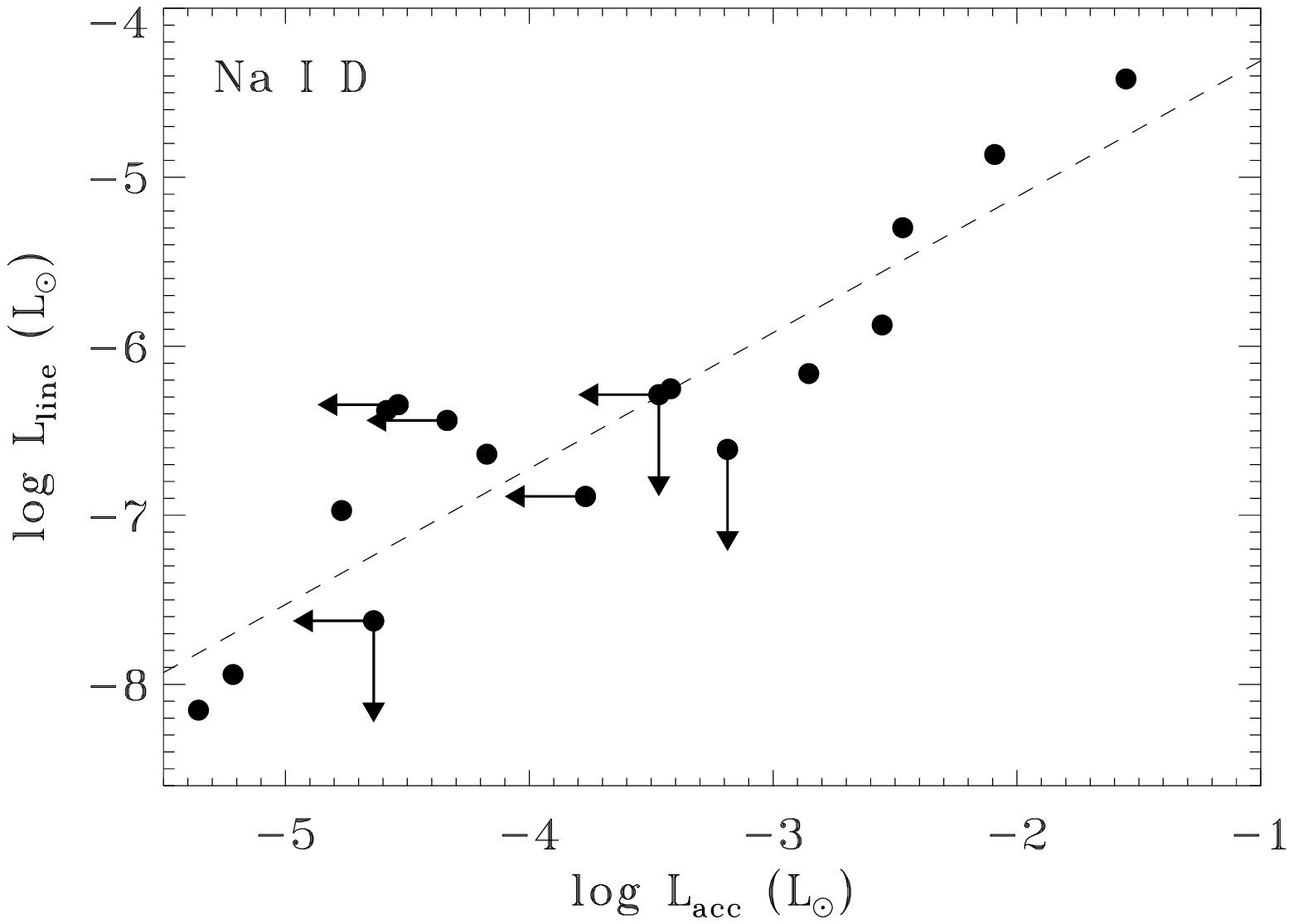}
\end{figure}

\begin{figure}[!h]
\plottwo{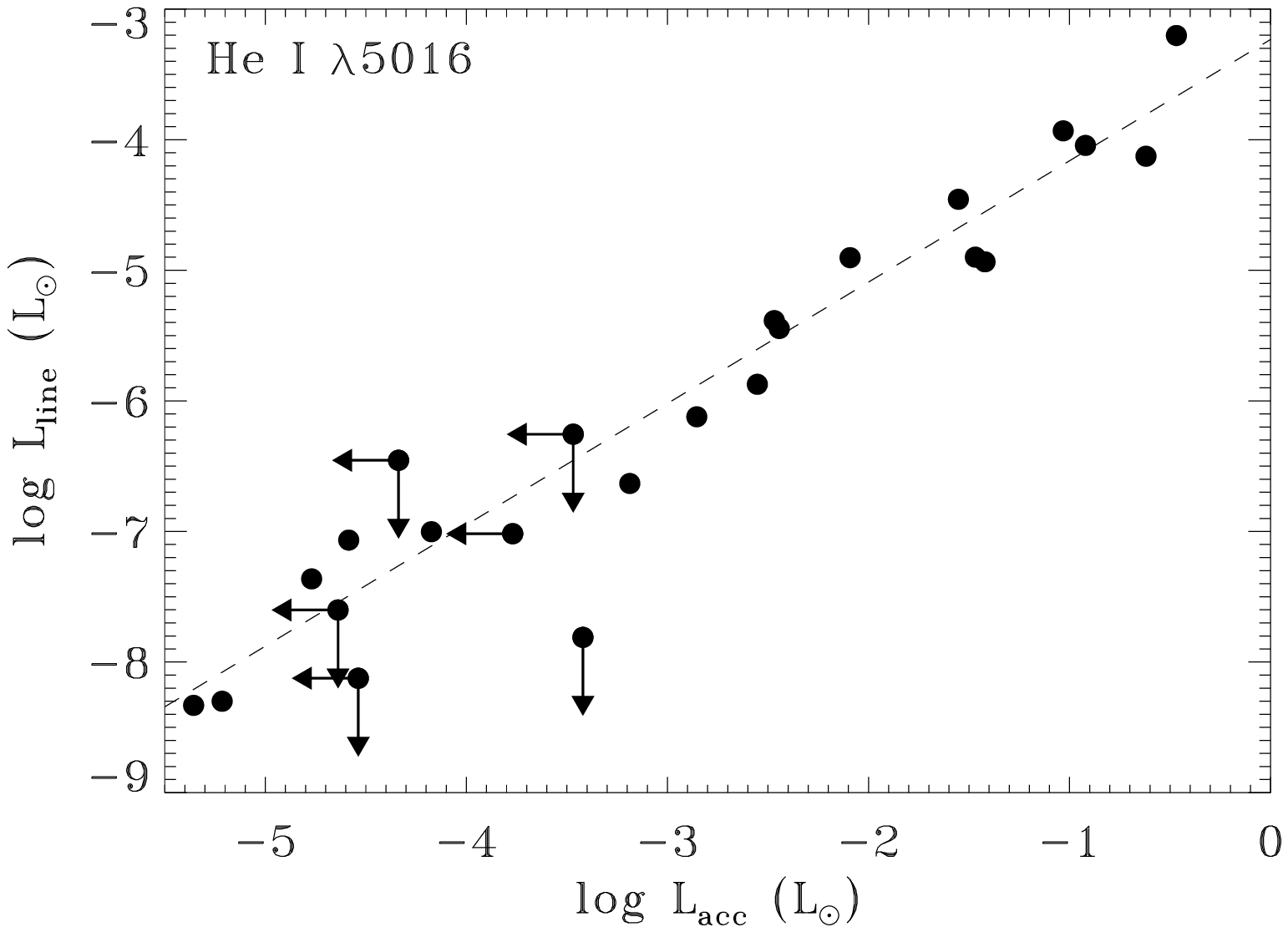}{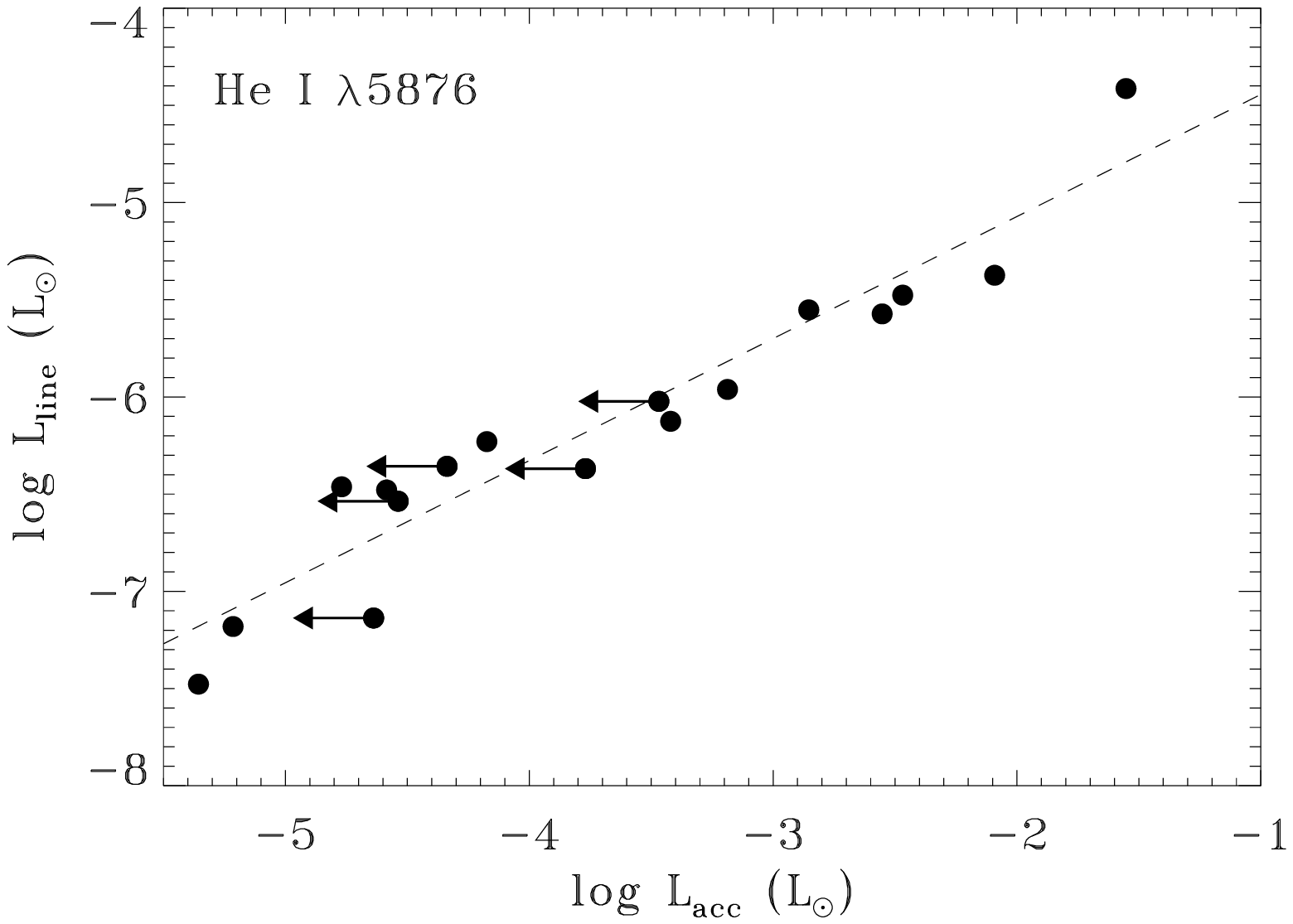}
\end{figure}

\begin{figure}
\plottwo{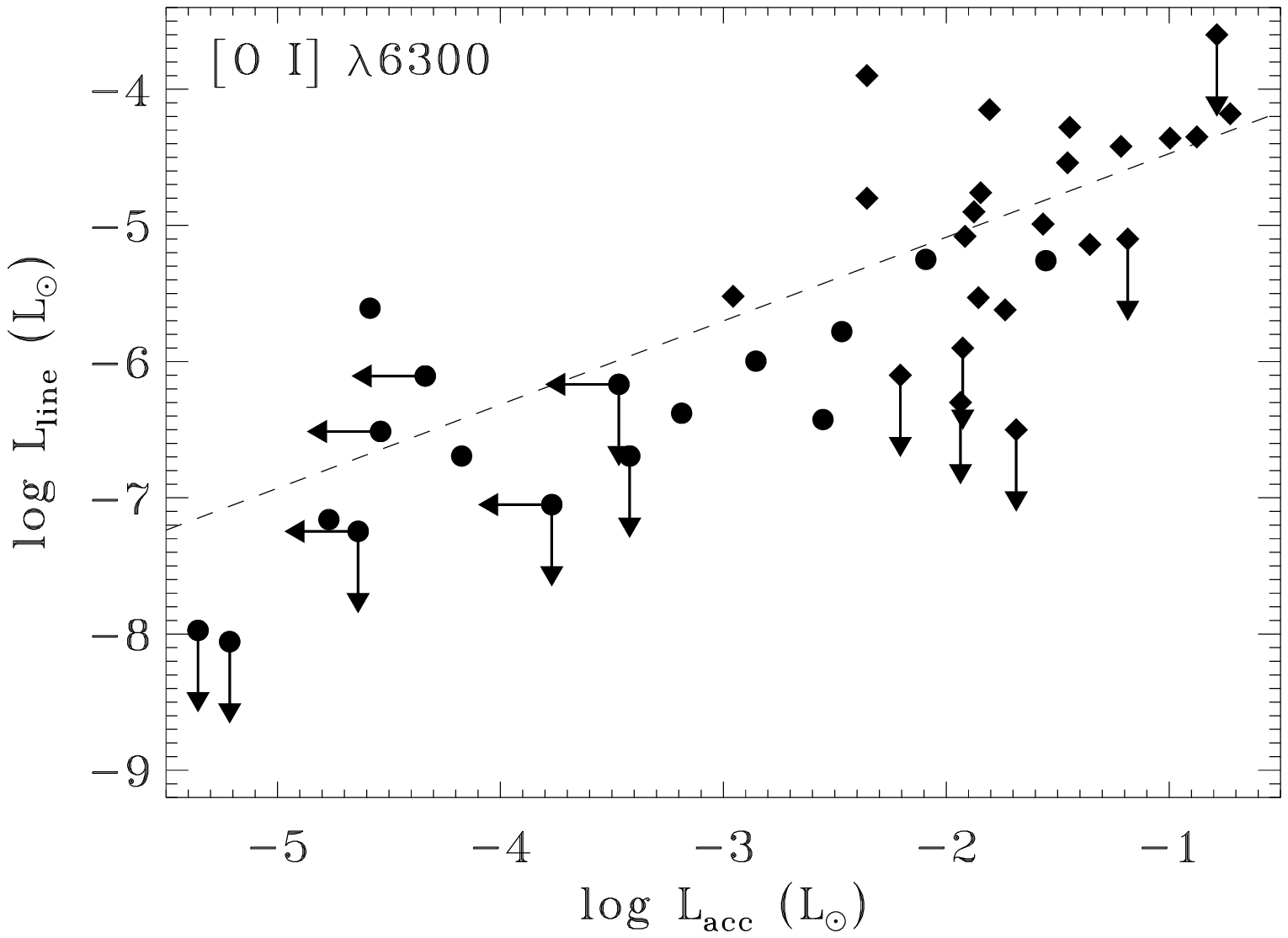}{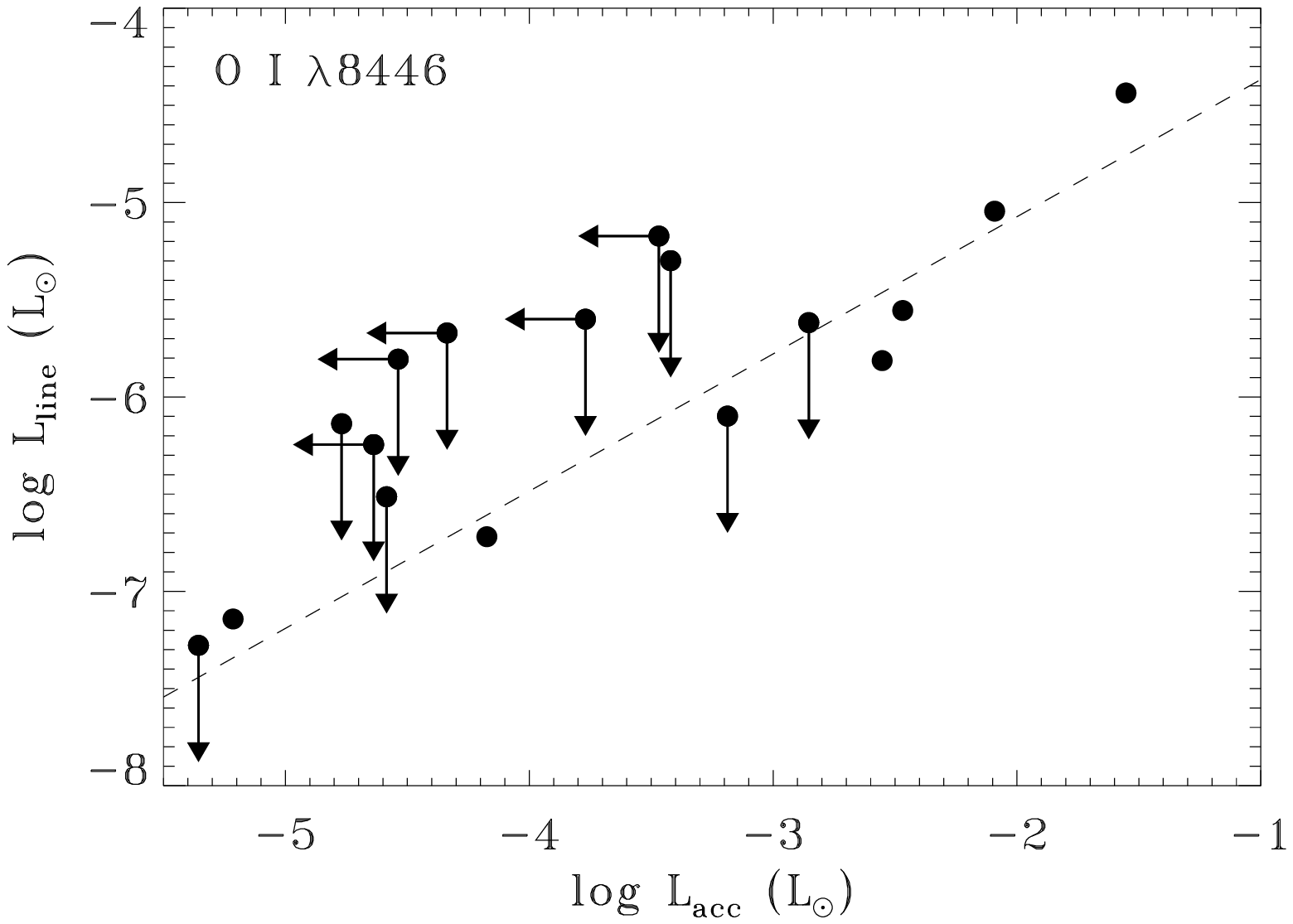}
\caption{The relationship between selected line luminosities and the accretion continuum luminosity.  The filled circles represent the accretors with
  strong excess Balmer continuum emission that is produced by 
  accretion.  The upper limits in $L_{acc}$ are the five possible accretors, which have weak Balmer continuum emission that may be attributed to either accretion or chromospheric activity.  We include results from Valenti et al.~(1993, empty diamonds and corrected for the 140 pc distance to Taurus)  and
  Gullbring et al.~(1998, filled squares) for fits to the H$\beta$, H$\gamma$, and \ion{Ca}{2} K
  line luminosities, and Hartigan et al.~(2003, filled triangles) for fits to the H$\alpha$, \ion{Ca}{2} $\lambda8542$, and [\ion{O}{1}] line luminosities.  The dashed lines are linear fits to the data as described in \S 5.3 and Table 15.} 
\end{figure}

\begin{figure}
\plottwo{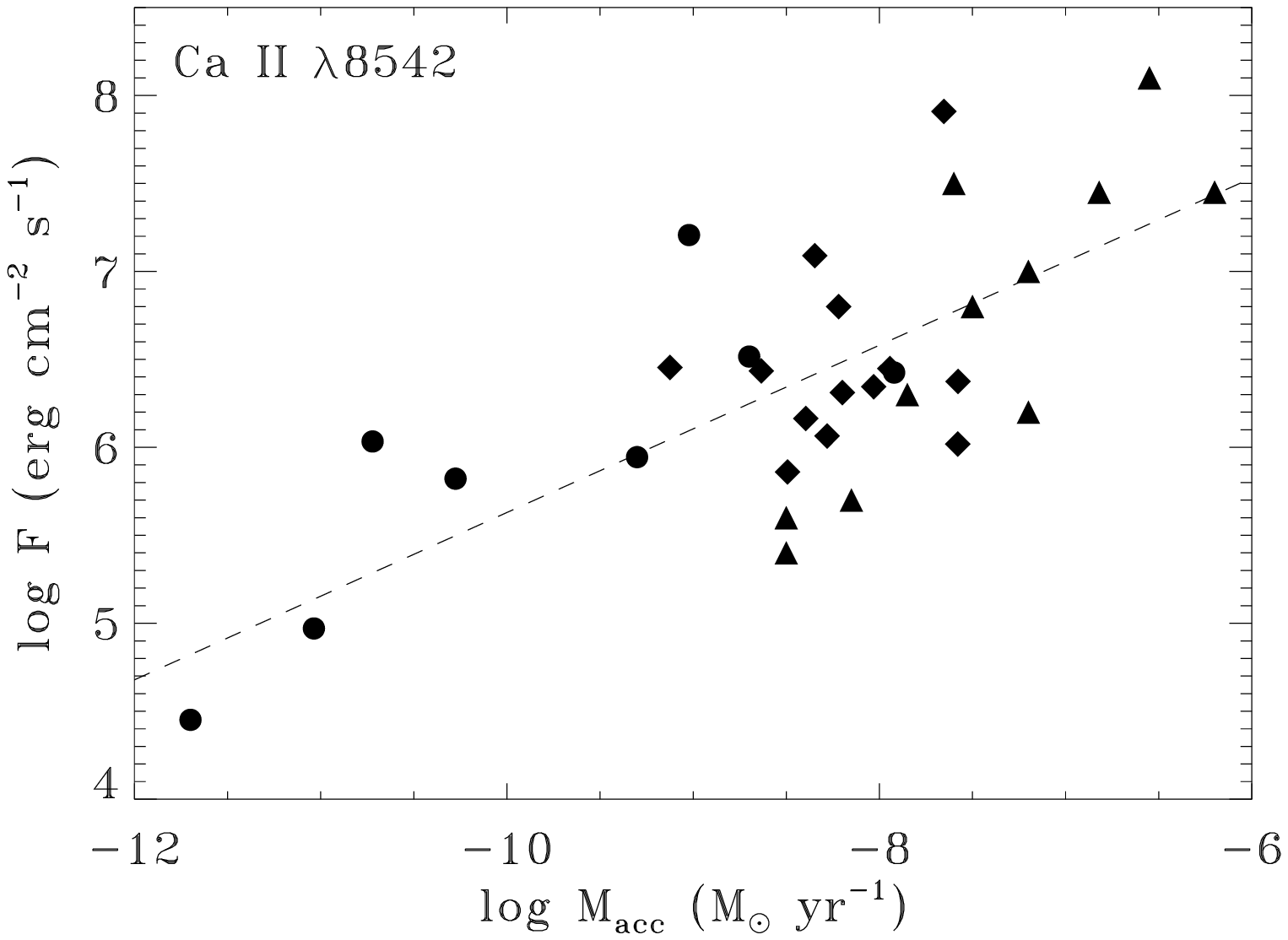}{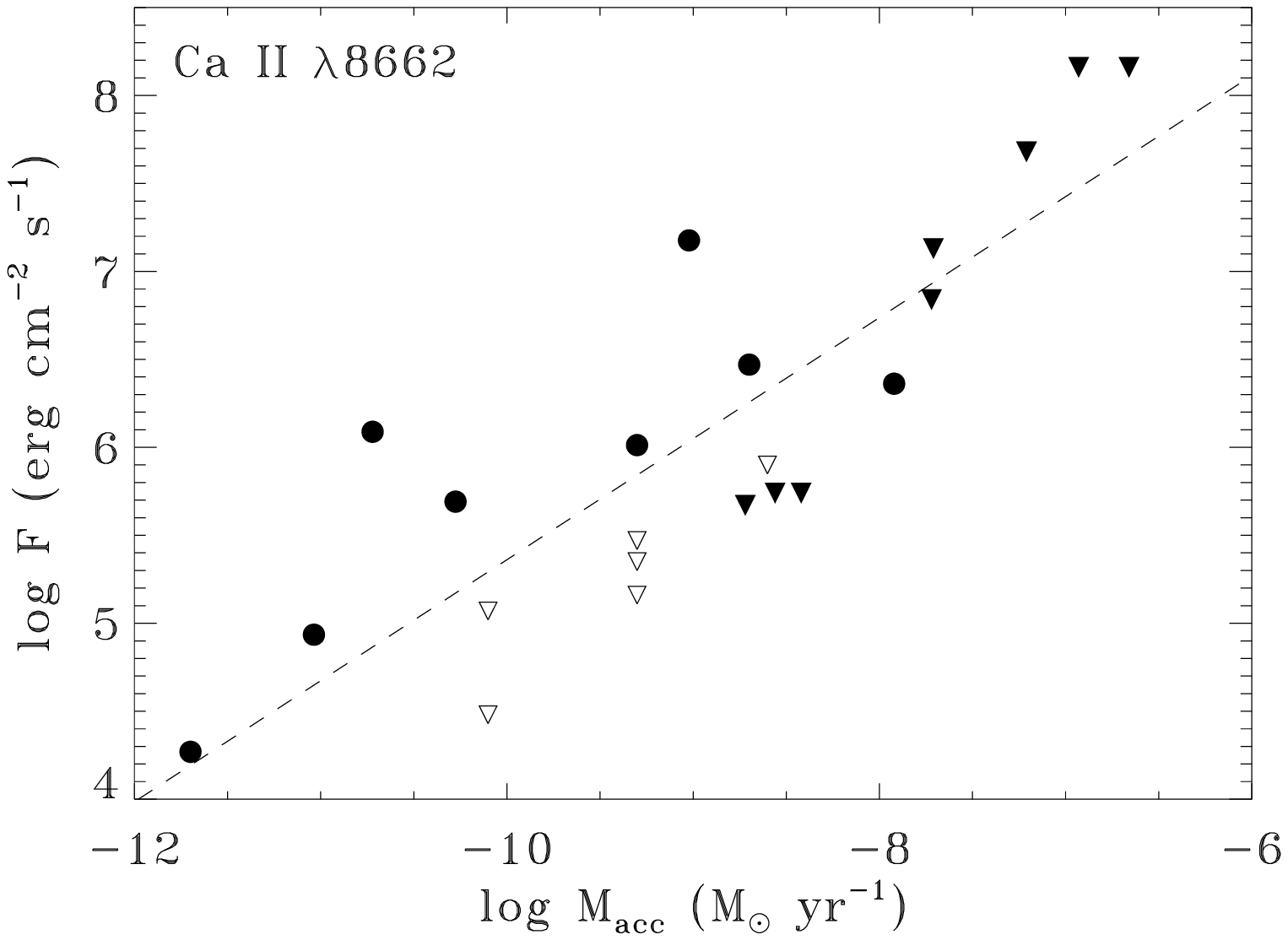}
\epsscale{0.5}
\plotone{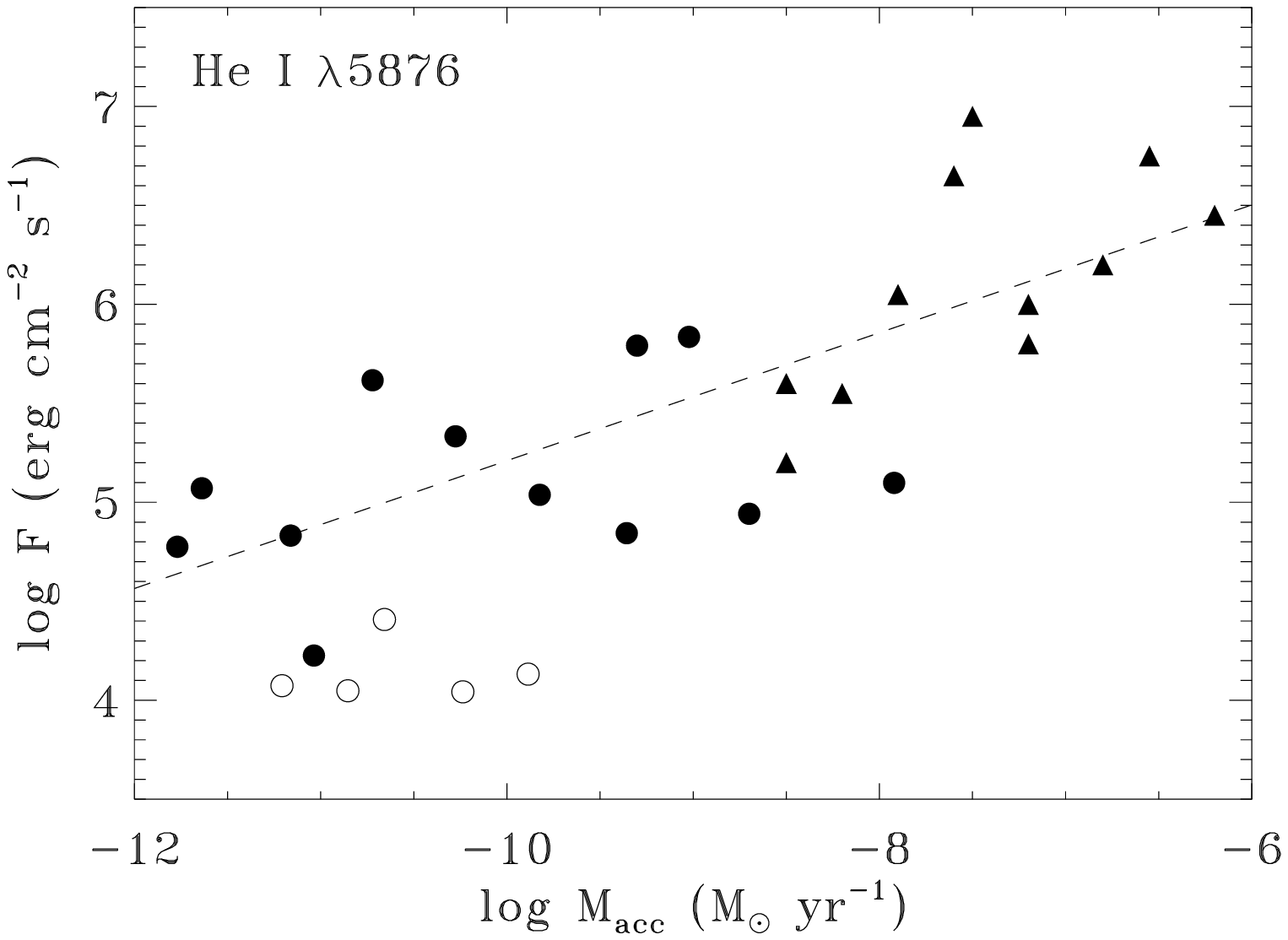}
\caption{\ion{Ca}{2} $\lambda8542$ (upper left) and \ion{Ca}{2} $\lambda8662$ (upper right) line flux versus mass accretion rate.  We compare our data (filled circles for accretors, empty circles for the five possible accretors) with that from Hartigan et al.~(2003, filled diamonds for \ion{Ca}{2} $\lambda8542$), Muzerolle et al. (1998, filled triangles for \ion{Ca}{2} $\lambda8542$ and \ion{He}{1} $\lambda5876$) and Mohanty et al. (2005, upside-down triangles for \ion{Ca}{2} $\lambda8662$).  The empty upside-down triangles on the upper right panel signify that the $\dot{M}$ was measured by H$\alpha$ modelling and is multiplied by 5.  The dashed lines in both panels show the best fit through the data.}
\end{figure}

\epsscale{1.0}


\begin{figure}
\plottwo{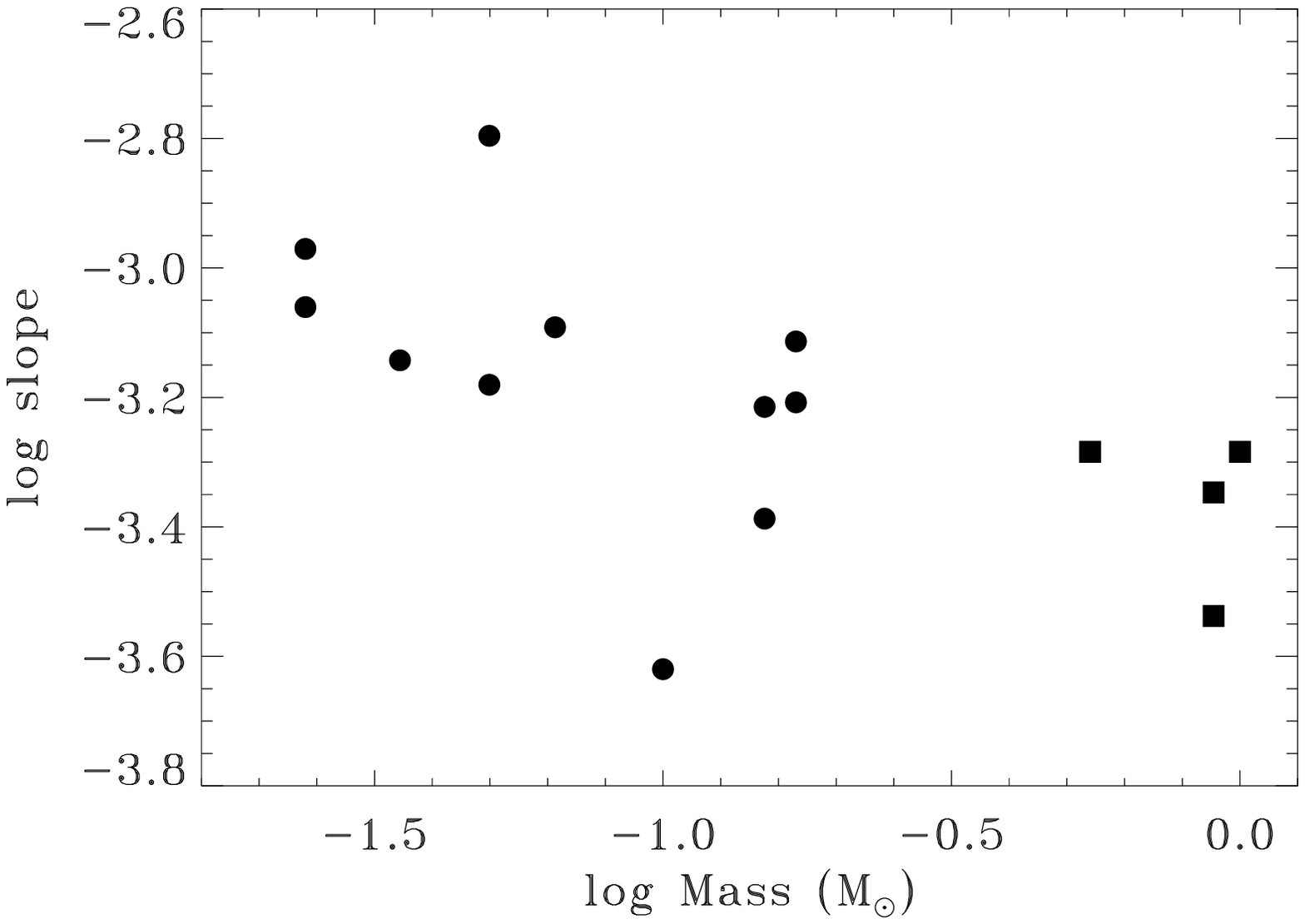}{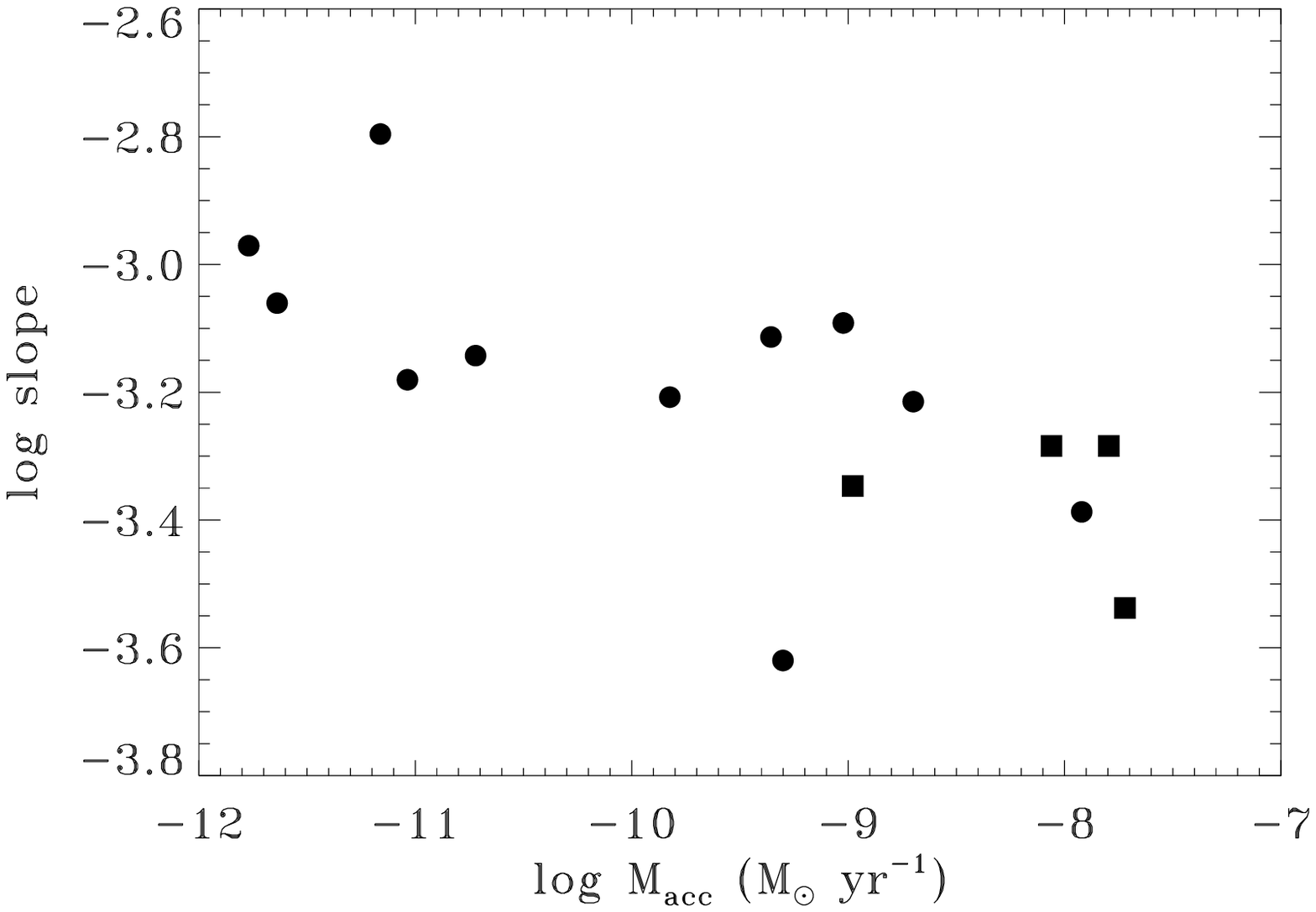}
\caption{The slope of the Balmer continuum versus stellar mass (left) and $\dot{M}$ (right).  The circles are from our {\it Keck I}/LRIS data, the squares from {\it HST}/STIS G430L spectra.  The slope of the Balmer continuum increases to smaller $M$ and $\dot{M}$, suggesting that it is produced at cooler gas.}
\end{figure}

\begin{figure}
\plottwo{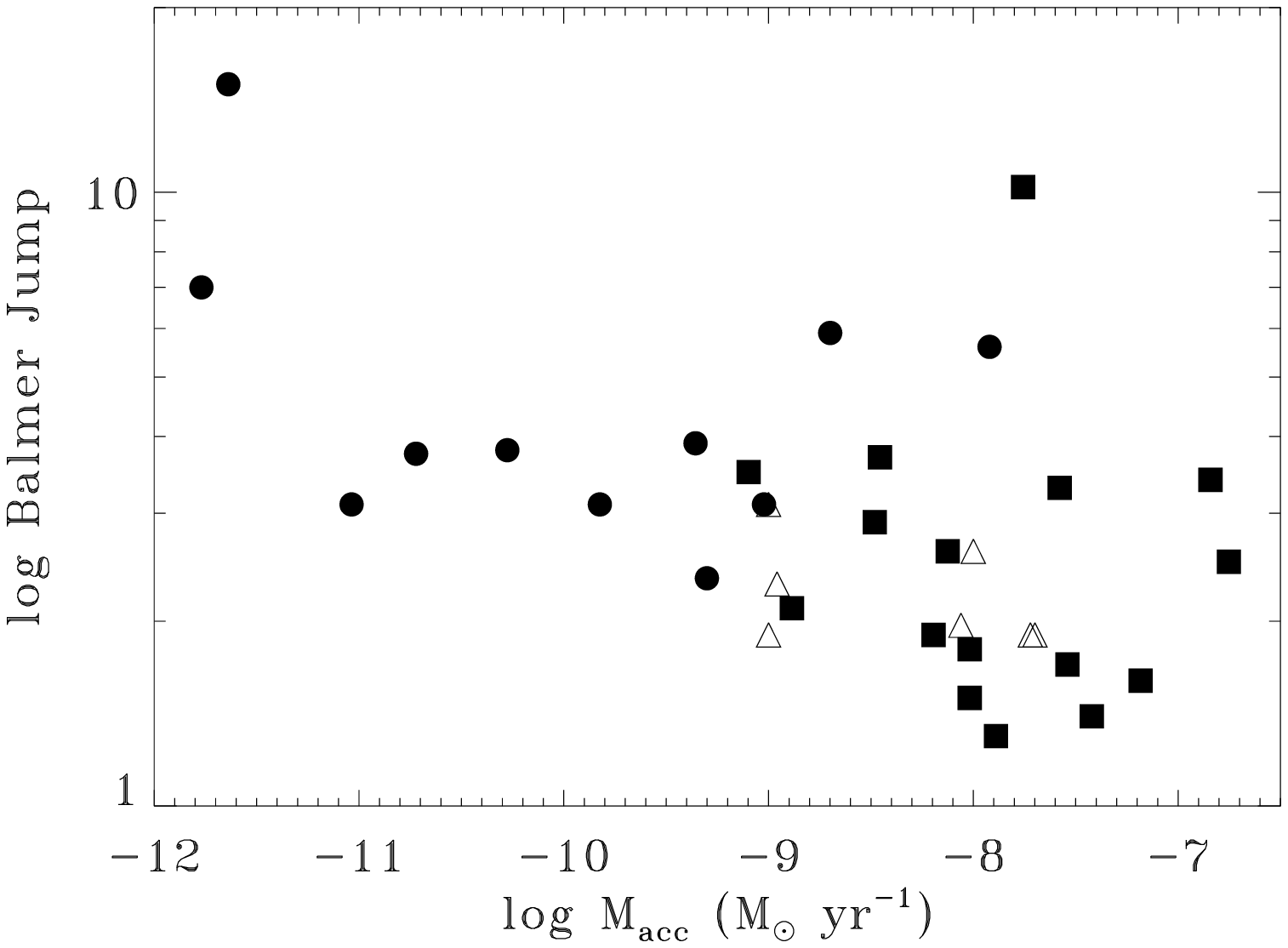}{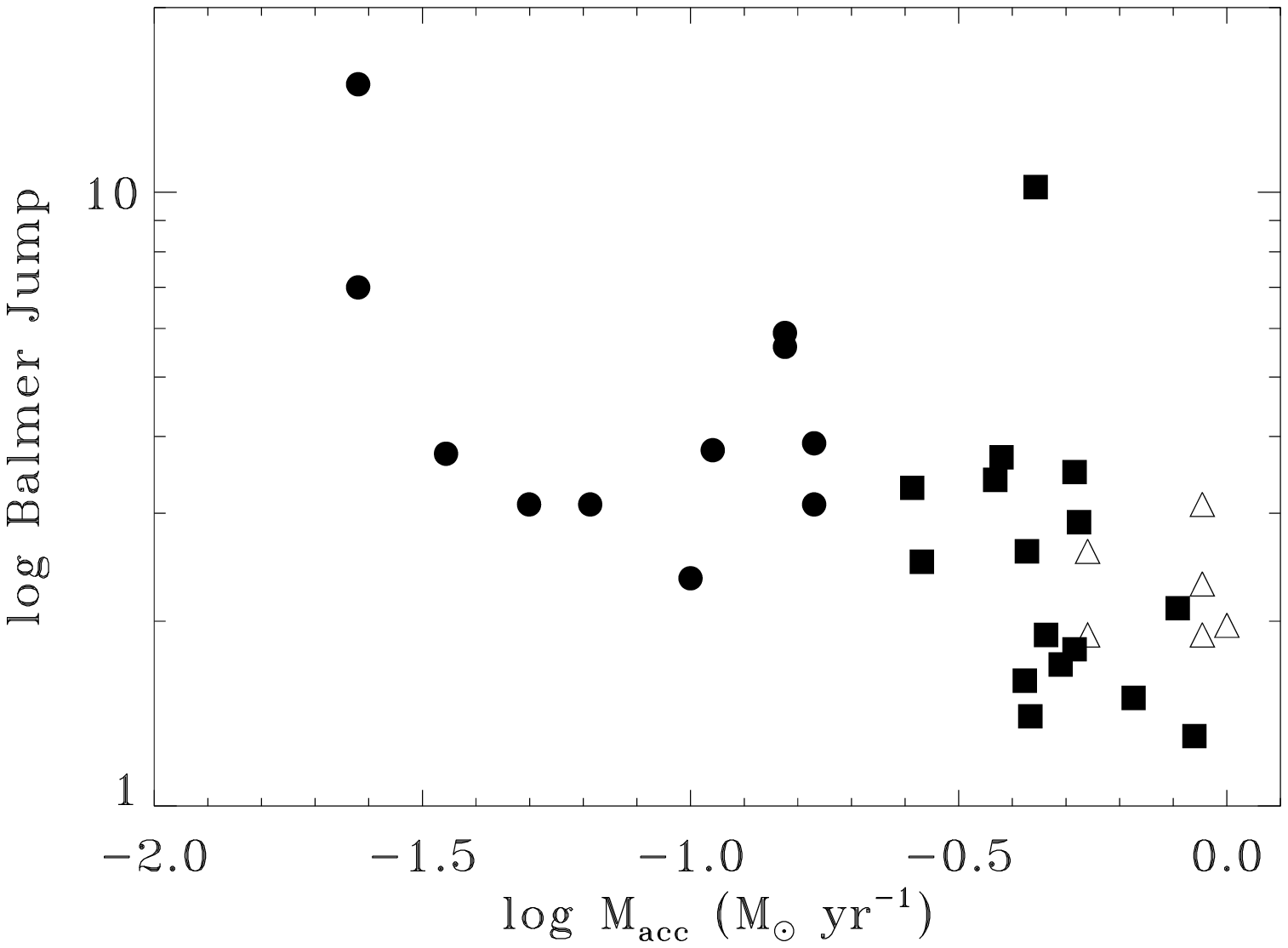}
\caption{The Balmer jump versus stellar mass (left) and $\dot{M}$ (right), as in Fig. 12.  The squares are from \citet{Gul98}.  The Balmer jump tends to increase to smaller $M$ and $\dot{M}$, suggesting that the accretion continuum is produced in gas with a lower electron density.}
\end{figure}

\clearpage
\pagebreak
\pagebreak

\begin{figure}[!b]
\epsscale{0.8}
\plotone{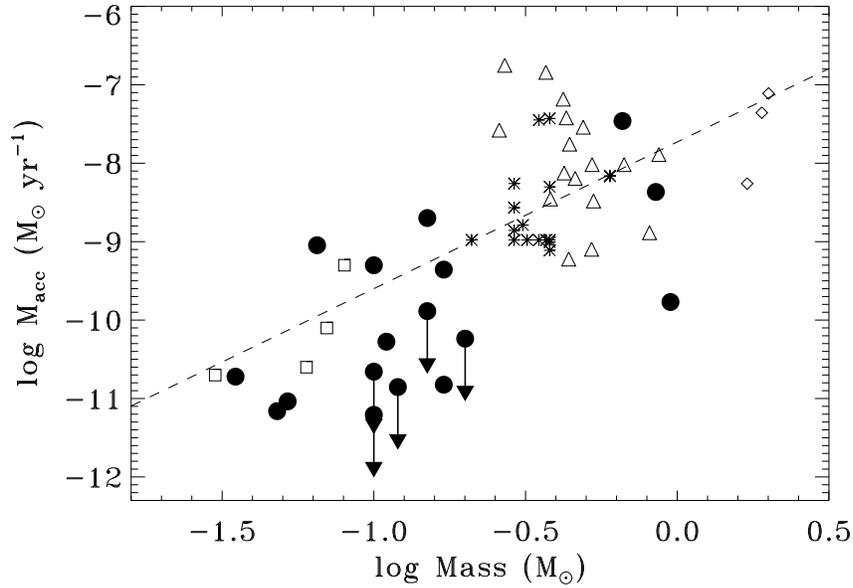}
\caption{Mass versus mass accretion rate for Taurus members with $A_V<2.5$ and measured accretion rates.  The data plotted are $\dot{M}$ measured from excess UV continuum emission in this work (filled circles), from excess UV continuum emission measured by Gullbring et al.~(1998, open triangles), optical veiling measured by Hartigan \& Kenyon (2003, asterisks), UV excess measured by Calvet et al. (2004, open diamonds), and H$\alpha$ profile modelling (open squares, multiplied by 5, Muzerolle et al. 2005).  The dashed line shows $\dot{M}\propto-7.7 M^{1.87}$. }
\end{figure}

\begin{figure}[t]
\plottwo{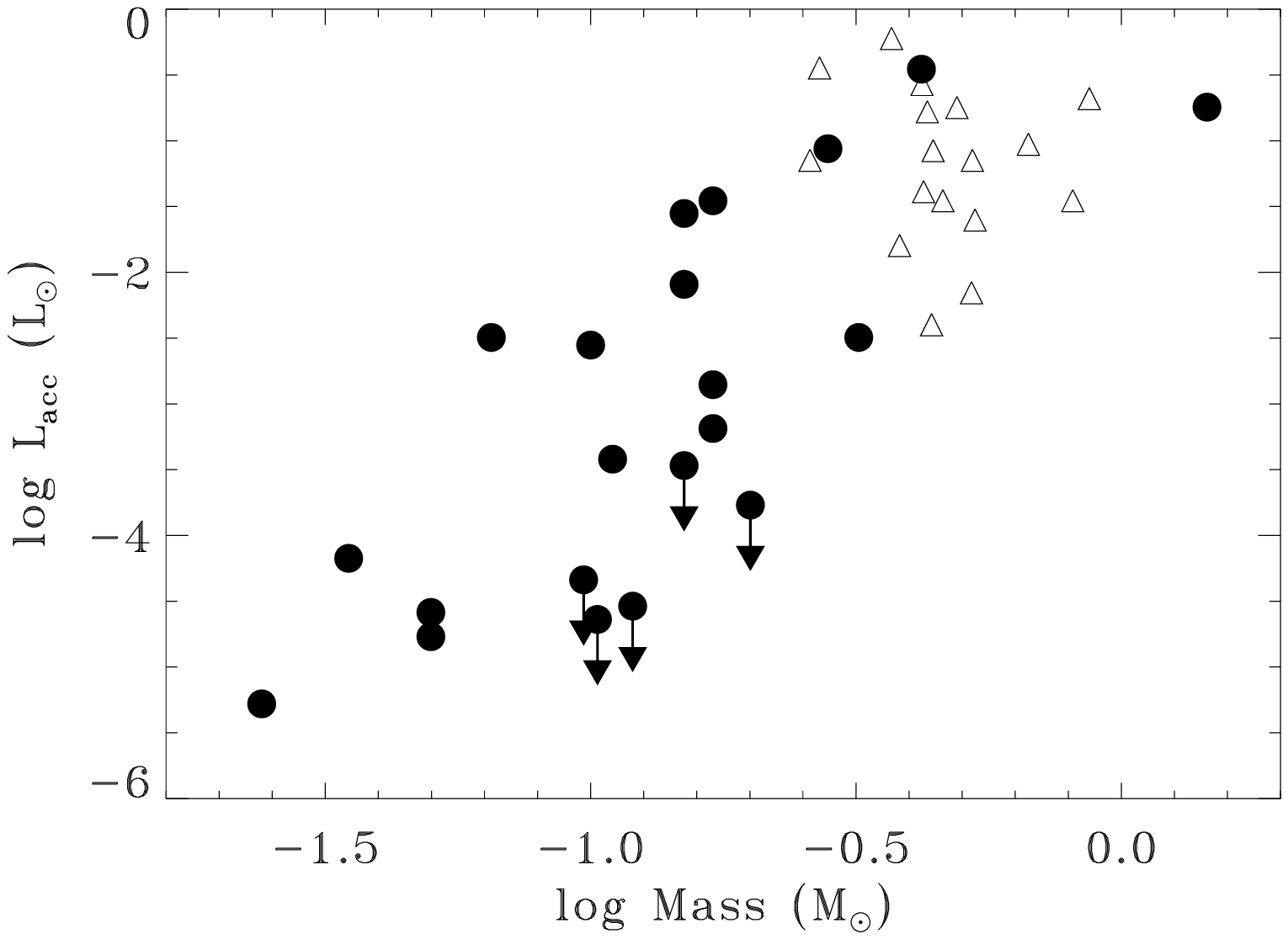}{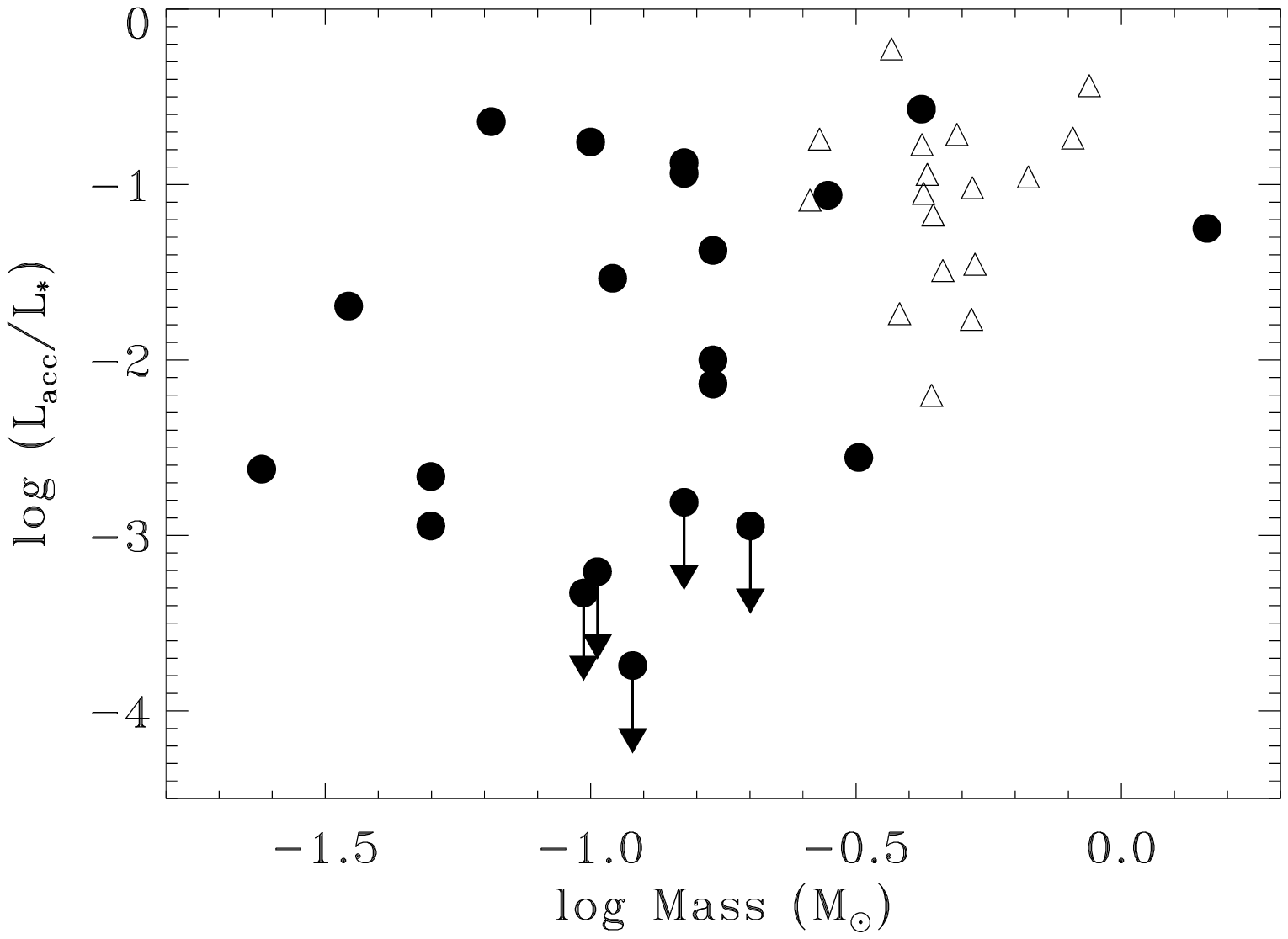}
\caption{The relationship between mass and $L_{acc}$ (left) and
  $\frac{L_{acc}}{L_{bol}}$ (right), with circles from our data and open triangles from \citet{Gul98}.  The sensitivity of measuring
  $L_{acc}$ and $\frac{L_{acc}}{L_{bol}}$  from the UV excess improves to lower
  masses.}
\end{figure}


\begin{thebibliography}{}
\bibitem[Alencar et al.(2005)]{Ale05}
Alencar, S.H.P., Basri, G., Hartmann, L., \& Calvet, N.  2005, A\&A, 440, 595

\bibitem[Allred et al.(2006)]{All06}
Allred, J.C., Hawley, S.L., Abbett, W.P, \& Carlsson, M.  2006, ApJ, 644, 484

\bibitem[Andrews \& Williams(2005)]{And05}
Andrews, S.M.,  \& Williams, J.P.  2005, ApJ, 631, 1134

\bibitem[Azevedo et al.(2006)]{Aze06}
Azevedo, R., Calvet, N., Hartmann, L., Folha, D.F.M., Gameiro, F., \& Muzerolle, J.  2006, A\&A, 456, 225

\bibitem[Baraffe et al.(1998)]{Bar98}
Baraffe, I., Chabrier, G., Allard, F., \& Hauschildt, P.H.  1998, A\&A, 337, 403


\bibitem[Basri \& Batalha(1990)]{Bas90}
Basri, G., \& Batalha, C.  1990, ApJ, 363, 654


\bibitem[Batalha et al.(1996)]{Bat96}
Batalha, C.C., Stout-Batalha, N.M., Basri, G., \& Terra, M.A.O.  1996, ApJS, 103, 211

\bibitem[Batalha et al.(2002)]{Bat02}
Batalha, C., Batalha, N.M., Alencar, S.H.P., Lopes, D.F., \& Duarte, E.S.  2002, ApJ, 580, 343

\bibitem[Beristain et al.(2001)]{Ber01}
Beristain, G., Edwards, S., \& Kwan, J.  2001, ApJ, 551, 1037

\bibitem[Bertout \& Genova(2006)]{Ber06}
Bertout, C., \& Genova, F.  2006, A\&A, 460, 499

\bibitem[Biller \& Close(2007)]{Bil07}
Biller, B.A., \& Close, L.M.  2007, ApJ, 669, L41

\bibitem[Bohlin et al.(1978)Bohlin, Savage, \& Drake]{Boh78}
Bohlin, R.C., Savage, B.D., \& Drake, J.F.  1978, ApJ, 224, 132

\bibitem[Bouvier et al.(2007)]{Bou07}
Bouvier, J., Alencar, S.H.P., Boutelier, T., Dougados, C., Balog, Z., Grankin, K., Hodgkin, S.T., Ibrahimov, M.A., Kin, M., Magakian, T. Yu., Pinte, C.  2007, A\&A, 463, 1017

\bibitem[Briceno et al.(1993)]{Bri93}
Briceno, C., Calvet, N., Gomez, M., Hartmann, L.W., Kenyon, S.J., \& Whitney, B.A.  1993, PASP, 105, 686


\bibitem[Briceno et al.(1998)]{Bri98}
Briceno, C., Hartmann, L., Stauffer, J., \& Martin, E.  1998, ApJ, 115, 2074

\bibitem[Briceno et al.(1999)]{Bri99}
Briceno, C., Calvet, N., Kenyon, S., \& Hartmann, L.  1999, ApJ, 118, 1354


\bibitem[Briceno et al.(2002)]{Bri02}
Briceno, C., Luhman, K.L., Hartmann, L., Stauffer, J.R., \& Kirkpatrick, J.D.  2002, ApJ, 580, 317


\bibitem[Calvet \& Gullbring(1998)]{Cal98}
Calvet, N., \& Gullbring, E.  1998, ApJ, 509, 802

\bibitem[Calvet et al.(2004)]{Cal04}
Calvet, N., Muzerolle, J., Briceno, C., Hernandez, J., Hartmann, L.,
Saucedo, J.L., \& Gordon, K.D.  2004, \aj, 128, 1294

\bibitem[Cardelli et al.(1989)Cardelli, Clayton, \& Mathis]{Car89}
  Cardelli, J. A., Clayton, G. C., \& Mathis, J. S. 1989, \apj, 345,
  245

\bibitem[Cardini \& Cassatella(2007)]{Card07}
Cardini, D., \& Cassatella, A.  2007, ApJ, 666, 393

\bibitem[Carpenter et al.(2006)]{Car06}
Carpenter, J.M., Mamajek, E.E., Hillenbrand, L.A., \& Meyer, M.R.
2006, ApJ, 651, L49

\bibitem[Chauvin et al.(2004)]{Cha04}
Chauvin, G., Lagrange, A.-M., Dumas, C., Zuckerman, B., Mouillet, D., Song, I., Beuzit, J.-L., Lowrance, P.  2004, A\&A, 425, L29

\bibitem[Cieza et al.(2005)]{Cie05}
Cieza, L.A., Kessler-Silacci, J.E., Jaffe, D.T., Harvey, P.M., \&
Evans, N.J.  2005, ApJ, 635, 422

\bibitem[Clarke \& Pringle(2006)]{Cla06}
  Clarke, C.J., \& Pringle, J.E.  2006, MNRAS, 370, L10

\bibitem[D'Antona \& Mazzitelli(1998)]{Dan98}
D'Antona, F., \& Mazzitelli, I.  1998, ASPC, 134, 442

\bibitem[Eason et al.(1992)]{Eas92}
Eason, E. L. E., Giampapa, M. S., Radick, R. R., Worden, S. P., \& Hege, E. K. 1992, AJ, 104, 1161

\bibitem[Edwards et al.(2006)]{Edw06}
Edwards, S., Fischer, W., Hillenbrand, L.A., \& Kwan, J.  2006, ApJ, 646, 319

\bibitem[Feigelson \& Kriss(1981)]{Fei81}
Feigelson, E.D., \& Kriss, G.A.  1981, ApJ, 248, L35

\bibitem[Feigelson \& Nelson(1985)]{Fei86}
Feigelson, E.D., \& Nelson, P.I.  1985, ApJ, 293, 192

\bibitem[Giampapa et al.(1982)]{Gia82}
Giampapa, M.S., Worden, S.P., \& Linsky, J.L.  1982, ApJ, 258, 740

\bibitem[Gizis et al.(2000a)]{Giz00}
Gizis, J.E., Monet, D.G., Reid, I.N., Kirkpatrick, J.D., Liebert, J., Williams, R.  2000, AJ, 120, 1085

\bibitem[Gizis et al.(2000b)]{Giz00a}
Gizis, J.E., Monet, D.G., Reid, I.N., Kirkpatrick, J.D., Burgasser, A.J.  2000, MNRAS, 311, 385


\bibitem[Gizis(2002)]{Giz02}
Gizis, J.E.  2002, ApJ, 575, 484

\bibitem[Gizis et al.(2002)]{Giz02a}
Gizis, J.E., Reid, I.N., \& Hawley, S.L.  2002, AJ, 123, 3356

\bibitem[Gizis et al.(2007)]{Giz08}
Gizis, J.E., Jao, W.-C., Subasavage, J.P., \& Henry, T.J.  2007, ApJ, 669, L45

\bibitem[Grosso et al.(2007)]{Gro07}
Grosso, N., et al.  2007, A\&A, 468, 391

\bibitem[Guieu et al.(2006)]{Gui06}
Guiei, S., Dougados, C., Monin, J.-L., Magnier, E., \& Martin, E.L.  2006, A\&A, 446, 485

\bibitem[Guieu et al.(2007)]{Gui07}
Guieu, S., et al.  2007, A\&A, 465, 855

\bibitem[Gullbring et al.(1998)]{Gul98}
Gullbring, E., Hartmann, L., Briceno, C., \& Calvet, N.  1998, \apj, 492, 323

\bibitem[Gullbring et al.(2000)]{Gul00}
Gullbring, E., Calvet, N., Muzerolle, J., \& Hartmann, L.  2000, ApJ,
544, 927


\bibitem[Hamann \& Persson(1989)]{Ham90}
Hamann, F., \& Persson, S. E. 1989, ApJ, 339, 1078 


\bibitem[Hamann \& Persson(1992)]{Ham92}
Hamann, F., \& Persson, S. E. 1992, ApJS, 82, 247 

\bibitem[Hamann et al.(1994)]{Ham94}
Hamann, F.  1994, ApJS, 93, 485



\bibitem[Hartigan et al.(1991)]{Har91}
Hartigan, P., Kenyon, S.J., Hartmann, L., Strom, S.E., Edwards, S., Welty, A.D., \& Stauffer, J.  1991, ApJ, 382, 617

\bibitem[Hartigan et al.(1995)]{Har95}
Hartigan, P., Edwards, S., \& Ghandour, L.  1995, 452, 736

\bibitem[Hartigan \& Kenyon(2003)]{Har03}
Hartigan, P., \& Kenyon, S.J.  2003, ApJ, 583, 334

\bibitem[Hartigan et al.~(2004)]{Har04}
Hartigan, P., Edwards, S., \& Pierson, R.  2004, ApJ, 609, 261

\bibitem[Hartmann et al.(1994)]{Har94}
Hartmann, L., Hewett, R., \& Calvet, N.  1994, ApJ, 426, 669


\bibitem[Hawley \& Pettersen(1991)]{Haw91}
Hawley, S. L., \& Pettersen, B. R. 1991, ApJ, 378, 725 


\bibitem[Henize(1954)]{Hen54}
Henize, K.G.  1954, ApJ, 119, 459


\bibitem[Herbig et al.(1972)]{Her72}
Herbig, G.H., \& Kameswara, R.N.  1972, ApJ, 174, 401

\bibitem[Herbig \& Soderblom(1980)]{Her80}
Herbig, G.H., \& Soderblom, D.R.  1980, ApJ, 242, 628

\bibitem[Herbig et al.(1986)]{Her86}
Herbig, G.H., Vrba, F.J., \& Rydgren, A.E.  1986, AJ, 91, 575

\bibitem[Herczeg et al.(2004)]{Her04}
Herczeg, G.J., Wood, B.E., Linsky, J.L., Valenti, J.A., Johns-Krull, C.M.
2004, \apj, 607, 369

\bibitem[Herczeg et al.(2005)]{Her05}
Herczeg, G.J., et al.  2005, ApJ, 129, 2777

\bibitem[Herczeg et al.(2006)]{Her06}
Herczeg, G.J., Linsky, J.L., Walter, F.M., Gahm, G.F., \& Johns-Krull, C.M.  2006, ApJS, 165, 256

\bibitem[Hillenbrand et al.(2007)]{Hil07}
Hillenbrand, L.A., Bauermeister, A., \& White, R.J.  2007, proc. of {\it Cool Stars, Stellar Systems, and the Sun XIC}, APS Conference Series, eds. G. van Belle.

\bibitem[Jayawardhana et al.(2003)]{Jay03}
Jayawardhana, R., Mohanty, S., \& Basri, G.  2003, ApJ, 592, 282

\bibitem[Johns-Krull \& Basri(1997)]{Joh97}
Johns-Krull, C.M., \& Basri, G.  1997, ApJ, 474, 433


\bibitem[Johns-Krull et al.(2000)Johns-Krull, Valenti, \& Linsky]{Joh00}
  Johns-Krull, C. M., Valenti, J. A., \& Linsky, J. L.  2000, \apj,
  539, 815

\bibitem[Johns-Krull(2007)]{Joh07}
Johns-Krull, C.M.  2007, ApJ, 664, 975

\bibitem[Kastner et al.~(2002)]{Kas02}
Kastner, J.H., Huenemoerder, D.P., Schulz, N.S., Canizares, C.R., \& Weintraub, D.A.  2002, ApJ, 567, 434

\bibitem[Kenyon et al.(1994)]{Ken94}
Kenyon, S.J., Dobrzycka, D., \& Hartmann, L.W.  1994, AJ, 108, 1872


\bibitem[Kenyon \& Hartmann(1995)]{Ken95}
Kenyon, S.J., \& Hartmann, L.  1995, ApJS, 101, 117



\bibitem[Kirkpatrick et al.(1993)]{Kirk93}
Kirkpatrick, J.D., Kelly, D.M., Rieke, G.H., Liebert, J., Allard, F., \&
Wehrse, Rainer.  1993, ApJ, 402, 643

\bibitem[Kirkpatrick et al.(1999)]{Kir99}
Kirkpatrick, J.D., et al. 1999, ApJ, 519, 802


\bibitem[Kraus et al.(2006)]{Kra06}
Kraus, A.L., White, R.J., \& Hillenbrand, L.A.  2006, ApJ, 649, 306

\bibitem[Kurosawa et al.(2006)]{Kur06}
  Kurosawa, R., Harries, T.J., \& Symington, N.H.  2006, MNRAS, 370, 580

\bibitem[Leinert et al.(1993)]{Lei93}
Leinert, C., Zinnecker, H., Weitzel, N., Christou, J., Ridgway, S.T., Jameson, R., Haas, M., Lenzen, R.  1993, A\&A, 278, L129


\bibitem[Luhman et al.(2003)]{Luh03}
Luhman, K.L., Stauffer, J.R., Muench, A.A., Rieke, G.H., Lada, E.A.,
Bouvier, J., \& Lada, C.J.  2003, ApJ, 593, 1093

\bibitem[Luhman(2004)]{Luh04}
Luhman, K.L.  2004, ApJ, 617, 1216

\bibitem[Luhman et al.~(2006)]{Luh06}
Luhman, K.L., et al.  2006, ApJ, 647, 1180


\bibitem[Luhman et al.(2007)]{Luh07}
Luhman, K.L., Adame, L., D'Alessio, P., Calvet, N., McLeod, K.K., Bohac, C.J., Forrest, W.J., Hartmann, L., Sargent, B., \& Watson, D.M.  2007, ApJ, 666, L1219

\bibitem[Mamajek(2005)]{Mam05}
Mamajek, E.E.  2005, ApJ, 634, 1385

\bibitem[Massey et al.(1988)]{Mas88}
Massey, P., Strobel, K., Barnes, J.V., \& Anderson, E.  1988, ApJ, 328, 315

\bibitem[Mauas \& Falchi(1996)]{Mau96}
Mauas, P.J.D., \& Falchi, A.  1996, A\&A, 310, 245

\bibitem[McCabe et al.(2006)]{McC06}
  McCabe, C., Ghez, A.M., Prato, L., Duchene, G., Fisher, R.S., \& Telesco,
  C.  2006, ApJ, 636, 932

\bibitem[McCarthy et al.(1998)]{McC98}
McCarthy, J. K., Cohen, J. G., Butcher, B., Cromer, J., Croner, E., Douglas, W. R., Goeden, R. M., Grewal, T., Lu, B., Petrie, H. L., Weng, T., Weber, B., Koch, D. G., \& Rodgers, J. M. 1998 SPIE, 3355, 81


\bibitem[Mohanty et al.(2005)]{Moh05}
Mohanty, S., Jayawardhana, R., \& Basri, G.  2005, ApJ, 626, 498

\bibitem[Muzerolle et al.(1998)]{Muz98} 
 Muzerolle, J., Hartmann, L., \& Calvet, N.  1998, AJ, 116, 2965

\bibitem[Muzerolle et al.(2000)]{Muz00} 
 Muzerolle, J., Calvet, N., Briceno, C., Hartmann, L., \& Hillenbrand, L.  2000, ApJ, 535, L47


\bibitem[Muzerolle et al.(2001)]{Muz01}
Muzerolle, J., Calvet, N., \& Hartmann, L.  2001, ApJ, 550, 944


\bibitem[Muzerolle et al.(2003)]{Muz03} 
Muzerolle, J., Hillenbrand, L., Calvet, N., Briceno, C., \& Hartmann, L.  2003, ApJ, 592, 266


\bibitem[Muzerolle et al.(2005)]{Muz05}
Muzerolle, J., Luhman, K.L., Briceno, C., Hartmann, L., \& Calvet, N.  2005, ApJ, 625, 906

\bibitem[Natta et al.(2004)]{Nat04}
Natta, A., Testi, L., Muzerolle, J., Randich, S., Comeron, F., \&
Persi, P.  2004, A\&A, 424, 603

\bibitem[Natta et al.(2006)]{Nat06}
Natta, A., Testi, L., \& Randich, S.  2006, A\&A, 452, 245

\bibitem[Oke et al.(1990)]{Oke90}
Oke, J.B.  1990, AJ, 99, 1621

\bibitem[Oke et al.(1995)]{Oke95}
Oke, J.B., Cohen, J.G., Carr, M., Cromer, J., Dingizian, A. \&
Harris, F. H. 1995, PASP, 107, 375O

\bibitem[Rauscher \& Marcy(2006)]{Rau06}
Rauscher, E., \& Marcy, G.W.  2006, PASP, 118, 617

\bibitem[Reipurth et al.(1996)]{Rei93}
Reipurth, B., Pedrosa, A., \& Lago, M.T.V.T.  1996, A\&AS, 120, 229

\bibitem[Rucinski \& Krautter(1983)]{Ruc83}
Rucinski, S.M., \& Krautter, J.  1983, A\&A, 121, 217

\bibitem[Rydgren \& Vrba(1983)]{Ryd83}
Rydgren, A.E., \& Vrba, F.J.  1983, AJ, 88, 1017

\bibitem[Schaefer et al.(2006)]{Sch06}
Schaefer, G.H., Simon, M., Beck, T.L, Nelan, E., \& Prato, L.  2006, ApJ, 132, 2618

\bibitem[Scholz et al.(2005)]{Sch05}
Scholz, A., Jayawardhana, R., \& Brandeker, A.  2005, ApJ, 629, L41

\bibitem[Scholz \& Jayawardhana(2006)]{Scho06}
Scholz, A., \& Jayawardhana, R.  2006, ApJ, 638, 1056

\bibitem[Short et al.(1997)]{Sho97}
Short, C.I., Doyle, J.G., \& Byrne, P.B.  1997, A\&A, 324, 196

\bibitem[Simon \& Prato(1995)]{Sim95}
  Simon, M., \& Prato, L.  1995, ApJ, 450, 824

\bibitem[Slesnick et al.(2006)]{Sle06}
Slesnick, C.L., Carpenter, J.M., Hillenbrand, L.A., \& Mamajek, E.E.  2006, ApJ, 132, 2665

\bibitem[Smak(1964)]{Smak64}
Smak, J.  1964, ApJ, 139, 1095

\bibitem[Stelzer et al.(2007)]{Ste07}
Stelzer, B., Scholz, A., \& Jayawardhana, R.  2007, astro-ph:/0707.1754



\bibitem[Valenti et al.(1993)]{Val93}
Valenti, J.A., Basri, G., \& Johns, C.M.  1993, \apj, 106, 2024

\bibitem[Valenti et al.(2003)]{Val03}
Valenti, J.A., Fallon, A.A., \& Johns-Krull, C.M.  2003, ApJS, 147, 305

\bibitem[Walter \& Kuhi(1981)]{Wal81}
Walter, F.M., \& Kuhi, L.V.  1981, ApJ, 250, 254

\bibitem[Webb et al.~(1999)]{Web99}
Webb, R.A., Zuckerman, B., Platais, I., Patience, J., White, R.J., Schwartz, M.J., McCarthy, C.  1999, ApJ, 512, L63

\bibitem[Whelan et al.(2007)]{Whe07}
  Whelan, E.T., Ray, T.P., Randich, S., Bacciotti, F., Jayawardhana, R.,
  Testi, L., Natta, A., \& Mohanty, S.  2007, ApJ, 659, L45


\bibitem[White \& Ghez(2001)]{Whi01}
White, R.J., \& Ghez, A.M.  2001, ApJ, 556, 265


\bibitem[White \& Basri(2003)]{Whi03}
White, R.J., \& Basri, G.  2003, ApJ, 582, 1109


\bibitem[White \& Hillenbrand(2004)]{Whi04}
White, R.J., \& Hillenbrand, L.A.  2004, \apj, 616, 998

\bibitem[Whittet et al.(2001)]{Whittet01}
Whittet, D.C.B., Gerakines, P.A., Hough, J.H., Shenoy, S.S.  2001, ApJ, 547, 872

\bibitem[Wichmann et al.(1998)]{Wic98}
Wichmann, R., Bastian, U., Krautter, J., Jankovics, I., \& Ricinski, S.M.  1998, MNRAS, 301, L39

\bibitem[Yang et al.(2007)]{Yan07}
Yang, H., Johns-Krull, C.M., \& Valenti, J.A.  2007, AJ, 133, 73

\end{thebibliography}
\end{document}